\documentclass[12pt,preprint]{aastex}

\newcommand{\kpc}{$\, {\rm kpc}$}
\newcommand{\kms}{$\, {\rm km\,s^{-1}}$}

\newcommand{\msun}{\,$M_{\odot\,\,}$}

\newcommand{\isot}{$\rm{^{12}C/^{13}C}\,\,$}
\newcommand{\no}{$\rm{^{14}N/^{16}O}\,\,$}
\newcommand{\feh}{${\rm [Fe/H]\,\,}$}
\newcommand{\sgb}{$\rm SGB\,\,$}

\begin{document}

\def\newpage{\vfill\eject}
\def\vs{\vskip 0.2truein}
\def\pp{\parshape 2 0.0truecm 16.25truecm 2truecm 14.25truecm}
\def\fun#1#2{\lower3.6pt\vbox{\baselineskip0pt\lineskip.9pt
  \ialign{$\mathsurround=0pt#1\hfil##\hfil$\crcr#2\crcr\sim\crcr}}}
\def\core{{\rm core}}
\def\min{{\rm min}}
\def\max{{\rm max}}
\def\kpc{{\rm kpc}}
\def\esc{{\rm esc}}
\def\crit{{\rm crit}}
\def\pc{{\rm pc}}
\def\kms{\,{\rm km}\,{\rm s}^{-1}}
\def\cbh{{\rm cbh}}
\def\bh{{\rm bh}}
\def\df{{\rm df}}
\def\bulge{{\rm bulge}}

\title{Abundance Anomalies and Rotational Evolution of Low Mass Red
Giants: A Maximal Mixing Approach}

\author{Julio Chanam\'e, Marc Pinsonneault, \& Donald M. Terndrup}
\affil{{}Department of Astronomy, The Ohio State University, Columbus,
OH 43210, USA}


\begin{abstract}

We use a fully self-consistent evolutionary code to follow the
rotational evolution of low mass red giants, making a comprehensive
attempt to assess the role of rotationally induced mixing in the
development of abundance anomalies in giants with a range of masses
and metallicities in stellar clusters and the field.  We adopt a
maximal mixing approach with reasonable initial conditions of angular
momentum distribution and main sequence rotation rates as a function
of stellar type.  Unlike most previous work, we do not focus on the
determination of combinations of mixing rate and depth that reproduce
the data on a particular stellar type.  Instead, we concentrate on the
more fundamental problem of the simultaneous reproduction of the
patterns of CNO surface abundances in both Population I and Population
II giants using the same physics and models.  We follow and discuss
the essential physics of rotational mixing in terms of the structural
and angular momentum evolution along the red giant branch (RGB).

A general result of all our models is that rotational mixing, although
present in small amounts, is inefficient on the lower RGB
independently of any inhibiting effect of $\mu$-barriers.  Therefore,
the lack of well mixed stars before the luminosity of the RGB bump in
globular clusters and the field does not constitute unquestionable
evidence for the inhibition of mixing by $\mu$-barriers.  Instead, we
argue that the rapid disappearance of the RGB bump as soon as mixing
is allowed to penetrate $\mu$-barriers is what actually constitutes
the first solid evidence of such inhibition.

Maximal mixing models with differentially rotating envelopes are able
to reproduce the \isot data on M67 giants with initial rotation rates
adequate to their progenitors, but fail to do so for open clusters of
larger turnoff mass as well as for metal-poor giants in the field and
globular clusters.  Possible solutions are discussed.  Our favored
scenario is one in which the overall strength of canonical extra
mixing has been underestimated by existent derivations, but which
additionally needs to be coupled with a much lower efficiency for
rotational mixing among the rapidly rotating open cluster giants than
in the slowly rotating ones in the field and globular clusters.  We
hypothesize that this last requirement is provided by the interaction
between convection and rotation in the envelopes of giants, in the
sense that rapidly rotating stars would develop much shallower angular
velocity profiles in their envelopes than do slowly rotating stars.

\end{abstract}

\keywords{Galaxy: Globular Clusters: General -- stars: abundances --
stars: evolution -- stars: interiors -- stars: rotation}

\setcounter{footnote}{0}
\renewcommand{\thefootnote}{\arabic{footnote}}

\section{Introduction}

Theoretical models of stellar structure and evolution have
traditionally neglected rotation, even though stars have been known to
rotate since the 17th century \citep{gal1613}.  The reasons for this
are essentially three: the remarkable agreement of
standard\footnote{By standard we mean non-rotating, spherically
symmetric models, where magnetic fields play no role, and the only
mixing agent is convection.} models with large quantities of
observational data; the fact that, typically, rotation amounts for
just a small perturbation on the structure of stars; and, finally,
because of the much larger complexity involved in modeling objects not
in spherical symmetry.  However, modern observations, particularly
regarding anomalous patterns of chemical abundances, clearly require
additional physics.  Convection, the only mechanism of internal mixing
contemplated by standard models, is not able alone to account for the
large and varied levels of mixing seen in stars across different
evolutionary stages, and stellar rotation probably constitutes the
most likely missing ingredient: because of the varying gravity on an
equipotential surface, a rotating star can not simultaneously satisfy
both hydrostatic and radiative equilibrium \citep{von24}, and
large-scale internal motions (i.e., mixing) inevitably arise in order
to compensate the imbalance \citep{edd25,edd29,vog25}.  In this work,
we address the problem of the surface abundance
anomalies\footnote{Abundance anomalies studied in this work involve
stellar material that has gone through nuclear reactions and is
transported to the stellar surface, and should not be confused with
anomalies that can be produced by gravitational settling and radiative
levitation.  For a recent account of such effects, which can be
important in turnoff and subgiant stars, see \citet{mich04}.} in
low-mass (0.8-2\msun) red giants using stellar models that
incorporate rotation and its extra mixing properties.  By combining
observational constraints on the angular momentum properties of the
Sun and other stars in various evolutionary stages, we develop a
``maximal rotational mixing'' scenario intentionally aimed to identify
the advantages and limitations of this mechanism to reproduce the well
documented anomalies in the CNO surface abundances of first-ascent
giants in the field and stellar clusters.

In standard models, the only opportunity for the surface convection
zone of first ascent red giants to reach nuclear processed material is
known as the first dredge-up \citep{iben65}.  As the star moves from
the main sequence (MS) to the red giant branch (RGB), the base of the
convective envelope, while staying roughly at the same physical
radius, deepens considerably in the mass coordinate.  At its point of
maximum penetration, the convective envelope encompasses about 75\% of
the total mass of a 1.3\msun star of solar metallicity.  This first
dredge-up ends when the base of the convective envelope has reached
its maximum depth and then starts to retreat (again, in the mass
coordinate).  After this point in the evolution, no changes in the
surface abundances are produced by standard models.

Observational evidence for mixing beyond the standard model is rich,
coming from measured abundances of elements such as helium, lithium,
carbon, nitrogen, oxygen, sodium, etc., and is constantly and rapidly
growing.  Although in varying amounts, giants in both the cluster and
field environments show clear signs of non-standard mixing.  Red
giants in different stellar systems and environments provide
information and clues on the mixing processes that are complementary
to each other and, therefore, a serious attempt to understand the
evolution of their surface abundances must always account for the
entire phenomenology.  Reviews on this subject can be found in
\citet{kra94}, \citet{pin97}, and \citet{sal03}.

We defer to \S\,2 a review of the observational evidence for extra
mixing in giants in stellar clusters and the field, which the reader
familiar with the subject may skip.  Here we only point out the marked
contrast between the anomalous surface abundances seen in red giants
that belong to different Galactic populations: while Population I
giants in open clusters ($1.3$\msun and above) exhibit very mild
levels of extra mixing, the much older, less massive Population II
giants in globular clusters (0.8 to 0.9\msun) show largely anomalous
patterns of surface abundances, in clear disagreement with standard
models.

Abundance anomalies developing as a function of increasing luminosity
along the RGB, primarily displayed by the lightest species and now
seen both in clusters and the field, offer no doubt that some of the
non-standard changes must be evolutionary in nature and intrinsic to
stars.  Also, trends of the CNO abundances with metallicity are
consistent with an evolutionary scenario: due to their lower
opacities, more metal-poor stars are hotter at any given depth than
their metal-rich counterparts, placing the regions of CN and
ON-processing closer to the surface and hence making it easier to
transport nuclear processed material to the surfaces.

On the other hand, the heavier species involved in anomalies in red
giants, Na, Mg, and Al, do not show a clear evolution with position
along the RGB.  Furthermore, the large star-to-star abundance
variations among MS stars and early subgiants in globular clusters (in
contrast with field stars), as well as trends involving species that
burn at temperatures most likely higher than those attainable by first
ascent giants, indicate the presence of inhomogeneities in the cluster
material which must be primordial to the current giants and dependent
on the environment.

Hence, it has become increasingly evident that these two scenarios
(evolutionary and primordial; \citealt{kra94}) for the origin of the
abundance anomalies are not mutually exclusive, but rather they must
be simultaneously at work, i.e., that initial abundance differences
from star-to-star in the same globular cluster are further modified by
a non-standard mixing mechanism interior to the stars that operates
during the giant branch evolution (see \citealt{bri02} for the case of
M13).

Two broad classes of models of extra mixing have been developed in
order to account for the patterns of abundance anomalies in red
giants, the difference between them being whether they incorporate or
not, in a self-consistent manner, the underlying physical mechanism
responsible for the additional mixing.  The first class of models,
which could be generally referred to as {\it parametric diffusion
models}, simulate the extra mixing by adding a diffusion term to the
equation for the temporal variation of the abundances of chemical
species at any given mass shell inside the star.  In these models the
problem is reduced to the determination of two parameters that control
the mixing: depth and rate
\citep{swe79,smi92,cha95,was95,den96,den98,boo99,dentout00,wei00,
den03}.  Since these are essentially standard models on top of which
ad-hoc extra mixing is simulated, their fundamental shortcoming is
either one or both of the following: (1) they do not implement the
physical mechanism(s) driving the extra mixing, or (2) they do not
account for the feedback effects that the redistribution of chemical
composition can have on the structure and evolution of the star.  On
the other hand, they are helpful since they do identify the
temperature to which mixing must be able to reach inside the star.

The second class of models is based on the actual implementation of
mechanisms responsible for the non-standard mixing, all of which are
associated with different ways of transporting angular momentum in the
stellar interior.  These mechanisms are: magnetic fields
\citep{mes87,charb92,charb93,spruit99,spruit02,mae03,mae04}, gravity
waves
\citep{press81,kumar97,zahn97,kumar99,talon98,talon02,talon03,talon04},
and hydrodynamic instabilities triggered by differential rotation.
All these are known to interact with the angular velocity field inside
stars and thus are potential mixing agents.  However, while their
relative importance in different evolutionary stages is still a work
in progress, there are reasons to expect that both magnetic fields and
gravity waves do not play an important role in the transport of
angular momentum in red giants.

Being capable of exerting torques, magnetic fields have the potential
to be efficient agents in the redistribution of angular momentum
(although not necessarily chemicals).  However, the existence of fast
rotators among horizontal branch stars in globular clusters and the
field provides solid evidence that a high degree of differential
rotation is preserved during the RGB evolution \citep{sip00}, thus
arguing against not only magnetic fields, but any means of efficient
angular momentum transport.  Furthermore, the timescale for angular
momentum evolution (tied to the timescale for magnetic field
evolution), as indicated by observations of open clusters of different
ages, is of the order of $\sim 100\,$Myr, comparable to the entire
duration of the RGB phase.  Therefore, there is not enough time for
any magnetically induced mixing to operate significantly on this
evolutionary phase.  The importance of waves, on the other hand, is
always determined by the balance between excitation and damping.
Waves are excited near boundary layers, such as the base of the
convective envelope, but would hardly be able to reach the required
depths since, due to the large physical sizes of evolved stars, they
would be easily damped.

Stellar rotation was found long ago to be a promising non-standard
mechanism to explain various of the abundance anomalies introduced
before.  Extra mixing induced by rotation appears in two forms, one
associated with the global departure from spherical symmetry, and
another due to internal angular velocity gradients.  In the first, the
latitude-dependent centrifugal acceleration makes the effective
gravity to be smaller at equatorial zones than at the poles,
generating large-scale flows of matter in meridional planes to
compensate for the pressure and temperature imbalance between poles
and equator \citep{edd25,edd29,vog25}.  This meridional circulation
also transports chemicals and therefore mixes the deep interior of
stars \citep{swe79}.  The second type of rotational mixing arises when
large internal gradients of angular velocity inside radiative regions
trigger hydrodynamic instabilities that redistribute the angular
momentum and, in the process, mix the material \citep{end78,pin89}.
While meridional circulation can take place in any evolutionary state,
mixing via hydrodynamic instabilities is largely favored by the large
internal shears that develop with the structural changes in post-MS
stages.

The efficiency of both types of rotational extra mixing is an
increasing function of the rotation rate and, therefore, have
different impacts on different types of stars.  The dependence of
rotation rate on stellar type is commonly known as the Kraft curve
\citep{kra67}, and can be summarized as follows.  Early-type stars
show a wide range of surface rotation rates and can reach velocities
of $200\kms$, while late-type stars all rotate very slowly, almost
never exceeding $10\kms$.  The transition between these two regimes is
extremely steep and occurs throughout stellar type F, a fact probably
associated with the development of a deep surface convection zone (CZ)
and magnetized winds that carry away angular momentum very
efficiently.

The direction of this dichotomy between the rotation levels of early
and late-type stars, therefore, presents a fundamental puzzle for the
theory in light of the patterns of abundance anomalies discussed
before: how is it that the open cluster giants (whose progenitors
belonged to the rapidly rotating part of the Kraft curve) display much
milder levels of extra mixing than the globular cluster giants (whose
progenitors rotated very slowly), if the mechanisms for extra mixing
are increasing functions of the rotational velocity?  This basic
question, as far as we know, has not been addressed yet in the
literature available on this subject, and constitutes one of the goals
of the present work (\S\,6.4 and \S\,7.2).

Fully self-consistent models of rotational extra mixing are few, and
all of them are restricted to the lower MS
\citep{pin90,cha92,chab95a,chab95b,talchar98,pal03} and high-mass
stars \citep{meynet97,talon97,mae00}.  Such models are only now being
applied to the case of low-mass giants, although there have been two
studies that got close to that goal and are worth mentioning.  The
first of these is the seminal work of \citet{swe79}, who studied
classical meridional circulation as a mechanism of mixing
CNO-processed material into the envelope following a procedure that,
while being physically motivated, can be viewed as a variation of
parametric diffusion models.  Although not able to account for the
evolution of the angular momentum profile or feedback effects,
\citet{swe79} were capable of relating the interior rotation rates
that are needed to produce the necessary mixing with the rotation at
the surface.  This already made them realize the need for the
convective envelopes of low-mass giants to have differential rotation
in order to achieve any extra mixing through rotational effects, a
fact more recently demonstrated through the comparison of the rotation
rates of HB stars in globular clusters with those of their MS
progenitors \citep{sip00}.  The second of these semi self-consistent
models of rotational mixing in low-mass giants is that of
\citet{den03}, who adopted the theoretical framework of anisotropic
turbulent mixing of \citet{chabzahn92} and \citet{zahn92} to compute
the strength of extra mixing associated to rotation, although they
neglected the transport of angular momentum in radiative regions.
However, \citet{den03} adopt what in our view is an unrealistic
initial distribution of angular momentum (strong differential rotation
during the MS) and, even then, find that their diffusion coefficients
are still too small to account for the levels of mixing exhibited by
the data.  Finally, and most importantly, none of these works
addresses the issue of the existence of different patterns and levels
of extra mixing among stars of different populations.

In the present work, we make use of a stellar evolutionary code that
includes rotation and treats the transport of angular momentum and
chemical species in a fully self-consistent way, i.e., accounting for
the impact of such redistributions on the structure of the star at
each timestep.  Our main goals are the study of the evolution of the
surface abundances in rotating low-mass giants of all types and
environments using the same underlying physics, as well as the
understanding of the roles of the various ingredients that determine
the angular momentum and mixing histories in these stars.  Similarly,
since stars of different types show quite different rotational
properties, special emphasis is placed in the adoption of initial
conditions and assumptions in agreement with observational
constraints.  To do this, we evolve models of both Pop I and Pop II
stars from the MS turnoff to the tip of the RGB using initial rotation
rates adequate to their corresponding MS progenitors, and compare our
results to the data on the surface abundances of giants in open and
globular clusters as well as the field.

In particular, we ask ourselves the following question: assuming the
most favorable (though physically plausible) set of conditions for
extra mixing driven by rotation, is it possible to reproduce the CNO
abundance patterns seen in Pop I and Pop II red giants of all
environments (\S\,2)?  To answer this question we will study the
consequences of rotational mixing under conditions for {\it maximal
mixing}, among which we identify no transport of angular momentum and
neglecting any inhibiting effect of composition gradients (\S\,3).
Thus, rotational mixing will be essentially driven by the evolution of
the angular velocity field and hydrodynamic instabilities arising in
response to the structural changes occurring during the post-MS
evolution.  In \S\,3 we outline our treatment of stellar rotation and
the induced mixing, the input physics, and the initial conditions of
our models with maximal mixing.  In \S\,4 and \S\,5 we discuss the
structural and angular momentum evolution of the models and the
overall properties of the induced extra mixing.  The comparison with
observations is done in \S\,6, where we also introduce a working
hypothesis that could account for the different levels of mixing
between Pop I and Pop II giants simultaneously and without conflict
with the contrasting rotational regimes of their progenitors.
Finally, we summarize our findings and present our conclusions in
\S\,7.

\section{A review of the evidence for extra mixing in red giants}

Open clusters in the Galactic disk provide us with giants with a range
of ages (from a few tens of Myr to several Gyr) and, therefore, trace
the dependence of mixing on stellar mass.  The primary disadvantage of
using open cluster giants is that nature has not provided us with open
clusters that span the full ranges of metallicity and age.
Furthermore, open clusters typically contain relatively few stars, and
by extension there will be few members in short phases of evolution
such as the upper RGB.

Globular clusters, on the other hand, are very populous and in
principle can trace the entire RGB evolution of the surface abundances
of single-mass giants, providing information as a function of
luminosity.  Because of their distances from the Sun, however,
spectroscopic studies of globular cluster giants were typically
restricted to the most luminous stars, those close to the RGB tip,
although this situation has been progressively changing with the use of
8-10m class telescopes.  In addition, globular cluster giants are the
oldest and most metal-poor cases, thus probing the domains of the
lowest-mass and longer-lived giants that are possible to find.

Things are more difficult in the field, but nevertheless some authors
have managed to assemble fair samples of stars with well determined
evolutionary states and studied their abundance patterns, usually
relying on parallax data from Hipparcos and the work of previous
investigators that determined accurate physical parameters for these
stars.  Field giants sample predominantly old and low mass MS
precursors.  Such giants are relatively bright and provide data across
a range of abundances, but little direct information on the mass of
the precursors.  Nevertheless, field giants give us indications on the
evolution of the surface abundances as a function of environment
through the comparison of Population I field giants with those in the
sparse open clusters and of Population II field giants with those in
dense globular clusters.

Typical indicators of mixing have been lithium and the CNO species,
but also Na, Mg, Al, and others have been studied.  Burning at
temperatures higher than $\sim 2.5\times 10^6\,$K, Li is by far the
most fragile among these tracers, and measurements of its abundance
are a powerful test of models of MS stars of different spectral types
\citep{bal95,pin97}.  However, Li exhibits a complex MS depletion
pattern in Population I stars below about 1.5\msun, and more massive
stars are too hot to permit its measurement on the MS.  Thus, we focus
most of our attention on the surface abundances of the CNO species,
which are not expected to evolve during the MS lifetimes of low-mass
stars.  Since the reactions of the CNO cycle of hydrogen burning are
highly temperature sensitive (heavier species burn at higher
temperatures), any internal mixing reaches regions of CN-processing
more easily than regions of ON-processing, and, therefore, anomalies
in the ratios of \isot, C/N, and (C+N)/O probe, in that order,
progressively deeper mixing.

Finally, at even higher temperatures, proton-capture reactions in the
neon-sodium and magnesium-aluminum cycles can potentially produce
enhancements of Na and Al, but it is still a matter of debate whether
some of the anomalies among these elements are due to internal mixing
or environmental effects
\citep{den90,lan93,arm94,was95,pag97,den98,den01,denherwig03,den04}.

The surface abundance patterns among open cluster giants was
thoroughly studied by \citet{gil89} and can be summarized as follows.
The most sensitive indicator of CNO-processed material, the ratio of
the carbon isotopes, is consistent with standard first dredge-up
expectations (\isot $\sim 30$) only among massive red giants, that is,
those in open clusters with turnoff masses above $\sim$ 2\msun.  Red
giants in clusters below this turnoff mass exhibit values of $10
\lesssim$ \isot $\lesssim 20$, considerably lower than the standard
predictions.  Faint giants (i.e., close to the base of the RGB) in the
$\sim 4.5$ Gyr-old open cluster M67 \citep{gil91}, with masses around
1.3\msun, show \isot ratios consistent with standard models, but
anomalously low \isot values are obtained for all clump giants
(He-burning horizontal branch stars past the He-flash), suggestive of
a non-standard mixing episode occurring sometime along the upper RGB
(see also \citealt{tau00}, who obtained systematically larger values
of \isot but reported similar overall trends).

The overall pattern found in field stars is consistent with the
properties of old open cluster stars, but also indicating a
metallicity dependence.  Population I giants in the field (\feh
$\gtrsim -0.5$) show a pattern similar to M67, with stars on the lower
RGB showing ratios \isot $\sim 40$, while stars brighter than some
luminosity showing \isot $\sim 13$, again, much lower than expected
just from standard first dredge-up \citep{she93}.  More metal-poor
field giants ($-2.2 \leq\,$ \feh $\leq -0.2$) experience a similar
evolution but reach systematically lower \isot ratios than their Pop I
counterparts in the field \citep{pil97,kel01}, evidencing a trend of
stronger mixing with decreasing metallicity.  In a thorough
analysis of the abundances of a large sample of metal-poor field (halo
and thick disk) stars restricted to a narrow mass range, \citet{gra00}
found good agreement with standard first dredge-up predictions among
stars in the lower giant branch, but confirmed a second, non-standard
mixing episode kicking in just after the occurrence of the RGB bump
\footnote{The luminosity function (LF) bump, or RGB bump, is a
standard feature of the color-magnitude diagram (CMD) that is also
consequence of the deepening of the convective envelope in first
ascent giants and which will be important throughout this work.  This
feature is due to a slowdown of the evolution of first ascent giants
produced when the hydrogen burning shell (HBS) crosses, and erases,
the discontinuity in mean molecular weight left behind by the
envelope at its point of deepest penetration.  It
observationally appears as an excess of stars in the differential LF
or as a change of slope in the cumulative one, its exact position
being metallicity dependent, and has been detected in several Galactic
globular clusters \citep{kin85,fus90,fer99,zoc99} as well as some
Local Group galaxies \citep{maj99,bel01,bel02,mon02}.}.  There is
evidence for changes in both the \isot ratio and the C/N ratio
consistent with in situ mixing above the bump: carbon decreases and
nitrogen increases with luminosity and there is a steady drop in the
\isot ratio from its post first dredge-up value of $\sim 30$ to values
$\lesssim 10$.  Furthermore, the \citet{gra00} data also extend to the
subgiant branch (SGB) as well as unevolved MS field stars, and these
do not display any anomalies or changes in their surface abundances
with evolutionary state.  This will be an important point when we turn
to the origin of the RGB abundance anomalies.

Giants in globular clusters display all the phenomenology of their
counterparts in open clusters and the field, but with two key
differences: the non-standard mixing is even deeper, and there are
clear signs of abundance variations already in place before the stars
evolve to the giant branch.  Deeper mixing is evidenced not only by
\isot ratios even lower than the most mixed field or open cluster
giants, usually reaching as low as the nuclear equilibrium value of
\isot $\approx 4$ \citep{sun91,she96b,bri97,smith02,pil03}, but also
by anomalies likely involving oxygen depletion, which only occurs at
temperatures higher than those of the CN-cycle \citep{pil88,bro91,
sne91,min96,smith96,cohen05}.  Although known for a long time to be
the case in field giants, a clear evolution of the surface \isot
ratios with increasing luminosity along the RGB of globular clusters
was seen only recently by \citet{she03} in the clusters M4 and NGC
6528, also finding that the transition luminosity corresponds well to
the location of the LF bump in these clusters.  A clear decline in
total carbon with increasing luminosity is seen in globular clusters
such as M3, M13, NGC 6397, and M92 \citep{bellman01,smi03}, and
whenever nitrogen abundances are available, they are typically
anticorrelated with carbon \citep{kra94}.  More recently, large
star-to-star variations in the carbon and nitrogen abundances have
been found among subgiants and MS turnoff stars in various globular
clusters \citep{ram02,coh02,bri02,cohen05,eugenio04} showing also the
strong anticorrelation expected from C$\rightarrow$N cycling of this
material.  Some of these data may also require the occurrence of
O$\rightarrow$N cycling.

Finally, one of the most ubiquitous trends seen among globular cluster
giants is a striking anticorrelation between oxygen and sodium, first
noticed by \citet{pet80} in M13 giants and since then confirmed in a
large number of clusters
\citep{sne92,sne94,sne97,kra93,kra95,kra97,kra98,bro92,min96,she96a,
iva99,iva01,gru02,ram02,ram03,yon03,cohen05,eugenio04}.  This O-Na
anticorrelation is often accompanied by a similar behavior between
magnesium and aluminum \citep{she96b,iva99,yon03}.  In contrast, field
stars do not show any signs of O or Na abundance variations, much less
any anticorrelation, all the way from the MS to the RGB tip
\citep{gra00}.  These proton-capture anomalies have now been traced
back to the subgiant branch (SGB) and all the way to the MS turnoff
\citep{ram02,ram03,cohen05}, thus probably indicating that they did
not originate in the current generation of stars.

\section{Method}

\subsection{On the treatment of rotation and the extra mixing}

We use the Yale Rotating Evolution Code (YREC; \citealt{gue92}) to
model the evolution of low-mass stars from the pre-main-sequence up to
the tip of the red giant branch (RGB), just before the onset of the
helium flash.  Although stellar rotation is an inherently three
dimensional problem, YREC solves for the structure of the rotating
star in one dimension by following the scheme initially proposed by
\citet{kip70} and later modified by \citet{end76}.  Under this
treatment, one defines an effective potential composed by the sum of
the usual (spherically symmetric) gravitational potential, and a
centrifugal potential.  The latter depends on the distance to the axis
of rotation, rather than the distance to the center, thus breaking
spherical symmetry.  All variables are computed on the resulting
non-spherical equipotential surfaces, and an effective gravity can be
defined as the gradient of the effective potential.  As a result, the
effective gravity varies with latitude on equipotential surfaces, it
is still perpendicular to them everywhere, but, except for poles and
equator, it does not point to the center of the star anymore.  Of the
four equations of stellar structure, only those of hydrostatic
equilibrium and radiative energy transport need to be modified to
account for the new form of the potential (see \citealt{sil00} for the
modified equations).  An important restriction on this scheme is that
it assumes that the potential is conservative, and this is not valid
when the star is changing in size.  Hence, during the giant phase,
when the star's envelope expands by over two orders of magnitude, it
is important to take sufficiently small time steps so as to minimize
the error introduced by this assumption.  For details of this
implementation see \citet{end76} and \citet{pin89}.  Here, we only
describe the two issues that are most relevant to this work, namely,
the angular momentum distribution and transport across the star, and
the treatment of the mixing processes.

Two factors determine the evolution of the internal rotational profile
along the giant branch: the efficiency of angular momentum transport
across the star, and angular momentum loss due to mass loss and/or
stellar winds.  Since both gravity waves and magnetic fields are not
expected to be important during this evolutionary stage (\S\,1), we
only consider meridional circulation and hydrodynamic instabilities
driven by the star's rotation as mechanisms for angular momentum
transport.  After the structure of the star is calculated at each time
step in the evolution, three things determine the new angular velocity
profile: local conservation of angular momentum in radiative regions,
the assumption for the rotation law in convective regions, and the
amount of angular momentum loss.  Once these are determined, the
gradients of angular velocity are computed and their stability is
analyzed.  When an unstable angular velocity gradient appears, the
corresponding hydrodynamic instability whose stability criterion is
not being fulfilled is triggered, rearranging the angular momentum
profile until it meets the stability conditions again.
Simultaneously, the redistribution of angular momentum is accompanied
by a redistribution of the chemical composition, i.e., mixing.  The
effect of typical mass loss rates is small \citep{sip00}.

Also, the diffusion of elements may be affected by gradients of
chemical composition, and this can happen in two ways.  First, the
pole-to-equator pressure imbalance resulting from the different
centrifugal accelerations as a function of latitude on equipotentials,
which is what drives meridional circulation on the first place, can be
actually accommodated without the need for circulation if the chemical
composition on equipotentials has the right distribution.  This can be
immediately realized from the equation of state of a perfect gas if
allowing the pressure $P$ and the mean molecular weight $\mu$ to vary
and keeping the other thermodynamic quantities fixed.  Under the
picture of anisotropic turbulence of \citet{zahn92}, it is the strong
horizontal turbulence what homogenizes the composition on
equipotential surfaces, thus maintaining the latitudinal pressure
imbalance that drives large-scale circulation.  The second effect of
composition gradients may be appreciated with an energy argument, by
considering that in order to lift a mass element of molecular weight
$\mu_1$ into a region of $\mu_2 < \mu_1$, one actually needs to do
some work on the mass element.  \citet{kippwei90} illustrate this with
the help of stability arguments similar to the Schwarzschild criterion
for convection.  Consider for example Eddington circulation, and
imagine a small element of stellar material being lifted by uprising
meridional currents.  Suddenly, this volume element finds itself
surrounded by material of lower molecular weight, thus requiring a
similarly positive temperature difference between the element and its
surroundings in order to maintain mechanical equilibrium (i.e., no
pressure or density difference).  Being hotter than its surroundings,
it has to radiate the excess energy, which, under pressure
equilibrium, necessarily leads to an increase in density.  Therefore,
the mass element sinks back again, reacting on the meridional currents
that made it rise in the first place.  First realized by
\citet{mes53}, this feedback is modeled as an additional velocity, a
{\it $\mu$-current}, that opposes the meridional ones.  For large
enough {\it $\mu$-gradients}, this mechanism could potentially inhibit
circulation, and one speaks of {\it $\mu$-barriers}.

Such a $\mu$-barrier actually appears during the giant branch phase
when the convective envelope reaches its maximum depth at the end of
the first dredge-up, retreating afterward and leaving behind a sharp
hydrogen discontinuity, i.e., effectively a $\mu$-barrier.
\citet{swe79} argued that, assuming that circulation currents can not
penetrate this $\mu$-barrier, and given that the temperature
immediately above it is not high enough even for C$\rightarrow$N
processing to occur, any mixing of CNO processed material into the
envelope would be prevented as long as this $\mu$-barrier stays.
Observationally, the abundance patterns along the RGB seem to indicate
that the non-standard extra mixing sets in only after the location of
the RGB bump \citep{gil91,gra00,kel01,she03,cha94,cha98a}, that is,
once this $\mu$-barrier has been erased by the advancing HBS, hence
apparently confirming the \citet{swe79} picture.  There is, however,
the case of M92: \citet{bellman01} showed that, despite the
appreciable scatter in the carbon abundance at any luminosity along
the RGB, the onset of carbon depletion in this cluster occurs well
before its LF bump (see also \citealt{smi03}), the location of which
has been determined by observations \citep{fus90}.  We discuss
inhibition of mixing by $\mu$-barriers in \S\,5.

Parametric diffusion models have typically imposed this inhibition of
mixing before the LF bump based on the aforementioned observations as
well as theoretical expectations \citep{cha95,was95,boo99,wei00}.
However, a self-consistent treatment of non-standard mixing must
obtain such inhibition, if any, from first principles rather than
imposing it as a universal property of stars.  Since the purpose of
this work is to investigate the properties of rotational mixing under
the most favorable conditions for it, we do not consider the opposing
effect of $\mu$-barriers.  

We anticipate that the above decision leads to the realization of
several facts on the angular momentum evolution and mixing properties
of red giants that have been overlooked before.  For example, we find
that mixing induced by rotation is inefficient in the lower RGB (i.e.,
before the LF bump) without any regard to the effects of
$\mu$-barriers, which has rather important implications on the
interpretation of the observational data (see \S\,5.2 and \S\,5.3).

Extra mixing is completely characterized by how fast it proceeds and
how deep into the star it reaches, that is, by its rate and depth.  At
this point we want to stress the fact that, unlike all the currently
available literature on the subject of abundance anomalies in red
giants, the goals of the present work are not to determine
combinations of rate and depth of extra mixing that can reproduce the
data.  Such an empirical way of determining the necessary parameters
has been extensively explored already (see references in \S\,1) and,
on the other hand, the computation of rate and depth from first
principles is still uncertain.  Instead, our goal is to determine if
rotational mixing is a viable solution to the problem of abundance
anomalies across all types of stars, i.e., those that are observed to
be extremely mixed as well as those observed to be mildly mixed or not
mixed at all.
 
In our models, the extra mixing occurs through the action of
rotationally-induced hydrodynamic instabilities.  According to the
timescale in which they act, hydrodynamic instabilities are separated
into two classes: dynamical and secular.  Dynamical timescales are
much shorter than any evolutionary timescale, and therefore dynamical
instabilities typically act on a free-fall timescale.  Secular
instabilities, on the other hand, are related to the local properties
of the stellar material, and therefore typically act on a
Kelvin-Helmholtz timescale or longer.

Because of these different timescales, the transport of angular
momentum and chemicals by dynamical and secular instabilities must be
treated differently.  When a dynamical instability appears, it gets
smoothed instantaneously until the particular condition for stability
is marginally matched.  In the actual star, this is accomplished by
mass motions that redistribute the angular momentum as well as mix the
material.  Hence, if a dynamically unstable region also had a
composition gradient, this will be reduced by the same amount as the
angular velocity gradient.  The only dynamical instability considered
in this work is the dynamical shear \citep{end78}.  Note that,
although not related to angular velocity gradients and actually an
ingredient of standard models, convection is also a dynamical
instability.

Secular instabilities, on the other hand, can not be instantaneously
erased, and the changes in angular velocity and composition are
treated as simultaneous (i.e., coupled) diffusion processes in which
the diffusion coefficients depend on the time scales, velocities, and
path lengths associated to each process.  The equations for the rates
of change of angular velocity and chemical abundances ($X_i$) are
then,

\begin{equation}
\rho\,r^2\,\frac{I}{M}\,\frac{d\Omega}{dt} = \frac{d}{dr}\biggl(\rho\,r^2\,\frac{I}{M}\,D\,\frac{d\Omega}{dr}\biggr),
\end{equation}

\noindent and

\begin{equation}
\rho\,r^2\,\frac{dX_i}{dt} = \frac{d}{dr}\biggl(\rho\,r^2\,D\,\frac{dX_i}{dr}\biggr),
\label{eqn:diffusion}
\end{equation}

\noindent where $I/M$ is the moment of inertia per unit mass, $D$ is a
diffusion coefficient, and the rest of the symbols have their usual
meaning.  The velocity associated to the diffusion coefficient is the
combination of the velocities associated to all the secular
instabilities, which in turn depend on the gradients of angular
velocity and composition.  As a result, $D$ is also a function of
these gradients, and equations (1) and (2) must be solved
simultaneously.

Our expressions for the diffusion coefficients are taken from
published estimates in the literature.  We consider three secular
mechanisms for mixing: the secular shear, the
Goldreich-Schubert-Fricke (GSF) instability, and meridional
(Eddington-Sweet) circulation.  Because we are neglecting angular
momentum redistribution in radiative regions we only need to solve
equation (\ref{eqn:diffusion}) above, with the total diffusion
coefficient given by $D = D_{\rm SS} + D_{\rm GSF} + D_{\rm eff}$,
where $D_{\rm SS}, D_{\rm GSF}$, and $D_{\rm eff}$ are the diffusion
coefficients for chemical mixing arising from the secular shear, GSF
instability, and meridional circulation, respectively.  See the
Appendix for the detailed expressions used in this work.

Due to the large structural changes experienced during post-MS stages,
instabilities triggered by angular velocity gradients become more
important during the RGB than what they were during the MS
\citep{pin89,chab95a}.  On the upper MS, where angular momentum loss
via magnetized winds is negligible ($> 1.3$\msun), the most important
instabilities are those associated to the absolute rotation
(meridional circulation).  For solar mass stars, which experience
magnetic braking during their MS lifetime, the relative importance of
the different instabilities is not well understood.

Finally, the maximum depth allowed for the action of extra mixing
needs to be specified as well.  Ideally, if mean molecular weight
barriers are the only agents that prevent mixing, the maximum depth
for mixing would be obtained internally from the particular profiles
of chemical composition, and there should be in principle a dependence
of this parameter with stellar mass and metallicity.  However, the
detailed dependence of the inhibition of mixing by $\mu$-gradients is
currently too uncertain and, therefore, we do not intend in this work
to explore the variation of the mixing depth with mass and
metallicity.  Advances in this particular issue are more likely to
come from careful analysis of the data, following reasonings similar
to that of the study of \citet{cha98a}.  See, however, our discussion
in \S\,5.3.

In our models, we set the maximum depth for extra mixing to the
location at which 99\% of the total stellar luminosity is generated,
very close to the top of the HBS.  Given that the star's luminosity is
generated by hydrogen being turned into helium inside the HBS,
parametrizing the mixing depth according to some fraction of the
luminosity should actually follow the development of a composition
gradient across the HBS, and therefore should map at least the form,
if not the absolute value, of the inhibition of mixing due to
$\mu$-gradients.

In parametric models, the maximum depth is typically parametrized by
$\Delta\log\,T \equiv \log\,T(M_{\rm core}) - \log\,T(M_{\rm mix})$,
the logarithmic difference between the temperature at the core's
surface and that at the mass coordinate of the maximum depth.
\citet{den03} find that $\Delta\log\,T = 0.19$ best matches both their
parametric and rotation-based models of a Pop II red giant to the data
of field giants.  This corresponds to slightly shallower mixing than
our chosen depth, and we discuss this in more detail in \S\,5.4.  We
only advance here that varying our mixing depth does not change any of
our conclusions regarding rotational mixing as a viable explanation of
the abundance anomalies in all types of giants.

\subsection{Input physics}

Using YREC, we investigate the RGB evolution of rotating stars with
masses in the range 0.8-2\msun for various metallicities.  All our RGB
runs start at the MS turnoff, which we define as the point where the
mass fraction of hydrogen in the star's core reaches 1\% $({\rm
i.e.,}\, X_{\rm core} \approx 0.01)$.  The starting turnoff models are
calculated by evolving the stars, without rotation but allowing for a
stellar wind, from a pre-MS, fully convective model on the upper
Hayashi track.  [Fe/H] metallicities are related to heavy-element
fractions by the equation ${\rm [Fe/H]}=\log(Z/X)-\log(Z/X)_{\odot}$.
The initial mixtures are computed from a solar mixture with
$Y_{\odot}=0.275$ and $Z_{\odot}=0.02003$, which are not the values
measured for the current Sun but instead account for gravitational
settling and diffusion \citep{gre98,bpb01}.  The Galactic enrichment
factor $\Delta Y/\Delta Z$ is computed using the above values of
$Y_{\odot}$ and $Z_{\odot}$ and a primordial helium fraction
consistent with recent observations of extragalactic HII regions, $Y_p
= 0.245$ \citep{izotov04}.  The mixing length parameter is set to
2.085, also taken from the solar models of \citet{bpb01}.  This value
of $\alpha$ is admittedly higher than the typical one used in RGB
models, but we verified that the extra-mixing properties of our models
are totally insensitive to the choice of this parameter, and its only
effect is in the position of the RGB on the color-magnitude diagram
(CMD).  We use the nuclear reaction rates of \citet{gruz98} and,
currently, YREC evaluates the nuclear burning and mixing
independently.  This approximation is valid as long as the mixing is
much slower than the rate at which the local composition changes
appreciably due to nuclear reactions.  Given the high temperatures
achieved during HBS burning, this approximation should hold during the
RGB, although we plan the simultaneous solution of burning and mixing
to be implemented in YREC in forthcoming work.  The low-temperature
opacities of \citet{alex94} are used for $T < 10^4\,K$ and the
interiors OPAL opacities \citep{opal96} otherwise.  Given that the
most recent and accurate equations of state \citep{rog96,sau95} and
atmospheres \citep{allard95} do not extend to the conditions near the
tip of the RGB, we use the Saha equation of state, complemented with
the Debye-Huckel correction for Coulomb interactions, and Krishna
Swamy atmospheres.  We do not consider convective overshoot nor
semi-convection.  Gravitational settling of helium and heavy elements
is implemented in the generation of the turnoff models, but not
considered for the RGB models.  Table 1 lists the turnoff age,
luminosity, and radius for the stellar masses investigated.

Regarding the ingredients related to rotation, we need to specify an
initial angular momentum budget, the angular momentum law in the
convective envelope, the details of the transport of angular momentum
and composition, and the rate of angular momentum loss.  Let us recall
at this point that the goal of this paper is to investigate the case
with the most favorable conditions for extra-mixing due to
hydrodynamic processes triggered by internal angular velocity
gradients.  If the transport of angular momentum by hydrodynamic
instabilities is too efficient, these gradients will be rapidly
smoothed and extra-mixing will consequently be minimal.  On the other
hand, in the case of inefficient angular momentum transport by these
processes, the velocity gradients can survive for longer times, hence
maximizing the mixing.  Therefore, the best case scenario for
additional mixing will be one in which no angular momentum transport
by hydrodynamic processes is allowed, and thus we will only let the
angular velocity profile adjust itself to the structural changes
during RGB evolution by requiring local conservation of angular
momentum (i.e., angular momentum transport will only occur following
changes in the structure of the star).  Mass loss and stellar winds
are not initially considered, but an exploration of their effects
under typical conditions produced no appreciable differences.  Also,
as part of our maximal-mixing conditions, we do not impose any
inhibition of extra-mixing due to composition gradients.

At this point, we would like to stress the importance of the choice of
initial angular momentum distribution for the final results of
rotational mixing during the RGB evolution.  Rotational mixing in red
giants occurs in the radiative region separating the core and the
convective envelope of the star, and thus, it is the content and
evolution of the angular momentum across this region what determines
the amount of extra-mixing to be experienced on the RGB.  As
beautifully illustrated in Fig.32.3 of \citet{kippwei90}, the total
mass of this radiative zone is actually very small, and at any given
time on the giant branch ascent it is comprised by material that is
falling from the convective envelope and that is on its way to be
burnt inside the HBS and finally deposited on the growing helium core.
{\it Thus, the angular momentum contained in the material that is
incorporated into the convective envelope during the first dredge-up
is the crucial quantity determining the amount of rotational energy
that will be available to drive extra-mixing.}  

Such a quantity naturally depends on the angular momentum profile
during the MS evolution.  Let us consider the two extreme cases of
angular momentum distribution typically discussed, namely, solid body
rotation and constant specific angular momentum, and let's label them
case A and B, respectively.  The specific angular momentum (angular
momentum per unit mass) of a uniformly rotating spherical shell is $j=
dJ/dm=(2/3)\,\Omega\,r^{2}$, where $\Omega$ is the angular velocity
and $r$ is the radius of the shell.  For a given surface angular
velocity $\Omega_s$, the specific angular momentum for case A is
$j_{\rm A}=(2/3)\,\Omega_{\rm s}\,r^{2}$, while for case B we have
$j_{\rm B}=(2/3)\,\Omega_{\rm s}\,(R/r)^{2}\,r^2$, where $R$ is the
total radius of the star.  Thus, if $r_{\rm cz}$ is the radius of the
bottom of the CZ at its point of maximum penetration during first
dredge-up, then the ratio of the total angular momentum incorporated
in the convective envelope for the two cases is

\begin{equation}
\frac{J_{\rm tot}^{\rm A}}{J_{\rm tot}^{\rm B}} = \frac{\intop^{R}_{r_{\rm cz}}\,\frac{2}{3}\,\Omega_{\rm s}\,r^2\,{\rm d}m}{\intop^{R}_{r_{\rm cz}}\,\frac{2}{3}\,\Omega_{\rm s}\,\biggl(\frac{R}{r}\biggr)^{2}\,r^{2}\,{\rm d}m} = \frac{\intop^{R}_{r_{\rm cz}}\,r^{2}\,{\rm d}m}{\intop^{R}_{r_{\rm cz}}\,R^{2}\,{\rm d}m} < 1\,.
\label{eqn:budget}
\end{equation}

Therefore, the post-MS convective envelope incorporates more angular
momentum from a MS progenitor that rotates differentially than from a
rigidly rotating one, and therefore, case B will experience more
mixing than case A.  This is the reason behind the conclusion of
\citet{den03} that ``the spin-up of the star's core while it is on the
MS is needed to get the right extra mixing depth on the upper RGB and
thus is very important''.  

However, the assumption of strong internal differential rotation
during the MS goes contrary to the only case of old MS star for which
such a measurement is available to us, the Sun, which is a solid-body
rotator down to at least 20\% of its radius \citep{tomczyk95}.
Instead, it is at this point safer to assume that turnoff stars in
stellar clusters as old as the Sun and older resemble the solar state
of rigid rotation.  This assumption is, furthermore, in accordance
with the current picture of the rotational evolution of solar- and
late-type stars, in which the angular momentum distribution in the
large majority of old, low-mass MS stars is determined by the coupling
of the radiative cores and convective envelopes of the stars through
the action of an ``interface'' magnetic field connecting them, and
which is responsible for the extraction of the internal angular
momentum via a magnetized wind \citep{barnes03}.

Hence, relying upon the helioseismic data on the Sun's current state
of rotation and the previous discussion, we choose to assign an
initial solid-body rotation profile throughout the entire MS turnoff
star, such that the rotational velocity at the surface of the star has
a value $V_{\rm TO}$, with the corresponding surface angular rotation
rate being

\begin{equation}
\Omega_{\rm initial} = 1.4\times10^{-5}\biggl(\frac{V_{\rm TO}}{10 \,{\rm km\,s}^{-1}}
\biggr)\biggl(\frac{R}{R_{\odot}}\biggr)^{-1}\,\,\,{\rm rad\,\,\,s}^{-1}.
\end{equation}

\noindent Here, $(R/R_{\odot})$ is taken from Table 1, and we run
models for a wide range of initial (turnoff) rotation rates, $V_{\rm
TO}$.

We study the angular momentum evolution for two extreme cases of
rotation law ($\Omega \propto r^{n}$) in convective regions: rigid,
solid-body rotation ($n = 0$), and constant specific angular momentum
$dJ/dm \equiv J/M$ ($n = -2$), and we show that they lead to very
different rotational and mixing histories.  As will be shown in the
next section, the key difference between these two cases is that they
concentrate the angular momentum content in different regions of the
star.  If the outer convective envelope of a star rotates as a solid
body, then the outer layers, those with large $r$, will concentrate
most of the envelope's angular momentum.  If, on the other hand, the
convective envelope rotates such that $j$ is a constant (in which case
$\Omega \propto r^{-2}$), then the deeper layers of the envelope,
those where the mass is mostly concentrated, will be the ones storing
most of the angular momentum of the envelope.  In other words, the
angular momentum content of a star is more centrally concentrated in
the case of differential rotation with depth than in the case of solid
body rotation.  The assumption of constant $j$ in the envelope also
minimizes the loss of angular momentum accompanied by mass loss.
Finally, note that, since the lower mass stars considered here
(M$\,\lesssim$ 1.1\msun) have large outer convective envelopes during
their MS lifetime, when enforcing a constant $J/M$ law to a convective
region at turnoff, the star's angular momentum profile will have a
discontinuity at the boundary between the radiative and convective
regions due of the assignment of rigid rotation to the initial turnoff
model.  Although small, this feature will be preserved in our models
because, as part of the maximal mixing approach, we will not allow the
hydrodynamic instabilities to transport any angular momentum.  We come
back to this point in more length in \S\,5.1 when discussing the
actual angular momentum profiles and diffusion coefficients.

\subsection{Calibrating Clusters and Sensible Rotation Rates}

We choose the clusters M67 and M92 as representative cases against
which we will test our results.  These two well studied clusters,
lying on opposite sides of the metallicity scale and with very
different turnoff masses, allow for a study of the consequences of
rotational mixing under very different conditions.  Also, we
complement the cluster observations with the data on field giants of
\citet{gra00}.

\begin{figure}
\vspace*{-1cm}
{\hspace{-15cm}
\plotfiddle{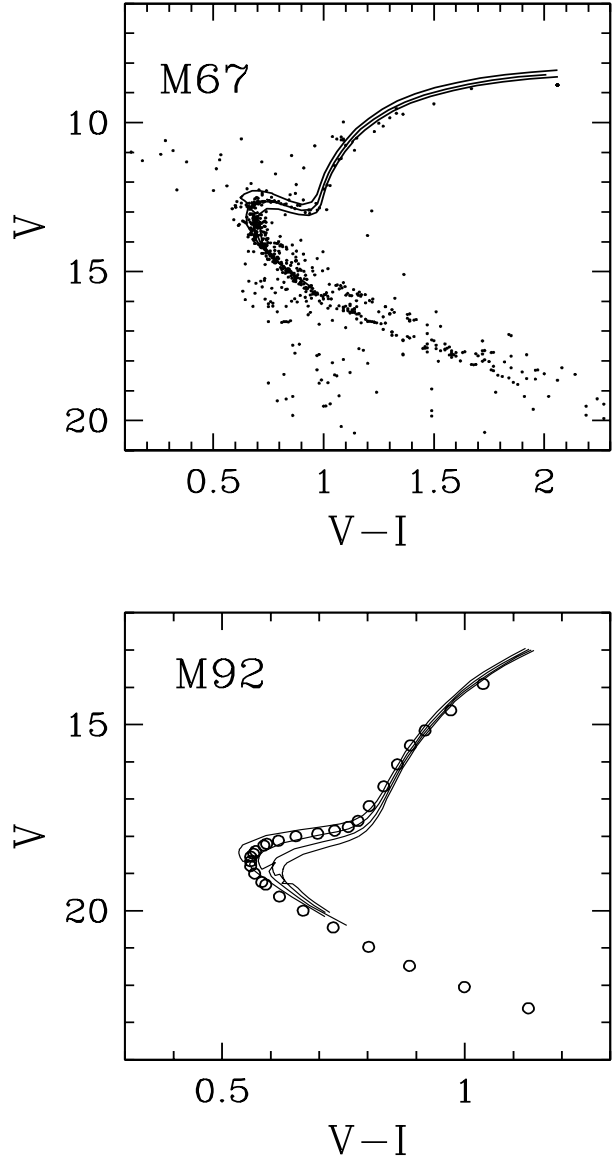}{1cm}{0}{400}{500}{-25}{0}}
\vspace*{-1cm}
\caption{\label{fig:cmds} Color-magnitude diagrams of representative
clusters used in the comparison with surface abundance data.  Data for
M67 is from Montgomery et al. (1993) and the WEBDA database, while for
M92 we plot the fiducial CMD given by \citet{jon98}.  The solid lines
are theoretical isochrones constructed with the same evolutionary code
used in this work, but without rotation.  For M67 we show solar
metallicity isochrones for ages of 3, 4, and 5 Gyr, using a distance
modulus of $(m-M) = 9.45$ and reddenings of $E(B-V)=0.04$
\citep{gro03} and $E(V-I)=1.5\,E(B-V)=0.06$.  The M92 isochrones are
for ages of 12, 14, 16, and 18 Gyr, have a metallicity of $[\rm{Fe/H}]
= -2.3$, and an initial mixture enhanced in $\alpha$-elements by
$\rm{[\alpha/Fe]=+\,0.3}$.  Distance modulus and reddening used to
shift the M92 isochrones are $(m-M) = 14.60$ and $E(V-I)=0.013$
\citep{har97}.  The transformation to magnitudes and colors was done
using the bolometric corrections and color-temperature calibrations of
\citet{lej98}. }\end{figure}

Figure \ref{fig:cmds} shows CMDs of M67 and M92.  The M67 CMD was
constructed using $BVI$ CCD data from \citet{mon93} and the WEBDA
database\footnote{http://obswww.unige.ch/webda/}, while for M92 we use
the fiducial $(V,V-I)$ CMD given in \citet{jon98}.  The solid lines
are standard-model isochrones computed using YREC, with the same input
physics as employed in the models with rotation.  Identifying the
bluest points in the 4 and 14 Gyr isochrones with the turnoff stars in
these clusters, we obtain turnoff masses of 1.25 and 0.75\msun for M67
and M92, respectively.  The models we use in the next sections for the
comparison with the data from these clusters actually have initial
masses of 1.3 and 0.85\msun.  However, this does not represent a
problem for our conclusions since, as will be seen in \S\,5.4, our
results are not too dependent on stellar mass.

\begin{figure}
\vspace*{-0cm}
{\hspace{-18cm}\plotfiddle{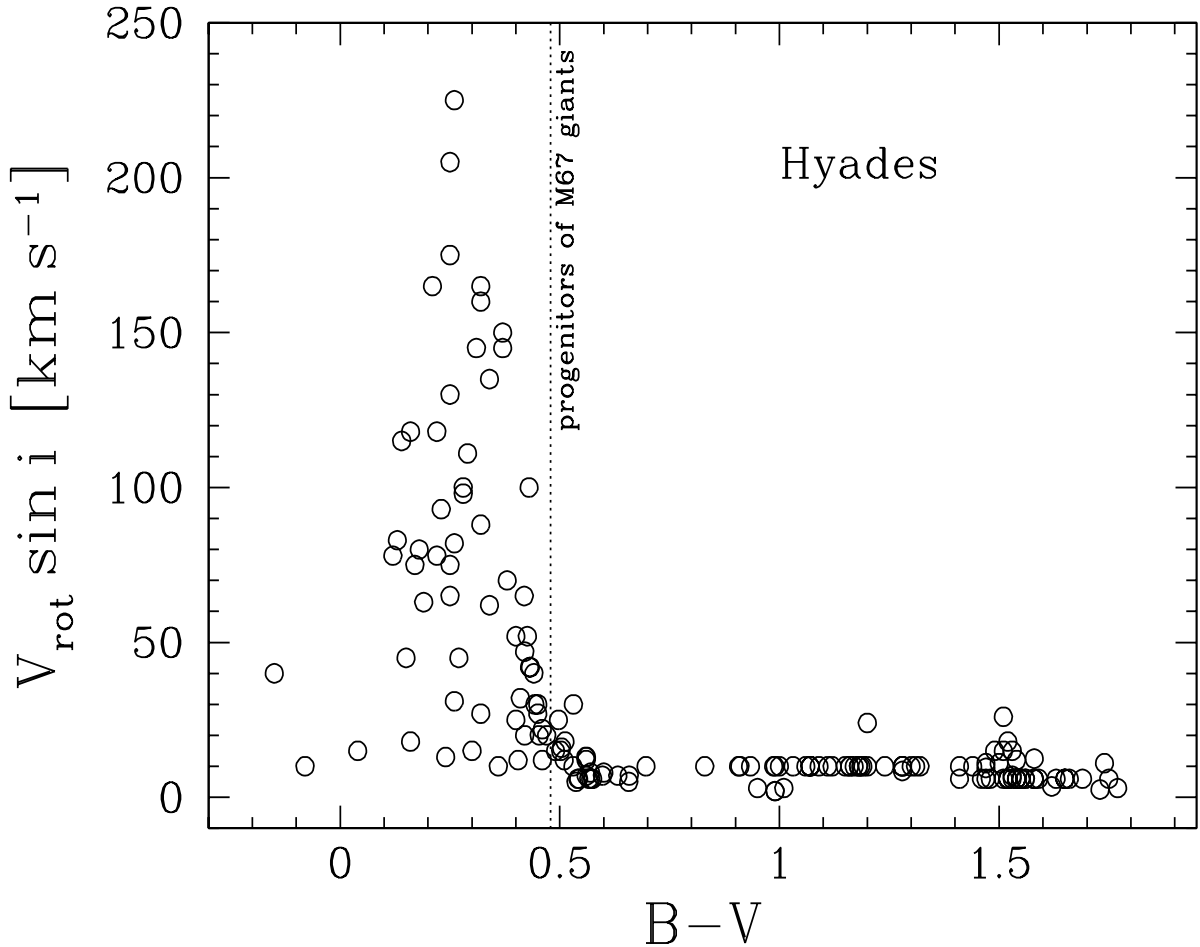}{1cm}{-90}{400}{500}{0}{0}}
\caption{\label{fig:hyades} Distribution of projected rotational
velocities for the Hyades cluster, obtained from the Open Cluster
Database (http://cfa-www.harvard.edu/~stauffer/opencl/).  The steep
break in the Kraft curve throughout spectral type F is clearly
displayed by the Hyades stars.  The vertical dotted line at $B-V=0.47$
marks the location of a 1.3\msun star of solar metallicity at the age
of the Hyades in our models (${\rm \log\,T_{\rm eff} = 3.82}$).  Such
star, a progenitor of current M67 first ascent giants, falls right on
the region of the diagram where rotational velocities are abruptly
decreasing, and it can be seen that appropriate rotational velocities
for this star are between 10 and $40\kms$.  See details in
\S\,3.3. }\end{figure}

We now need to determine the turnoff rotation rates for our models.
The rotation velocities of turnoff stars have been measured in M67
(Melo, Pasquini \& De Medeiros 2001) and the average rotation velocity
is low ($\langle v \sin i \rangle\, = 7\kms$).  However, the current
red giants were more massive than the current turnoff stars, and would
therefore have had higher turnoff rotation velocities.  This effect is
particularly important for clusters such as M67 with turnoffs in the
mid-F star range, where there is a transition (Kraft 1965) from very
rapid average rotation to very low average rotation (Figure
\ref{fig:hyades}).  In M67 in particular, there is good indirect
evidence from two methods that the current giants had a significantly
higher average rotation rate than the current turnoff stars.

The first method takes advantage of the fact that the ($T_{\rm
eff}-v\,\sin i$) relationship is independent of age for spectral types
earlier than F8 \citep{wol97}.  Therefore, a much younger cluster like
the Hyades ($\sim 600$ Myr), with the same metallicity as M67, can be
used to infer the rotation rates of the progenitors of the current
giants in M67.  For our M67 turnoff mass of 1.3\msun, Hyades analogs
with the same $T_{\rm eff}$ have a mean rotation velocity of $30\kms$
(more details in Figure 2).

An independent test is provided by the comparison of lithium data.
\citet{boes86} discovered the mid-F star lithium dip in the Hyades.
Lithium is very efficiently destroyed in stars with a narrow range of
effective temperatures, and thus there is a strong correlation between
the lithium abundance, $T_{\rm eff}$, and the stellar mass.
\citet{bal95} found that the current red giants at the RGB base in M67
had progenitors in the lithium dip region (see her Figs. 9 and 13).
Hyades lithium dip stars have rotation velocities in the $20-40\kms$
range (right on the break of the Hyades Kraft curve), which implies
that the current M67 giants had comparable ZAMS rotation rates,
consistent with the estimates derived above.

The same approach can not be applied to the case of M92 because there
is no younger cluster with measured rotation rates that shares such a
low metallicity, while well studied open clusters such as the Hyades
can not be used here because the relationship between color and
surface rotation rate is metallicity dependent.  There is, however,
some evidence that turnoff stars in globular clusters do not show
large rotational velocities.  \citet{luc03} find an average projected
rotational velocity of 3.1\,$\kms$ for 5 turnoff stars in the cluster
NGC 6397 ([$\rm{Fe/H}] = -2$), and old field halo stars are observed
to have small rotational velocities as well.  Hence, it is clear that
surface rotation rates larger than, say, 10\,$\kms$ are not observed
among old, metal-poor stars.

Having stressed the importance of adopting initial conditions that
reflect the observed properties of the different stellar types under
study, let us now argue why the generation of models with initial
surface rotation rates larger than realistic for any given type of
star is actually not a fruitless exercise but, instead, has important
advantages.  As discussed in \S\,3.2 and quantified by equation
(\ref{eqn:budget}), a red giant star whose MS progenitor had a core
more rapidly spinning than its envelope experiences much stronger
extra mixing than if the progenitor star, like the Sun, rotated
rigidly.  Since the reason behind this result is that the material
incorporated into the convective envelope during the first dredge-up
has a much higher angular momentum content when starting from a
differentially rotating turnoff star, we can mimic such a situation
but for a rigidly rotating turnoff star simply by starting with an
unusually large (for a given type of star) surface rotation rate.

Last, but not least, since the diffusion velocities used to model the
extra-mixing scale as $\Omega^2$, the same results of a rapidly
rotating model can be achieved with a slowly rotating one by enhancing
the diffusion coefficients associated with extra mixing by the
appropriate factor.  Therefore, the great utility of generating models
that start with unreasonably large rotational velocities is that these
are an excellent way to estimate by how large a factor do the
available theoretical estimates of the diffusion coefficients fail to
account for the observations when using sensible initial conditions,
and this approach will be used extensively in our analysis.  Such
direct proportionality is only possible in these models because we are
not allowing any angular momentum transport due to hydrodynamical
instabilities, thus preventing the feedback on the structure due to
different degrees of absolute rotation.  For these reasons, it is
worthwhile to run rotating models with initial surface rotation rates
that are too high for the stellar types studied here.

\section{Results: The Structural and Rotational Evolution}

\begin{figure}
\vspace*{-1cm}
{\hspace{-15cm}
\plotfiddle{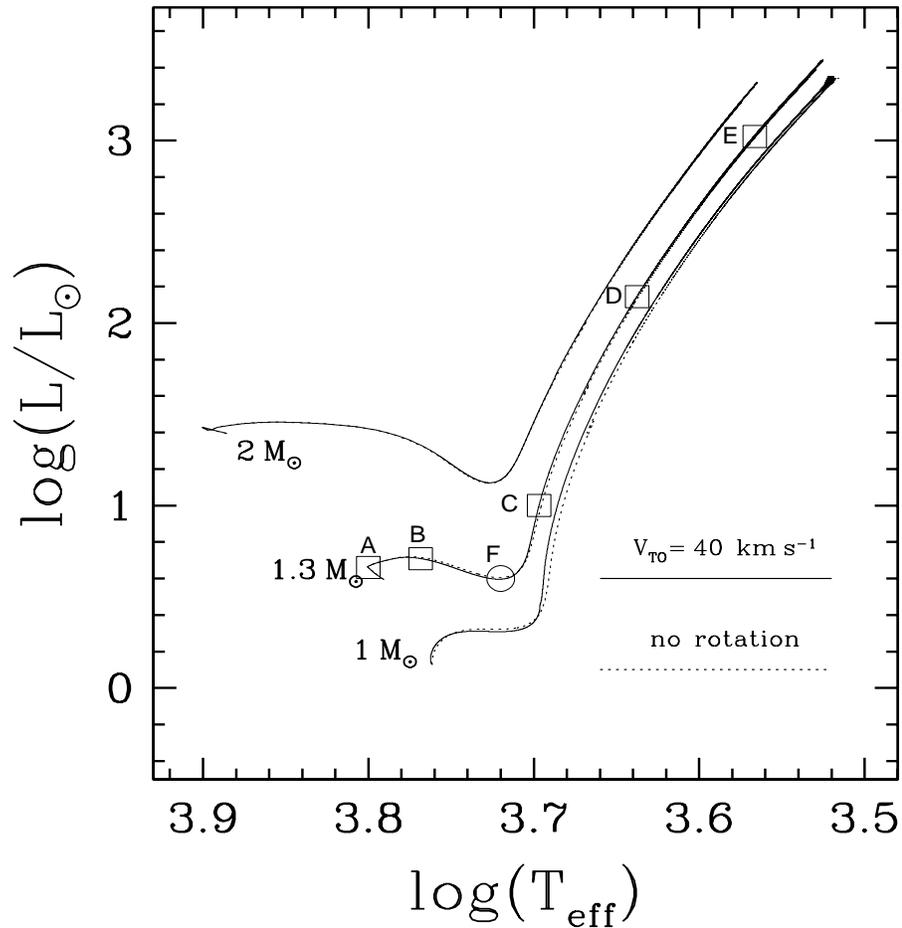}{1cm}{0}{400}{500}{-25}{0}}
\caption{\label{fig:rotnorot} Evolutionary tracks for stars with
masses of 1.0, 1.3, and 2.0\msun and solar metallicity, from the main
sequence turnoff and up to the tip of the giant branch.  Dotted lines
represent standard, non-rotating models, and solid lines are models
that started with a rotational velocity of $40\kms$ at the main
sequence turnoff.  The open squares along the track for the 1.3\msun
star represent the locations of the snapshots chosen to illustrate the
rotational evolution and shown in Figure \ref{fig:Jsnapshots}.  The
open circle labeled F marks the location of the bottom of the
RGB. }\end{figure}

Compared to its effects on the pre-MS and MS on the
Hertzsprung-Russell (HR) diagram \citep{sil00}, rotation has a small
effect on the giant branch tracks.  Figure \ref{fig:rotnorot} shows
post-MS evolutionary tracks for non-rotating and rotating models.  The
models shown start with a surface rotational velocity of 40\,$\kms$ at
turnoff, and do not include mass or angular momentum loss.  Models
with rotation are slightly brighter because of rotationally induced
mixing (which we allow to occur regardless of composition gradients)
rather than from direct structural effects.  The effects of rotation
increase at lower masses because of their longer RGB lifetimes.
Although not shown here, the choice of angular momentum law in the
convective envelope does not have any effect on the HR tracks but
leads to quite different mixing histories (\S\,5).

Figure \ref{fig:Jsnapshots} shows snapshots of the evolution of the
angular velocity and angular momentum profiles as a function of mass
and radius, for a star of 1.3\msun of solar metallicity that started
with a rotation rate at turnoff of 40 $\kms$.  The location of these
snapshots in the HR track is indicated in Figure \ref{fig:rotnorot} as
the open squares labeled with a capital character.  Solid lines
represent the case with constant specific angular momentum in
convective regions, and dotted lines show the profiles for solid body
rotation in convective regions.

The structural changes experienced during the giant branch generate a
contrast of more than 10$^{6}$ between the rotation rates of the core
and the surface of the star, a result impossible to achieve during the
much longer MS lifetime, and also independent of the rotation law
assumed for convective regions.  As the model ascends the giant
branch, its core contracts and spins up considerably, while the
envelope expands.  At the tip of the giant branch (E), the core
rotates more than 100 times faster than at the turnoff (A).  At the
same time, the model's radius has grown by about a factor of 50 and
the surface angular velocity is 4 orders of magnitude smaller,
becoming completely undetectable.

The angular velocity profiles in Figure \ref{fig:Jsnapshots} show that
almost all the contrast between the rotation at the surface and the
core is confined to a narrow region in mass coordinate that coincides
with the radiative zone between the core and the base of the
convective envelope.  This is crucial because angular velocity
gradients can generate extra mixing.

Initially (stage A), all the profiles are nearly equal to the initial
solid body rotation throughout the star.  There is little change
during the transition from core to shell hydrogen burning (the \sgb,
panels B).  At C, on the giant branch but before the LF bump, the
convective envelope has reached its maximum depth in both mass and
radius.  Thereafter, its base slowly retreats in mass, while staying
at the same location inside the star (see the position of the arrows
from panels C to E).  In the case of solid body rotation (dotted
lines), while the angular velocity stays constant, $J/M$ increases as
$r^{2}$.  

For constant specific angular momentum in the convective envelope
(solid lines), there is a discontinuity in the angular velocity and
angular momentum profiles which, only at the very beginning, is a
consequence of our choice of initial conditions, appearing when a
differentially rotating convective envelope encounters a rigidly
rotating interior.

Later, however, the angular momentum being incorporated into the
deepening envelope is redistributed such that $J/M$ is constant across
the envelope.  This effectively spins up the deep layers immediately
above the base of the convective envelope, therefore creating what is
now a large discontinuity in the angular momentum profiles.  We
emphasize that this latter effect is real, and has nothing to do with
the initial (small) discontinuity introduced by the initial
configuration of angular momentum at the turnoff.  Furthermore, this
discontinuity does not get smoothed in our models because we are not
allowing any transport of angular momentum as an ingredient of the
maximal mixing scenario; we come back to this point in \S\,5.1.

In the case of constant specific angular momentum and after the end of
the first dredge-up, the material that was part of the (now)
retreating convective envelope and which now is falling into the
radiative regions remains with constant $J/M$ and an $r^{-2}$ angular
velocity profile (regions at the left of the arrows in panels D and
E).  As a result, the star stores much of its angular momentum in its
radiative interior, retaining a large angular velocity gradient.  In
the presence of efficient angular momentum transport, this gradient
would be smoothed by hydrodynamical instabilities.  Therefore, our
assumptions of large differential rotation in the envelope and
suppression of angular momentum transport by hydrodynamic
instabilities results in a case where rotational mixing is maximized.

As \citet{sip00} found, the fast observed rotation rates of HB stars
imply that red giants retain most of their MS angular momentum, even
after the mass loss experienced at the top of the RGB.  If the star
were effective in distributing its internal angular momentum to the
envelope (which would be the case if angular momentum transport were
effective in smoothing internal gradients), then mass loss episodes
would be accompanied by significant loss of angular momentum, and HB
stars would rotate slowly.

This result is illustrated by Figure \ref{fig:Jsnapshots}, where it
can be seen that the solid body profiles in the inner radiative region
get progressively lower relative to the constant $J/M$ profiles.  This
implies that, for the rigidly rotating case, angular momentum from the
interior is being deposited in the outer regions.  Simple arguments
explain this effect.  For stars with rigidly rotating convective
envelopes, the ratio of specific angular momentum at the surface to
that at the base of the convection zone scales as $(R/r_{\rm
cz})^{2}$, where $R$ is the radius of the star and $r_{\rm cz}$ is the
position of the base of the convection zone.  As the star expands, $R$
gets larger but $r_{\rm cz}$ remains constant, and the surface's
specific angular momentum increases.  Since there are no external
sources of angular momentum, this happens at the expense of angular
momentum from the interior, which is therefore redistributed within
the surface convection zone.  The opposite effect occurs for the case
with constant $J/M$, which concentrates most of the angular momentum
content where the bulk of the mass lies and, as the model star
expands, the interior spins up with respect to the outer layers.

\section{Results: The physics of mixing}

An internal mixing episode other than the standard first dredge-up is
now recognized to occur in first ascent giants.  One of its main
characteristics is a strong dependence with spectral type: while the
more massive (early type) giants in open clusters show mild levels of
extra mixing \citep{gil89}, the oldest (late type) giants in globular
clusters experience stronger and deeper extra mixing, typically
displaying abundance patterns characteristic of material in nuclear
equilibrium.  

Still, theoretical investigations on the subject have dealt almost
exclusively with the lowest-mass, most metal-poor end of the spectrum,
i.e., Pop II giants (in the field and globular clusters) with
progenitor masses between 0.8 and 0.9\msun.  Furthermore, such
non-standard mixing is widely assumed to work only once the
$\mu$-barrier left after first dredge-up has been erased by the
advancing hydrogen burning shell, and thus would only affect stars on
the upper RGB.  

This ``canonical'' extra mixing, to use a term coined recently by
\citet{den03}, is mostly believed to be associated with rotational
effects \citep{swe79,cha95,den03} that must be deep enough to have an
effect on the surface abundances of the CNO elements, but can not be
responsible for the anomalies involving heavier species like Na, Mg,
and Al.

A rapid summary follows of the overall features of our models, which
involve several differences with all previous theoretical studies of
this subject.  Then, in \S\,5.1 to \S\,5.4, we use rotating models for
M67 giants (Figures \ref{fig:Jsnapshots} to \ref{fig:mu_standard}) to
discuss in detail the physics of rotational mixing and the results
from our models.

We have a standard first dredge-up much similar to that of all other
works on the subject.  Post dredge-up abundance ratios can be
significantly altered only at high rotational velocities, due to deep
meridional mixing arising as a consequence of the significant
departure from spherical symmetry.  However, when exploring the impact
of differences in the initial composition from cluster to cluster, we
obtain that the post first dredge-up abundance ratios, particularly
that of \isot (which has always been assumed to start at the solar
system value), depend significantly on the adopted initial mixture.

While all previous theoretical studies have assumed that extra mixing
is only possible beyond the RGB bump, a generic result of all our
models is that rotational mixing, although present in small amounts,
is anyway inefficient on the lower RGB independently of any inhibiting
effect of $\mu$-barriers.  Therefore, the lack of well mixed stars
before the luminosity of the RGB bump in the field and globular
clusters (\S\,1 and \S\,6) should not be interpreted as evidence for
the inhibition of mixing by $\mu$-barriers.

Furthermore, our models, with no transport of angular momentum allowed
in radiative regions, actually erase the $\mu$-barrier left by the
deepening of the convective envelope even at very small rotation
rates.  This fact, rather than the absence of mixed stars before the
RGB bump, is what really constitutes evidence that $\mu$-barriers may
actually inhibit mixing.

On the upper RGB, only models with differentially rotating (constant
$J/M$) convective envelopes experience canonical extra mixing,
consistent with the early predictions of \citet{swe79}, although for
reasons other than the disappearance of the $\mu$-barrier (\S\,5.1).
The amount of extra mixing is a strong function of rotation rate and
shows a weaker dependence on metallicity.  On this evolutionary stage,
our work presents two main departures from that of previous
investigators.  First, canonical extra mixing does indeed start at the
approximate location of the RGB bump but, rather than being due to the
disappearance of the $\mu$-barrier, it occurs as a consequence of
simple angular momentum considerations which, not surprisingly, are
intimately related to the requirement of having differential rotation
in the envelope (\S\,5.1).  Second, we simultaneously treat both the
Pop I and Pop II regimes using the same models and adopting initial
rotation rates in accordance with data as a function of stellar type.
This comprehensive approach has led us to new insights on the problem
of the interaction between rotation and convection in the envelopes of
stars (\S\,6.4).

Mass loss rates typical of red giants in globular clusters ($\Delta
M\sim 0.1-0.15$\msun), since important only close to the tip of the
RGB, produce almost negligible effects in the surface abundance ratios
of the most luminous giants.  For extreme mass loss rates, however,
larger differences might be possible.

\subsection{Mixing and the angular momentum distribution in the 
convective envelope}

Different assumptions about the distribution of angular momentum
result in different amounts of rotationally induced mixing.  The mode
of rotational mixing related to differential rotation is that
associated to hydrodynamical instabilities, in contrast with
rotational mixing associated to departures from spherical symmetry
(\S\,5.2), which depends on whether rotation is fast or slow.

\begin{figure}
\vspace*{-1cm}
{\hspace{-18.5cm}
\plotfiddle{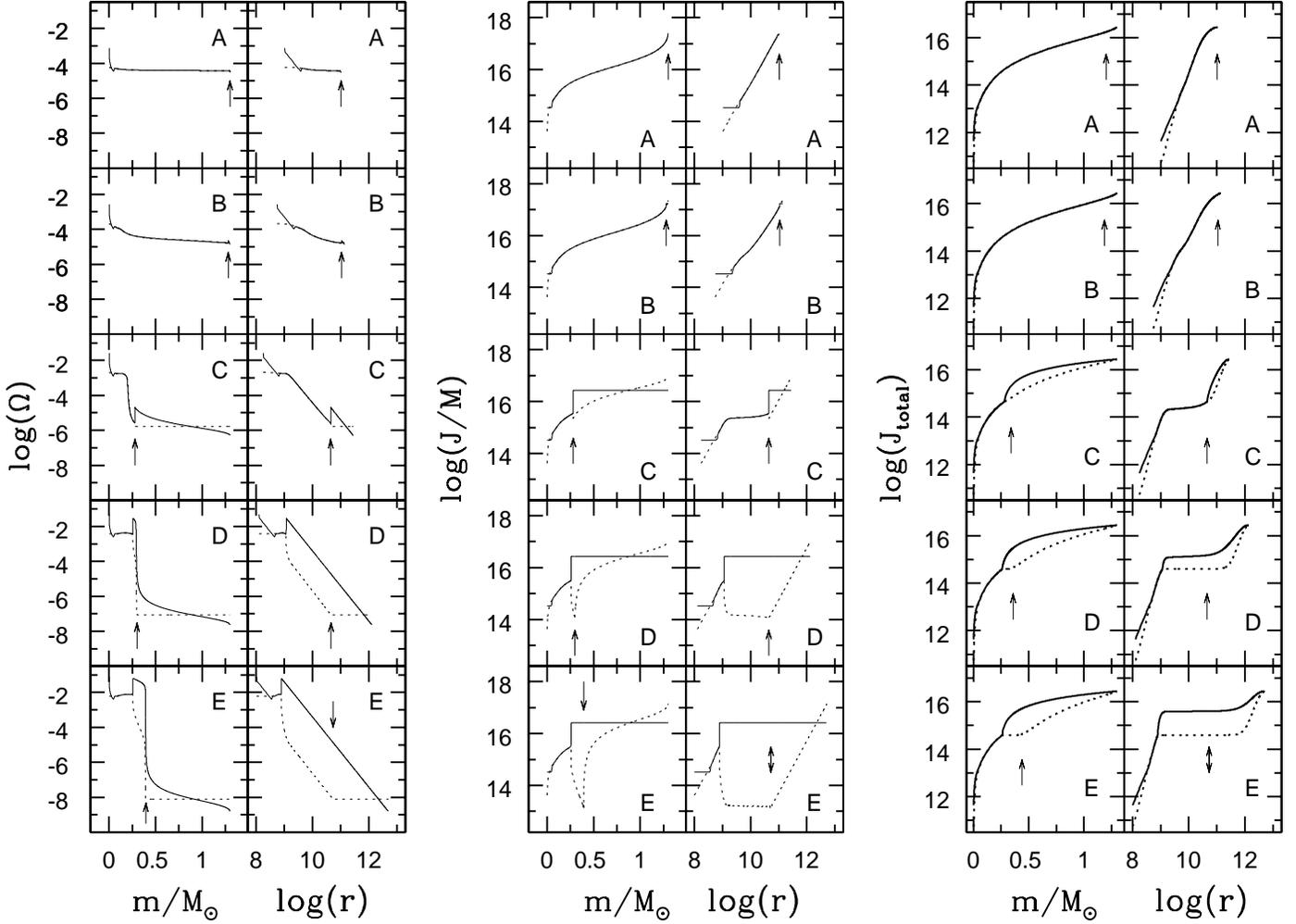}{1cm}{-90}{420}{550}{0}{0}}
\vspace*{-3cm}
\caption{\label{fig:Jsnapshots} Evolution of the profiles of angular
velocity ($\Omega$), specific angular momentum ($dJ/dm \equiv J/M$),
and total angular momentum (J$_{\rm total}$), as a function of mass
and radius (cm), for a star of 1.3\msun of solar metallicity that
started with a rotational velocity of $40\kms$ at turnoff.  Each
snapshot is labeled by a capital character (A, B, C, D, and E) that
indicates its location on the evolutionary track shown in Figure
\ref{fig:rotnorot}.  Solid lines represent constant specific angular
momentum in the convective envelope, and dotted lines represent solid
body rotation.  The location of the base of the outer convection zone
is indicated with arrows.}\end{figure}

The impact that different angular momentum profiles in the convective
envelope have on the core's rotation is well illustrated in Figure
\ref{fig:Jsnapshots} where, even though both cases of angular momentum
law in the convective envelope start with almost the same angular
momentum budget (A), at the RGB tip (E) the case of constant $J/M$ has
a specific angular momentum as well as a rotation rate at the base of
the convective envelope that are about two orders of magnitude larger
than those of the case with a rigidly rotating envelope.  In both
cases, the material falling from the convective envelope keeps its
angular momentum content (due to local angular momentum conservation),
{\em but the key difference is that the material that had constant
$J/M$ while being part of the convective envelope is spinning at a
much higher rate} (material immediately to the left of the arrows in
the radial profiles in panels D of Fig. \ref{fig:Jsnapshots}).  This
occurs because the assumption of a constant $J/M$ in the convective
envelope during the \sgb and lower RGB effectively redistributes the
envelope's angular momentum content such that most of it is stored in
the deeper layers, therefore spinning up those immediately above its
base.  Comparison of the mass profiles of the specific angular
momentum $J/M$ in panels A and C in Figure \ref{fig:Jsnapshots}
clearly illustrate this phenomenon.  Hence, hydrodynamical
instabilities are triggered in both cases of angular momentum
distribution in the convective envelope, but the constant $J/M$ case
will experience mixing velocities more than $10^4$ times larger ($D
\propto \Omega^2$) than those of the case with solid body rotation.
This is a solid result of all models of rotational mixing since the
work of \citet{swe79}, and constitutes the ultimate reason for the
departure of the extra mixing experienced by the two cases of rotation
law in the convective envelope (see also discussion of Figure
\ref{fig:vels} ahead).

\begin{figure}
\vspace*{1cm}
{\hspace{-19cm}
\plotfiddle{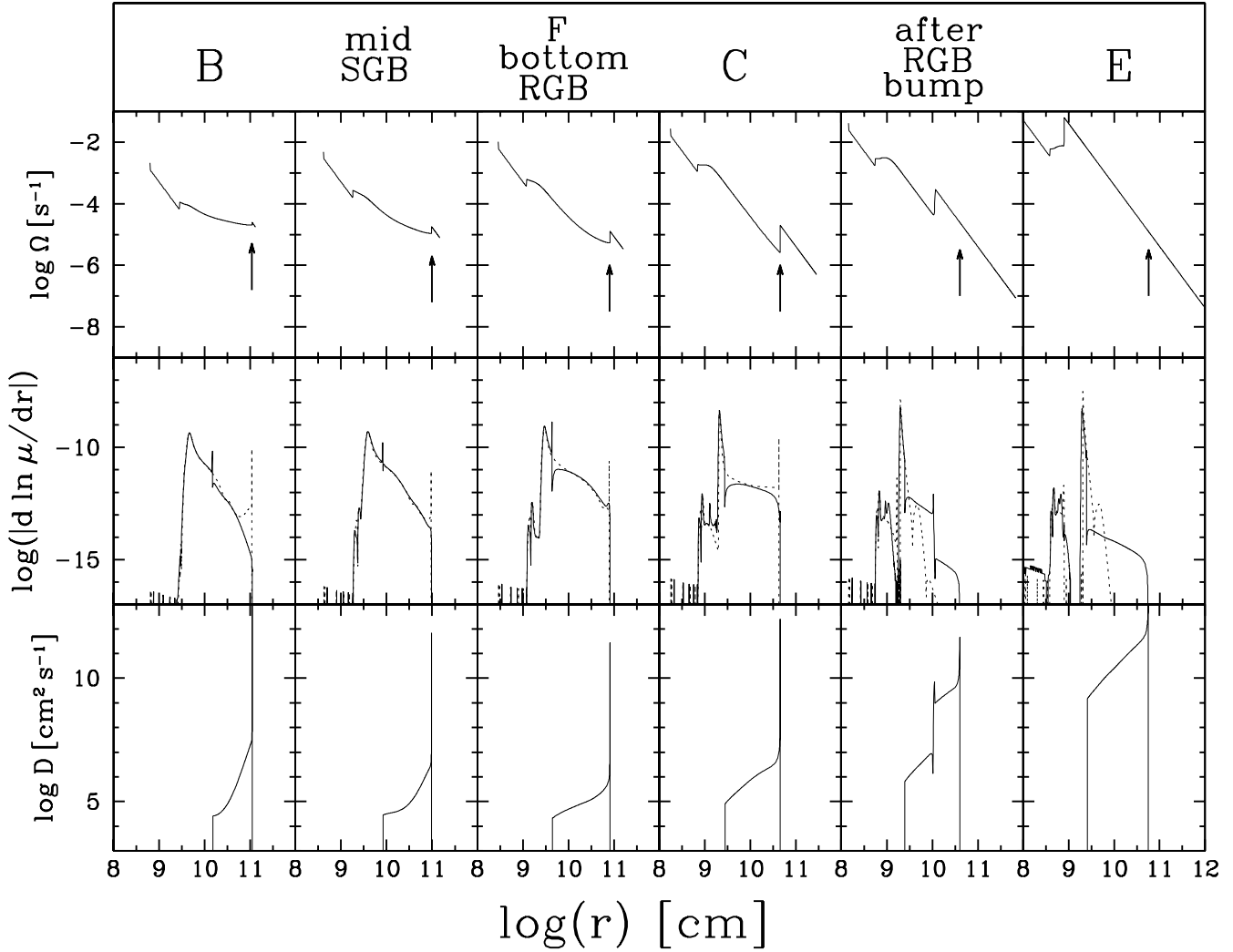}{1cm}{-90}{420}{550}{0}{0}}
\vspace*{-2cm}
\caption{\label{fig:coeffs} Radial profiles of angular velocity,
gradient of mean molecular weight, and the diffusion coefficient
associated with extra mixing at six different points in the post-MS
evolution for a rotating model of a 1.3\msun star of solar metallicity
that started with a rotational velocity of $40\kms$ at turnoff.  The
B, C, E, and F labels refer to the locations shown on the evolutionary
track of Figure \ref{fig:rotnorot}.  The base of the convective
envelope at each snapshot is indicated by arrows.  The dotted lines in
the $\mu$-gradient panels represent the standard model
profiles. }\end{figure}

\begin{figure}
\vspace*{0cm}
{\hspace{-18cm}
\plotfiddle{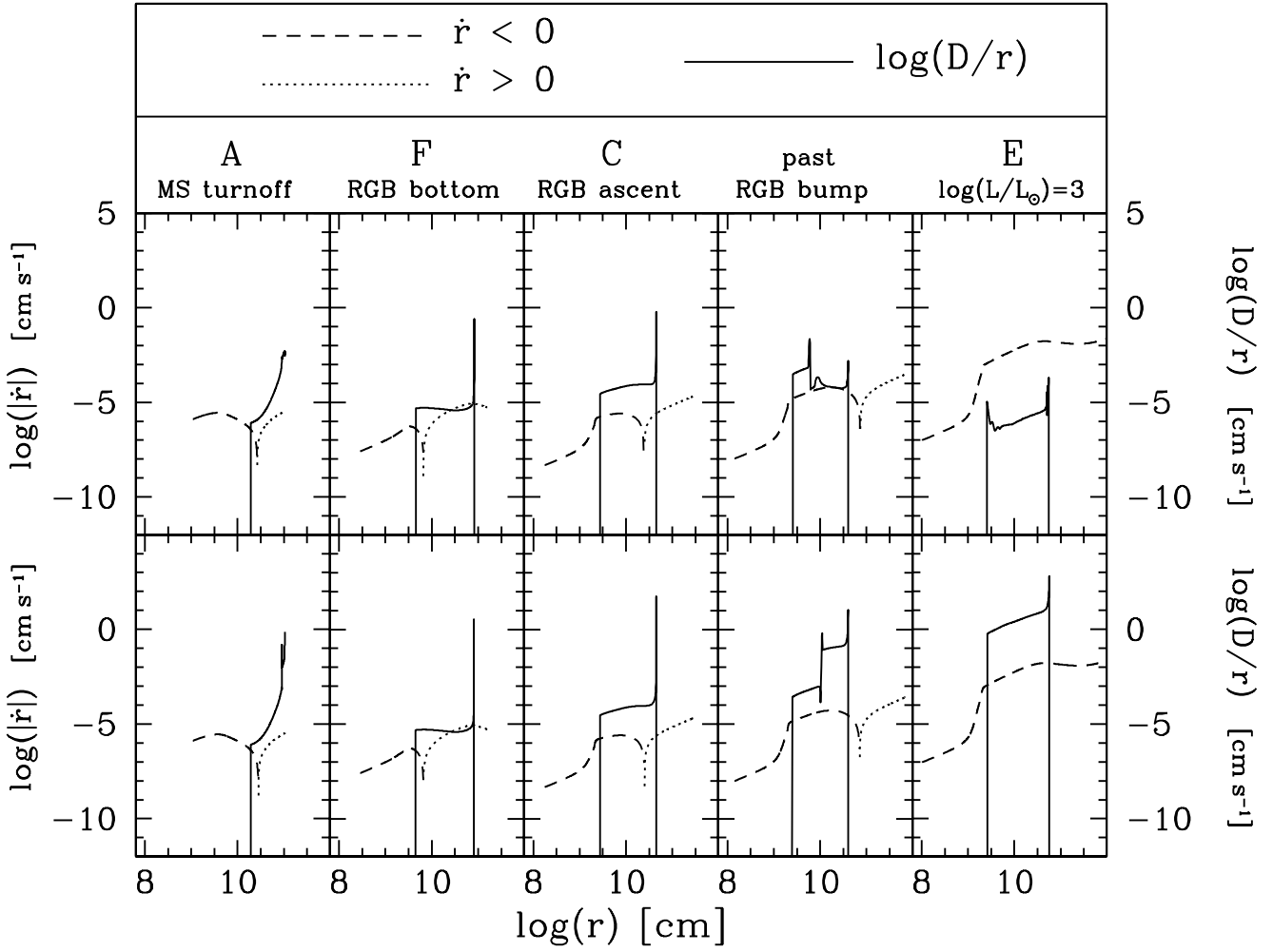}{1cm}{-90}{420}{550}{0}{0}}
\vspace*{-4cm}
\caption{\label{fig:vels} Diffusion velocities of extra mixing against
the instantaneous velocity field of matter inside a star of 1.3\msun
and solar metallicity from the MS turnoff to close to the RGB tip.
The A, C, E, and F labels refer to the locations shown on the
evolutionary track of Figure \ref{fig:rotnorot}.  Dashed lines
represent infalling velocities, dotted lines represent regions in
expansion, and the solid line is an order-of-magnitude estimate of the
velocity at which extra mixing proceeds in our models.  The upper
sequence of panels correspond to the case of solid body rotation in
the convective envelope, and the lower sequence to a constant specific
angular momentum distribution in the envelope.  Giants with any of the
angular momentum laws in the convective envelope behave identically
until the location of the RGB bump, after which only the case with a
differentially rotating envelope achieves diffusion velocities larger
than the velocity at which the material falls from the convective
envelope, and thus this is the case able to affect the surface
abundances. }\end{figure}

\begin{figure}
\vspace*{0cm}
{\hspace{-16cm}
\plotfiddle{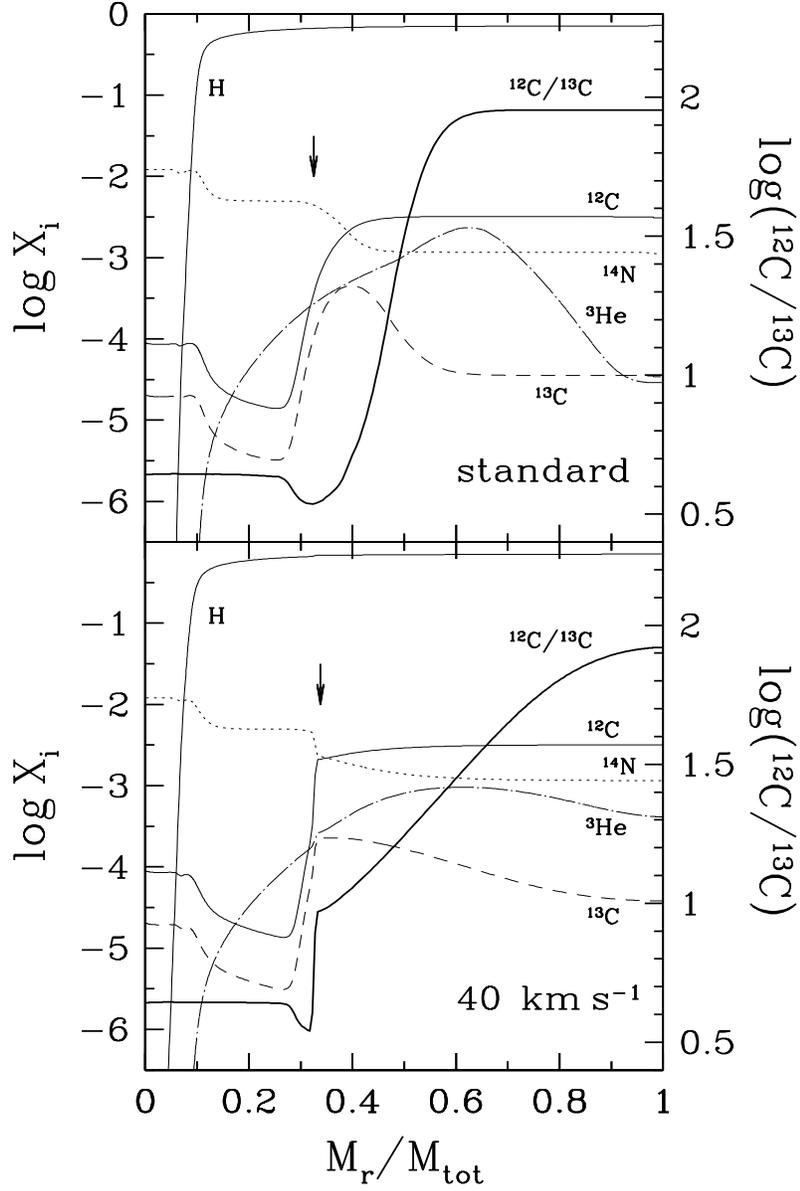}{1cm}{0}{400}{500}{0}{0}}
\vspace*{-0.5cm}
\caption{\label{fig:Xi_profiles} Internal composition profiles of some
relevant species, as well as the \isot ratio, for a 1.3\msun star of
solar metallicity on the \sgb (marked as B in Figure
\ref{fig:rotnorot}, i.e., $\log(L/L_{\odot}) \sim 0.7$).  The upper
panel shows a standard model, and the lower panel is a rotating model
that started with a rotational velocity of $40\kms$ at turnoff.  The
arrows indicate the location of the base of the convective envelope at
its point of maximum depth during the first dredge-up.  Comparison of
both panels clearly reveals the smoothing effect of the extra mixing
on the abundance profiles, which produces the slight difference of the
surface \isot ratio between the standard and the rotating cases even
before the onset of the first dredge-up.  }\end{figure}

Figure \ref{fig:coeffs} shows the evolution of the radial profiles of
angular velocity, composition gradient, and the diffusion coefficient
associated with the extra mixing for a 1.3\msun star of solar
metallicity (our M67-like case).  Only the case capable of extra
mixing, i.e., that with constant specific angular momentum in the
convective envelope, is shown.  The upper panels include some of the
snapshots of angular velocity profile also appearing in Figure
\ref{fig:Jsnapshots} (labeled B, C, and E), along with others chosen
because of their relevance to the present discussion, and again
illustrate the building of a large angular velocity gradient soon
after the MS turnoff.  The location of the convective envelope is
indicated with an arrow in each panel.  Note that, while the star is
progressively expanding, the base of the convective envelope remains
almost at the same physical position, $10.5 \lesssim \log\,r \lesssim
11$, even though it is at the same time incorporating an increasingly
larger fraction of the star's total mass until the first dredge-up
finishes around point C (see also the mass profiles in
Fig. \ref{fig:Jsnapshots}).

Until point C in Figure \ref{fig:coeffs}, the location of the arrow
coincides with a jump or step in the angular velocity profile already
noticed in \S\,4 and which, rather than being an artifact of our
choice of initial conditions, is instead a physical phenomenon that is
fundamental to our picture of rotational mixing.  This jump
corresponds to the same discontinuity present in the specific angular
momentum profiles of Figure \ref{fig:Jsnapshots} for the case of
constant $J/M$ in the convective envelope, and is the natural
consequence of three factors.

First, the condition of constant specific angular momentum in the
convective envelope ensures that the angular momentum content of the
material incorporated into the envelope during the deepening of the
convection zone is redistributed such that most of it is stored in the
deepest layers of the envelope, that is, where most of the mass lies.
We further assume local conservation of angular momentum in the
radiative region, which means that nothing happens to the angular
momentum profile of the region below the point of maximum depth of the
convection zone.  Finally, the absence of any transport of angular
momentum in regions of strong angular velocity gradients also
effectively decouples the rapidly spinning base of the convective
envelope from the slowly rotating radiative core.  If, on the other
hand, there was effective angular momentum transport in differentially
rotating regions, then this discontinuity would tend to be smoothed.
However, we argue that the real situation could actually closely
resemble that depicted here because $\mu$-barriers may have the effect
of suppressing circulation and inhibiting extra mixing.  If such were
the case, angular momentum transport by hydrodynamic means would also
be suppressed, and a discontinuity in the angular momentum
distribution similar to that seen in our models should appear.

The middle panels of Figure \ref{fig:coeffs} show the evolution of the
gradient of chemical composition from the \sgb to the tip of the RGB
for the rotating model of a 1.3\msun star of solar metallicity.  The
sharp edge of the profiles at large radii marks the base of the
convective envelope, where composition gradients vanish.  Note also
the evolution of the HBS, slightly broad past the MS turnoff in panel
B, and progressively narrowing thereafter.  Similar to the base of the
convective envelope, the physical location of the HBS remains
approximately constant (its peak always around $\log\,r \approx 9.5$),
as hydrogen-rich material falls from the convective envelope and is
afterward deposited in the growing core of the star.

In the bottom panels of Figure \ref{fig:coeffs} we show the evolution
of the diffusion coefficient associated with rotational mixing.  Note
that the profiles are always confined by the outer edge of the HBS on
the inner side (more exactly, where the luminosity reaches 99\% of the
total luminosity, see \S\,3.1) and the base of the convective envelope
on the outer side.  Inspection of panels B through C (SGB and lower
RGB) provides a clear demonstration of the direct proportionality
between the diffusion coefficient, i.e., the strength of extra mixing,
and the total luminosity of the star (as is expressed in equation
\ref{eqn:epsilon}).  On the one hand, since the locations of the HBS
and the base of the envelope remain approximately the same and,
furthermore, local angular momentum is conserved, this radiative
region does not experience major structural changes across these
evolutionary stages.  On the other hand, the luminosity does change by
a factor of $\sim 2.5$ from B to C, experiencing a local minimum at an
intermediate location (marked as F in Fig. \ref{fig:rotnorot}), and
one can verify in the sequence of lower panels of Figure
\ref{fig:coeffs} that the diffusion coefficient simultaneously
experiences the same minimum.

The rapid structural evolution of red giants makes rotational mixing a
threshold process.  To be reflected in the surface abundances,
rotational mixing must overcome the infalling velocity field inherent
to the region of the red giant where it occurs.  In Figure
\ref{fig:vels} we show, for our M67-like giant, both the instantaneous
velocity field of the material inside the star and a first-order
estimate of the diffusion velocity of extra mixing, for various
evolutionary stages from the MS turnoff to close to the tip of the
RGB.  The dashed lines represent infalling material ($\dot{r} < 0$),
the dotted lines represent material moving outward ($\dot{r} > 0$),
and the solid line is our estimate of the velocity of extra mixing.
The upper set of panels shows the case of a model with a rigidly
rotating convective envelope.  No extra mixing is experienced by this
model, and the diffusion velocities are always smaller than those
associated with the structural changes (either infall or expansion)
all the way from the MS turnoff (panel A) to the tip of the RGB (past
panel E).  For models with differentially rotating envelopes (lower
set of panels), the opposite occurs, and mixing eventually overcomes
the infalling velocity field.

As demonstrated by Figure \ref{fig:coeffs} and \ref{fig:vels}, our
scenario contemplates extra mixing at all evolutionary stages
immediately past the MS turnoff, including the \sgb and the lower RGB.
At the same time it reproduces the observed lack of well mixed stars
before the location of the RGB bump as a natural consequence of the
still small amount of energy available to drive rotational mixing
during those early stages (equation \ref{eqn:epsilon}), i.e., for
reasons different than so far claimed in the literature
\citep{cha98a,den03}.  

Finally, we have also shown that the key ingredient for driving
efficient extra mixing on the upper RGB is the angular momentum
content of the material falling from the convective envelope into the
radiative region above the HBS.  The conclusion that the $\mu$-barrier
is not the agent preventing the extra mixing is evidenced by the fact
that our models with solid-body rotation in the convective envelope
are not able to mix even after the $\mu$-barrier has already been
erased by the advancing HBS (\S\,5.2 and \S\,5.4).  

Once again, we stress that this constitutes a crucial theoretical
departure from all previous models of extra mixing in red giants: in
our models, canonical extra mixing becomes a vigorous process only at
the location of the RGB bump, not necessarily because a steep
$\mu$-barrier inhibits the mixing of chemicals to occur earlier in the
RGB evolution (which it may or may not be able to do), but, more
fundamentally, because a steep $\mu$-barrier would also inhibit the
transport of angular momentum between the rapidly rotating base of the
convective envelope and the radiative region, thus creating the
appropriate conditions of angular momentum distribution that permits
extra mixing to occur.

\subsection{Mixing on the SGB and lower RGB}

Figure \ref{fig:Xi_profiles} shows selected internal abundance
profiles for a 1.3\msun model of solar metallicity (M67-like case)
located on the \sgb (point B in Fig. \ref{fig:rotnorot}), where the
upper panel corresponds to a standard model and the lower panel to a
rotating model.  In contrast to the standard case, the small amount of
extra-mixing during the early post-MS stages in the rotating models
has the effect of flattening the internal profiles but, with the
exception of ${\rm ^{3}He}$, has little impact on the abundances at
the surface.  Indicated with arrows is the location of the base of the
convection zone at its point of maximum penetration during the first
dredge-up $(M_{\rm r}/M_{\rm tot} \approx 0.34)$.  The discontinuity
in all the composition profiles of the rotating model (lower panel) is
artificial, and is due to the maximum depth allowed for extra mixing
to occur in our models (approximately the outermost edge of the HBS).

\begin{figure}
\vspace*{-1cm}
{\hspace{-16cm}
\plotfiddle{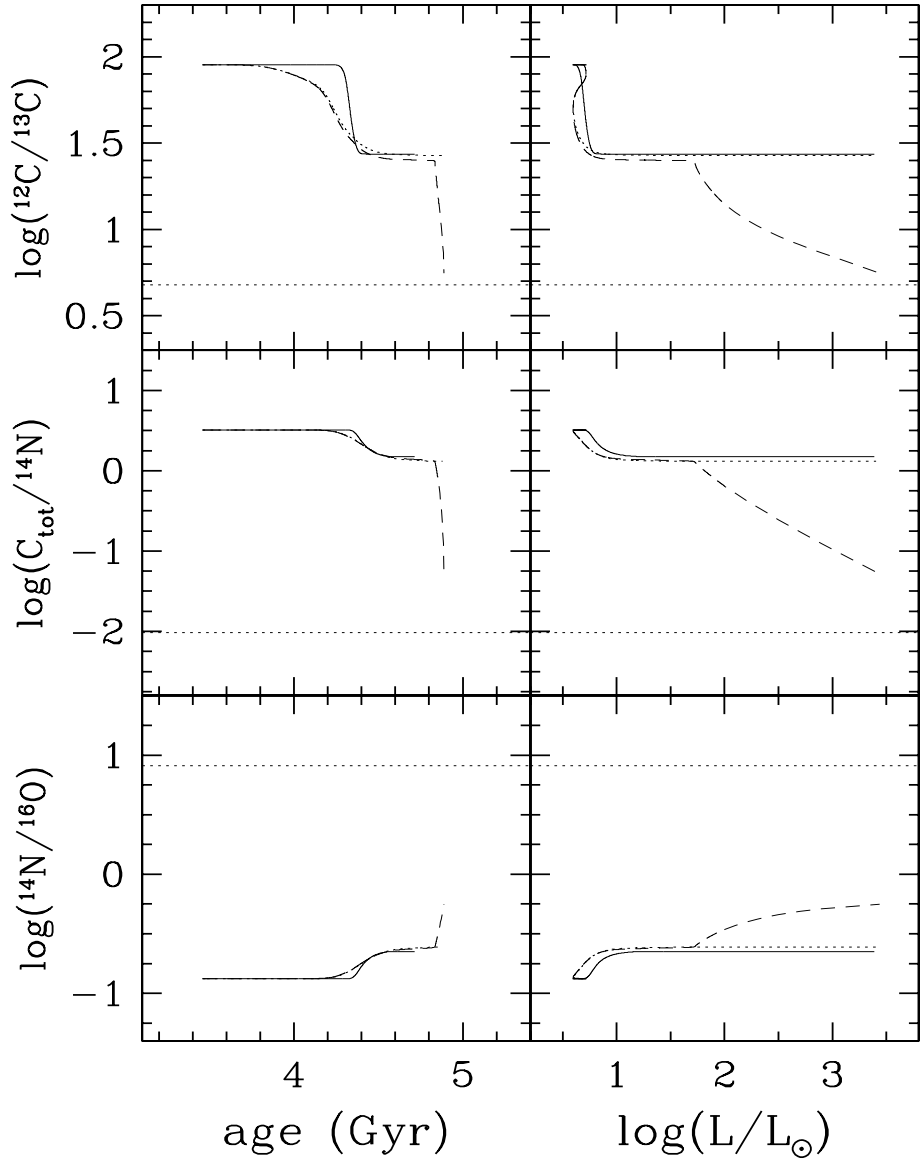}{1cm}{0}{400}{500}{0}{0}}
\caption{\label{fig:mix1st} The surface ratios of the carbon
isotopes, carbon to nitrogen, and nitrogen to oxygen, for the case of
a 1.3\msun star of solar metallicity, as a function of age and
luminosity since the MS turnoff.  Solid lines correspond to the
standard model, i.e., without any rotation.  The other two lines
represent rotating models that started with a rotational velocity of
$40\kms$ at turnoff, for the two different rotation laws in convective
regions studied: dotted lines = solid body, and dashed lines =
constant specific angular momentum.}\end{figure}

The evolution of the surface \isot ratio, as well as that of carbon to
nitrogen, and nitrogen to oxygen, is shown in Figure \ref{fig:mix1st},
for both standard (solid lines) and rotating (broken lines) models of
a 1.3\msun star of solar metallicity.  In the standard model, the
changes in surface abundance ratios only occur during the first
dredge-up, and take place early in the ascent up the giant branch
(almost completely done at $\log(L/L_{\odot}) \sim 1$, right panels).
By the end of the first dredge-up, the top panels show that the carbon
isotope ratio \isot has dropped from the initial (solar) value of
$\sim$ 100 to $\sim$ 25.  As one moves to ratios that are more
temperature sensitive the dredge-up effects decrease, and one can see
in the middle and bottom panels the characteristic overall net
conversion of carbon and oxygen into nitrogen, typical of CNO
processing.

The broken lines in all panels of Figure \ref{fig:mix1st} represent
the two extreme cases of angular velocity law in the convective
envelope: dotted lines correspond to rigid, solid-body rotation, while
dashed lines correspond to constant specific angular momentum $J/M$.
All rotating models started at the MS turnoff with a rotational
velocity of 40 $\kms$, which lies at the high end of the range of
rotation rates shown by the precursors of this type of star
(Fig. \ref{fig:hyades} and discussion in \S\,3.3).  In the rotating
models, the changes start even before the first dredge-up takes place
in the standard models.  This is the result of subsurface changes in
the composition profile of our models in early post-MS stages driven
by the departure from spherical symmetry.  

However, rotation does not significantly affect the post first
dredge-up ratios, regardless of the angular momentum law in the
convective envelope.  Then, after dredge-up, the abundance ratios in
the two rotating models remain constant throughout the entire lower
RGB, a general result of our models without invoking any inhibiting
effect from $\mu$-barriers.

The fact that extra mixing, although present in small amounts, is
inefficient on the \sgb and lower RGB is not surprising when one
considers that the energy $\epsilon$ available to drive circulation
can be expressed as a fraction of the total luminosity,

\begin{equation}
\epsilon \approx \chi\,{L} \approx \biggl(\frac{\Omega^2\,r}{g}\biggr)_{\rm eq}\,{L}\, ,
\label{eqn:epsilon}
\end{equation}

\noindent with $\chi$ usually expressed as the ratio between the
centrifugal acceleration and that of gravity at the equator, thus
measuring the departure from spherical symmetry.  While the fraction
$\chi$ does not change much all along the RGB, the star's luminosity
during the \sgb and the lower RGB is one or two orders of magnitude
smaller than during the upper RGB, and thus rotational mixing is
correspondingly inefficient in these early post-MS stages.  Note that
this happens independently of any inhibiting effect of $\mu$-barriers:
while most investigations suppress mixing on the lower RGB because of
the existence of a strong $\mu$-barrier, we do not enforce such
suppression, and obtain that our lower RGB models exhibit minimal
mixing anyway.

Even though the globular clusters, particularly the data on the carbon
isotopes (\S\,6.3), seem to show strongly mixed stars only in the
upper RGB and not before, this should not be taken as confirmation
that extra mixing ``only'' occurs after the RGB bump, nor as evidence
for any inhibiting effect of $\mu$-barriers.  Instead, if one were to
follow equation (\ref{eqn:epsilon}), the lack of well mixed stars on
the \sgb and the lower RGB is a natural consequence of the small
amount of energy available to perform mixing, given that the
luminosity of the star is still too low in comparison to its
luminosity once on the upper RGB.  Thus, it seems more appropriate to
talk about inefficient, albeit present, extra mixing on the lower RGB,
rather than extra mixing occurring only on the upper RGB (we also
refer the reader to the discussion of Figs. \ref{fig:coeffs} and
\ref{fig:vels} in \S\,5.1).

Even in our most mixed models, the vigorous extra mixing only kicks in
when the star is on the upper RGB, and this rather abrupt transition
happens approximately at the expected location of the RGB bump.  The
important difference is, however, that since we have not considered
any inhibiting effect of composition gradients in the models, {\em the
disappearance of the $\mu$-barrier when the HBS crosses it is not the
reason behind the onset of the strong mixing at that particular point
in the evolution}.  Instead, as discussed in \S\,5.1, what really
matters is the angular momentum content of the material falling from
the retracting (in mass coordinate) convective envelope into the
radiative interior above the HBS, where the extra mixing is at work.

Therefore, rotational mixing before and after the RGB bump may be
thought of as the difference between the two modes of mixing provided
by rotation.  In the absence of any inhibition by $\mu$-barriers,
rotational mixing is active at any time on the RGB in the form of
meridional circulation driven by the departure from spherical symmetry
(equation \ref{eqn:epsilon}).  On the other hand, rotational mixing
associated with strong differential rotation only becomes important
when the right conditions of angular momentum content are matched
(\S\,5.1), which in our models occurs at the approximate location of
the RGB bump.

\subsection{The RGB bump and the effect of $\mu$-barriers}

In \S\,5.2 we showed that the CNO abundances previous to the location
of the RGB bump are not sensitive to whether mixing is inhibited or
not by $\mu$-barriers, and therefore they should not be taken as
evidence for such inhibition.  However, here we find that it is
another indicator, the very existence of the RGB bump itself, the one
that indeed provides sound evidence for inhibition of mixing by
$\mu$-barriers.

\begin{figure}
\vspace*{-1cm}
{\hspace{-15cm}
\plotfiddle{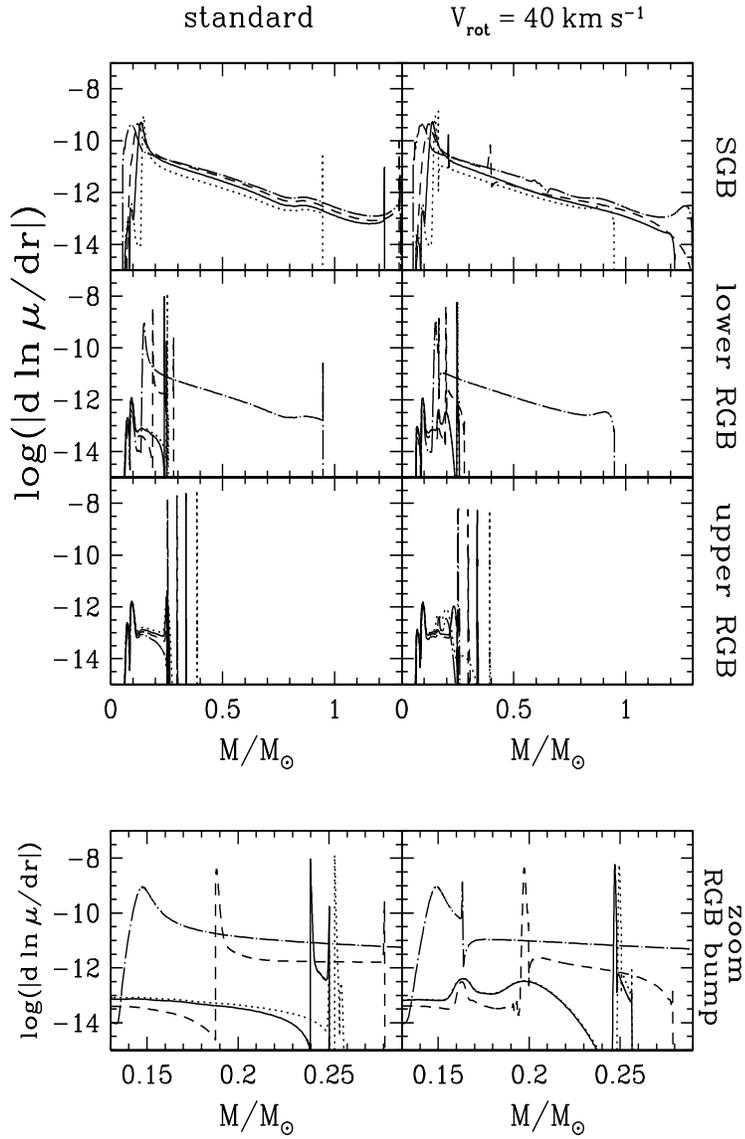}{1cm}{0}{300}{450}{0}{0}}
\caption{\label{fig:mu_standard} Evolution of the gradient of mean
molecular weight for a standard model (left hand panels) and a
rotating model (right hand panels) of a 1.3\msun star of solar
metallicity.  Each of the upper three pairs of panels shows the
profile of $\mu$-gradient at four closely consecutive evolutionary
stages located at the \sgb, the lower RGB, and the upper RGB, as
indicated by the labels at the right.  The order of the profiles in
each panel is, with increasing evolutionary state (age): dot-long dash
lines, dashed lines, solid lines, and finally dotted lines.  The last
profile on each pair of panels is then plotted again as the first one
on the panel immediately below.  The bottom pair of panels shows zooms
of the $\mu$-gradient profiles around the location of the RGB bump.
With respect to Fig. \ref{fig:rotnorot}, the different lines
represent: dot-long dash = F, dashed = C, solid = just before the RGB
bump ($\log(L/L_{\odot}) = 1.63$), and dotted = just past the RGB
bump, ($\log(L/L_{\odot}) = 1.66$). }\end{figure}

Our models, with no inhibition of mixing by $\mu$-barriers, erase the
discontinuity in the hydrogen profile that gives rise to the standard
$\mu$-barrier, and this happens even for very slow rotation rates.
Figure \ref{fig:mu_standard} shows the evolution of the gradient of
chemical composition from the MS turnoff to the tip of the RGB, for
both a standard model (left hand panels) and a rotating model (right
hand panels) of a 1.3\msun star of solar metallicity with a turnoff
rotation rate of $40\kms$.  Each of the upper three pairs of panels
shows the profile of $\mu$-gradient at four closely consecutive
evolutionary stages located at the \sgb, the lower RGB, and the upper
RGB, as indicated by the labels at the right.  The order of the
profiles in each panel is, with increasing evolutionary state (age):
dot-long dash lines, dashed lines, solid lines, and finally dotted
lines.  The last profile on each pair of panels is then plotted again
as the first one on the panel immediately below.  

The first peaks on the left edges of the profiles in Figure
\ref{fig:mu_standard}, very broad close to the turnoff and rapidly
narrowing with advancing evolutionary state, correspond to the HBS.
The growth of the convective envelope, which starts during the \sgb
and continues to the lower RGB, can be followed by tracking the motion
of the right edge of the profiles of $\mu$-gradient in the upper two
pairs of panels.  Below, the pair of panels corresponding to the upper
RGB show how the extremely narrow HBS gradually advances with
increasing luminosity, depositing the ashes of hydrogen burning into
the growing helium core.  Note that all the sharp edges of the
profiles of $\mu$-gradient in the standard case are smoothed by extra
mixing in the rotating models.

The bottom pair of panels in Figure \ref{fig:mu_standard} shows the
$\mu$-gradient profiles near the location of the RGB bump.  Note, in
the standard model case (left panel), the sharp right edge of the
$\mu$-gradient profiles, located at $M/M_{\odot} \approx 0.28$ for the
dashed line (point C in Fig. \ref{fig:rotnorot}), and $M/M_{\odot}
\approx 0.248$ for the solid line (just before the occurrence of the
RGB bump, $\log(L/L_{\odot}) \sim 1.6$): they correspond to the
standard $\mu$-barrier, created by the base of the convective envelope
as it penetrates during the first dredge-up.  The locations of these
two lines differ because at point C ($\log(L/L_{\odot}) = 1$) the
first dredge-up has not ended yet, and the base of the convective
envelope is still deepening in mass.  The model just after the RGB
bump is represented by the dotted line in the bottom pair of panels.
In the standard case (left panel) the $\mu$-barrier does not exist
anymore, since it was erased by the pass of the HBS (now being the
highest peak of the dotted line) through that location.

On the other hand, in the rotating case, the $\mu$-barrier is almost
entirely erased by the action of mixing well before the location of
the bump.  Given that the RGB bump is a real feature already seen in
the CMDs and LFs of various Galactic and extragalactic systems, it is
clear that stellar models should not get rid of $\mu$-barriers as
easily as ours.  

Therefore, our results actually provide the first convincing evidence
that $\mu$-barriers must indeed inhibit mixing across them, since
otherwise the RGB bump would easily disappear.  Note the large
qualitative departure of this claim from those of all previous works
on the subject, which mistakenly identified the lack of well mixed
stars on the lower RGB as evidence for the inhibiting effect of
$\mu$-barriers.

\subsection{Upper RGB - Canonical extra mixing}

Differential rotation provides an additional mixing mechanism which
becomes important only on the upper RGB, and different assumptions on
the distribution of angular momentum in the convective envelope result
in different amounts of rotationally induced mixing (\S\,5.1).

Figure \ref{fig:mix1st} shows that after first dredge-up the abundance
ratios remain constant for both cases of angular velocity distribution
in the envelope only until the approximate location of the RGB bump
($\log(L/L_{\odot}) \sim 1.6$ for our M67-like giant).  At this
position, the situation is very different for the case of rotation
with constant $J/M$ in the envelope: this model suddenly starts to
experience strong mixing that steadily continues up until the tip of
the RGB.  This is the canonical extra mixing, whose appearance has
been thoroughly discussed in \S\,5.1.

\begin{figure}
\vspace*{-1cm}
{\hspace{-16cm}
\plotfiddle{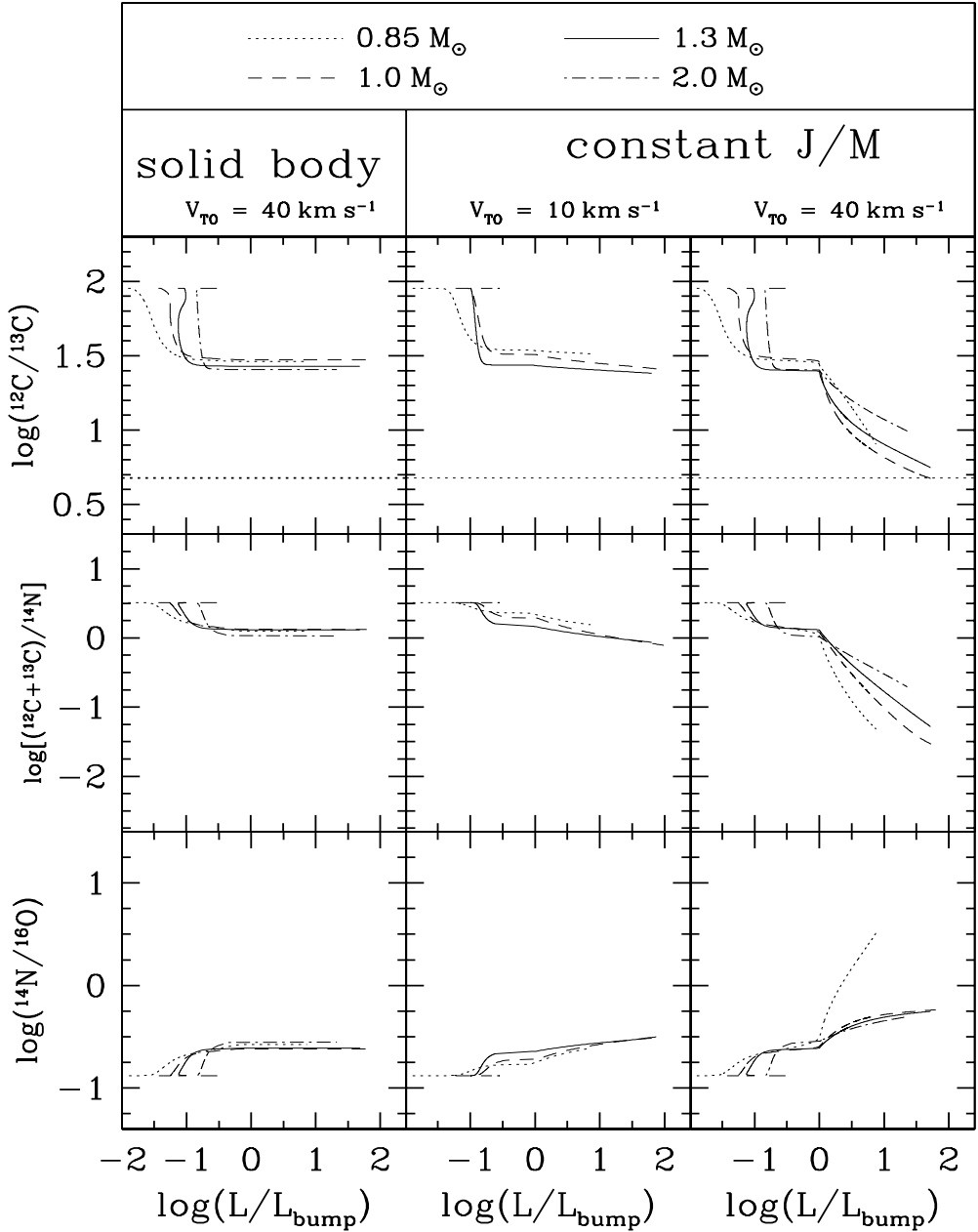}{1cm}{0}{400}{500}{0}{0}}
\caption{\label{fig:massdep} Evolution of the surface abundance ratios
as a function of stellar mass.  Luminosities are plotted with respect
to the luminosity of the RGB bump of the corresponding model in order
to separate metallicity effects due to the first dredge-up from those
of extra mixing.  The 0.85\msun model has ${\rm [Fe/H]=-2.3}$, while
the other three are solar metallicity models.  All masses display
similar behavior, with the differences among them being mainly due to
first dredge-up.  }\end{figure}

\begin{figure}
\vspace*{-1cm}
{\hspace{-16cm}
\plotfiddle{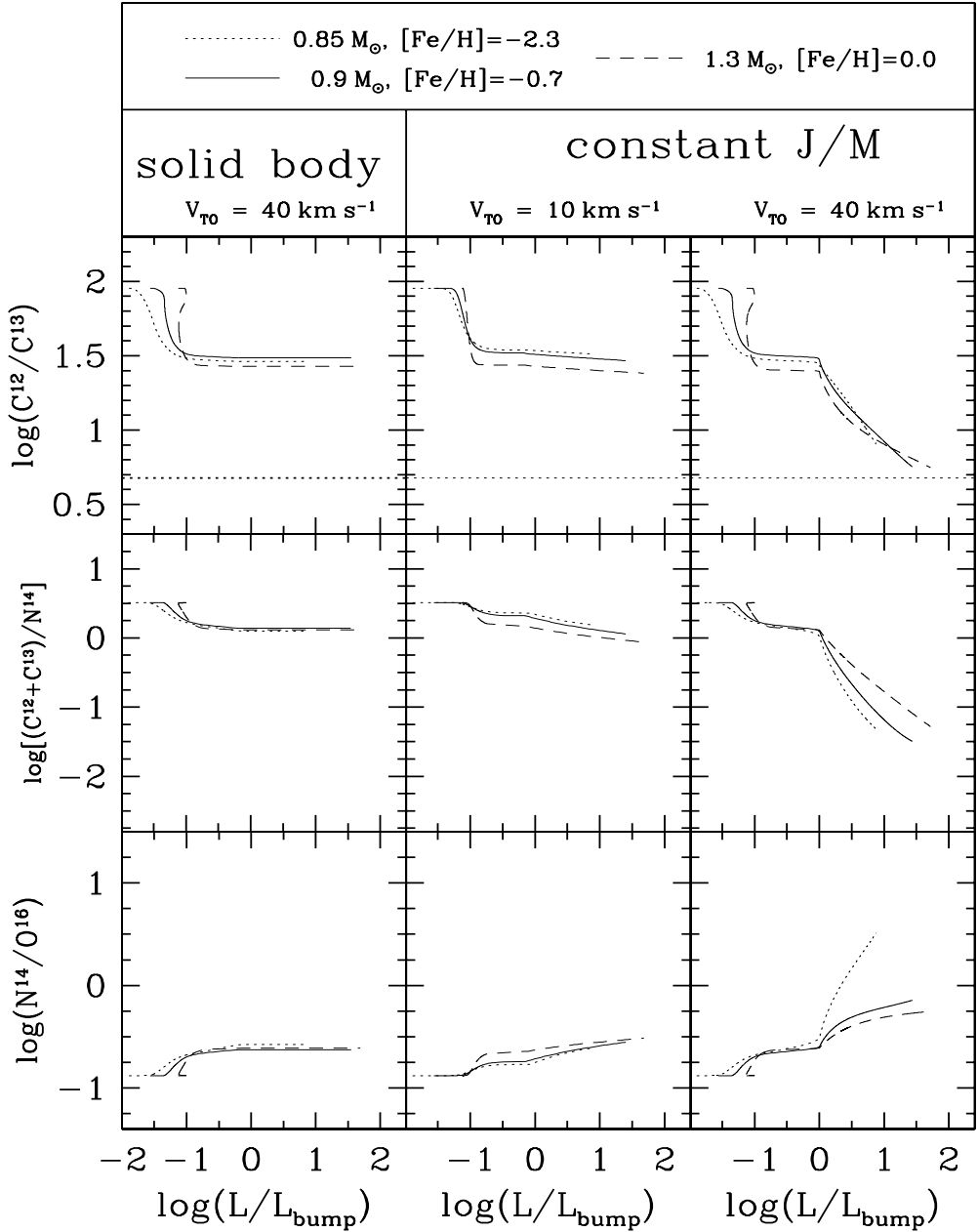}{1cm}{0}{400}{500}{0}{0}}
\caption{\label{fig:FeHdep} Metallicity dependence of the evolution of
surface abundance ratios for rotating stellar models.  Luminosities
are plotted with respect to the luminosity of the RGB bump of the
corresponding model in order to separate metallicity effects due to
the first dredge-up from those of extra mixing.  Left-hand panels
correspond to solid-body rotation in the convective envelope, and
right-hand panels to constant specific angular momentum.
}\end{figure}

The dependence of canonical extra mixing on rotation rate, mass, and
metallicity, are shown in Figures \ref{fig:massdep} and
\ref{fig:FeHdep}, respectively, in which luminosities are normalized
with respect to those of the corresponding RGB bump.  Such a
normalization is optimal for our purposes of looking for mass and
metallicity effects because, as has been extensively discussed before,
extra mixing only starts to be important once the star has passed the
location of this feature in the HR diagram.  Furthermore, in this way
we also separate the effects of extra mixing on the upper RGB from
those due to the first dredge up as well as the different \sgb
morphologies inherent to different stars.  In both figures, the left
hand panels show models with solid body rotation in the convective
envelope for an initial rotation rate of $40\kms$ at turnoff, while
the middle and right hand panels show the results of models with
differentially rotating envelopes and initial rates of 10 and
$40\kms$, respectively.

At all masses, metallicities, and initial rotation rates, only the
models with differential rotation in the convective envelope can mix,
and the strongest effect is that of initial rotation rate: the fastest
the rotation, the more extra mixing.

Once past their respective RGB bumps, the models with constant $J/M$ in
the envelope and initial rotation rate of $10\kms$ show very mild
extra mixing for all the stellar masses shown in Figure
\ref{fig:massdep}, but no obvious trend is distinguishable.  For the
differentially rotating case with larger initial rotation rate
($40\kms$), however, a slight trend can be noticed of more efficient
extra mixing for lower mass stars: all the right hand panels in Figure
\ref{fig:massdep} evidence a larger rate of change in the surface
abundance ratios as the stellar mass decreases, or, in other words,
the slope of the evolution of the abundance ratios is shallower for
higher mass stars.  Note that this trend of progressively more
efficient mixing is in general very mild, with the exception of the
\no ratio in the lowest mass, most metal-poor star, the M92-like case,
which shows a very large level of oxygen depletion for the model with
an initial rotation rate of $40\kms$ that is not shared by any of the
higher mass models.

Metallicity effects are explored in Figure \ref{fig:FeHdep}, where we
plot the evolution of the surface abundances for rotating models of
giants representative of clusters with progressively decreasing
metallicity: M67, M71, and M92.  The situation is very similar to that
of Figure \ref{fig:massdep}.  For an initial rotation rate of $10\kms$
(middle panels), the evolution of the surface abundances is only
slightly different than the standard case.  Although the three lines
are vertically displaced from one another (due to their different
standard first dredge up), the rates of change of the surface ratios
after the RGB bump for the three stars are essentially
indistinguishable.  For a larger initial rotation rate (right hand
panels), however, extra mixing is strong in all of them, and, although
it would be ambiguous to say anything from the evolution of the \isot
ratio, the other two abundance ratios, ${\rm (^{12}C+^{13}C)/^{14}N}$
and \no, clearly indicate more efficient extra mixing for the more
metal-poor stars.  

\begin{figure}
\vspace*{-1cm}
{\hspace{-16cm}
\plotfiddle{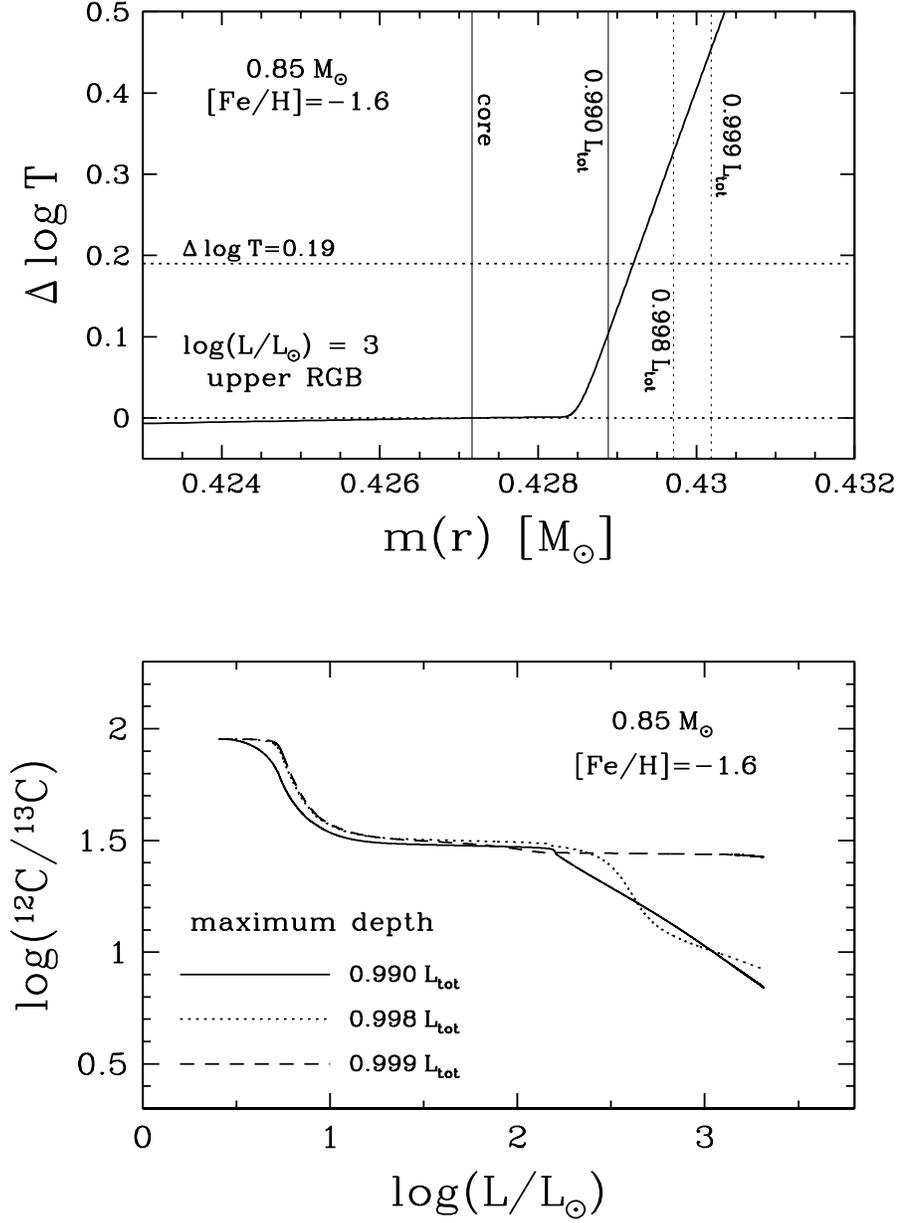}{1cm}{0}{400}{500}{0}{0}}
\vspace*{-0.1cm}
\caption{\label{fig:depth} Exploring the variation of the maximum
depth for extra mixing.  For a star of 0.85\msun and $\feh =
-1.6\,\,(Z=0.0005)$, the upper panel compares the parametrization of
the maximum depth for extra mixing in terms of the temperature inside
the hydrogen burning shell, and our parametrization in terms of some
fraction of the total stellar luminosity.  Here $\Delta\log\,T \equiv
\log\,T(M_{\rm core}) - \log\,T(M_{\rm mix})$, where $M_{\rm core}$
and $M_{\rm mix}$ are the mass coordinate of the core's surface and of
the maximum depth for mixing, respectively.  The dotted horizontal
line at $\Delta\log\,T = 0.19$ indicates the best value found by
\citet{den03} for the same stellar mass and metallicity, which in our
models lies in between the locations at which 99\% and 99.8\% of the
total luminosity is generated.  The lower panel shows that, for mixing
depths between these two mass coordinates, our models produce almost
the same amount of canonical extra mixing, as indicated by the
evolution of the surface \isot ratios.  }\end{figure}

Finally, we explore the effects of varying our choice of the maximum
depth for extra mixing.  As discussed in \S\,3.1, we parametrize the
mixing depth as the location where some (large) fraction of the total
luminosity is generated, that is, to some coordinate close to the top
of, but inside, the HBS.  For no strong reason in particular, we chose
this fraction to be 99\%.  We compare this choice with the recent work
by \citet{den03}, where they find that the mass coordinate at which
$\Delta\log\,T = 0.19$ produces a good match between their models and
the data of field giants.  The upper panel of Figure \ref{fig:depth}
shows the parameter $\Delta\log\,T$ in the region of interest for one
of our models of a 0.85\msun star of $\feh = -1.6$ ($Z = 0.0005$), the
same star studied by \citet{den03}.  It can be seen that our chosen
depth (marked by the vertical line close to $m(r) = 0.429$) is
slightly deeper than the location where $\Delta\log\,T = 0.19$.

In order to quantify the effect of changing the maximum depth, we run
models with shallower mixing, as indicated in the upper panel of
Figure \ref{fig:depth}, where the dotted vertical lines mark the
locations at which 99.8\% and 99.9\% of the total luminosity is
generated.  Note that the preferred mixing depth of \citet{den03},
$\Delta\log\,T = 0.19$, lies in between the depths set by 99\% and
99.8\% of the total luminosity.  The lower panel of Figure
\ref{fig:depth} shows the RGB evolution of the \isot ratio for the
models with maximum depths of 99.8\% (dotted line) and 99.9\% (dashed
line) of the total luminosity, in comparison with our adopted depth
(solid line).  While the model with the most shallow mixing does not
experience almost any extra mixing, the model with just 0.1\% deeper
in total luminosity produces as much extra mixing as that of our
adopted depth.  Therefore, we see that reasonable variations of the
maximum depth for extra mixing do not affect our results on the
evolution of the carbon isotopes.

\section{Results: Comparison to observations}

The general trends discussed in \S\,5 provide a set of theoretical
predictions that can be tested against the large observational
database.  We begin with our benchmark solar abundance case, the old
open cluster M67.  In this system the theoretical predictions match
the observed pattern.  However, we will see that there are some issues
related to the behavior of more massive Population I giants, as well
as with the metal-poor Population II giants in the field and globular
clusters, that depart from the theoretical expectations.  In our view
these two issues provide two distinct clues about the underlying
physical nature of the mixing process, namely, that the overall
efficiency of mixing must be higher than in our base case, and that
there must be a relative reduction in the efficiency of mixing in more
rapidly rotating giants.  We discuss the physical interpretation of
these empirical results in \S\,6.4.

\subsection{Population I giants}

\begin{figure}
\vspace*{-3cm}
{\hspace{-18cm}
\plotfiddle{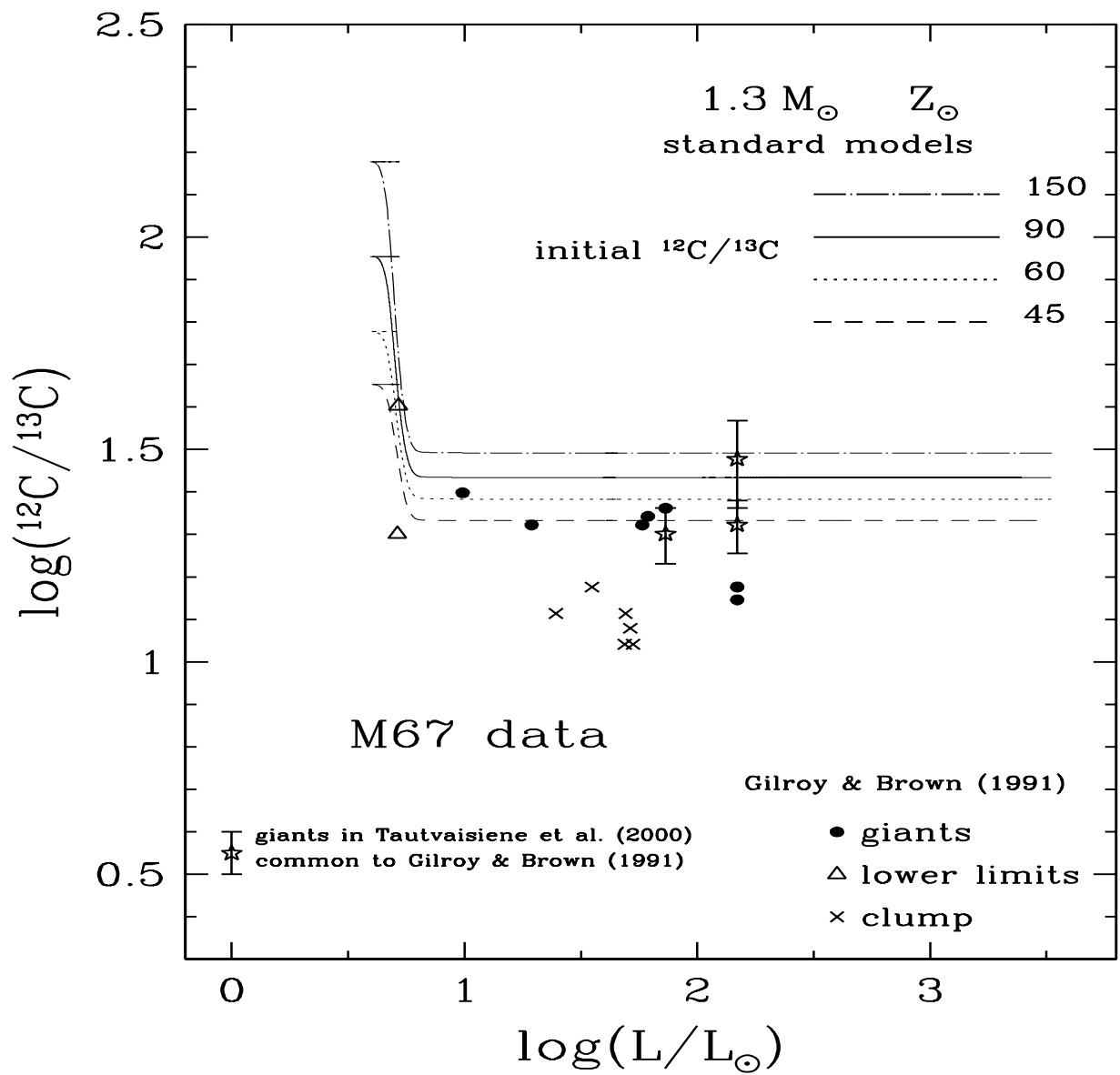}{1cm}{0}{550}{550}{0}{0}}
\vspace*{-1cm}
\caption{\label{fig:m67} Dependence of the post first dredge-up \isot
ratio on the adopted initial \isot value.  The four lines are standard
(i.e., non-rotating) models of M67-like giants with varying MS \isot
ratios, bracketing the range of values measured in the interstellar
medium of the Galaxy.  It can be seen that the surface \isot ratio at
the end of the first dredge-up depends on the initial \isot ratio.
Also shown are the \isot data for M67 evolved stars from two different
sources.  Black dots, open triangles, and crosses are taken from
\citet{gil91}, representing red giants with actual detections of both
carbon isotopes, lower limits, and clump stars, respectively.
Uncertainties in these \isot data are typically between 5 and 8\%.
Stars with error bars represent the measurements of \citet{tau00} of
three giants in common with \citet{gil91}, corresponding to the three
most luminous ones in the sample.  }\end{figure}

\begin{figure}
\vspace*{-6cm}
{\hspace{-17cm}
\plotfiddle{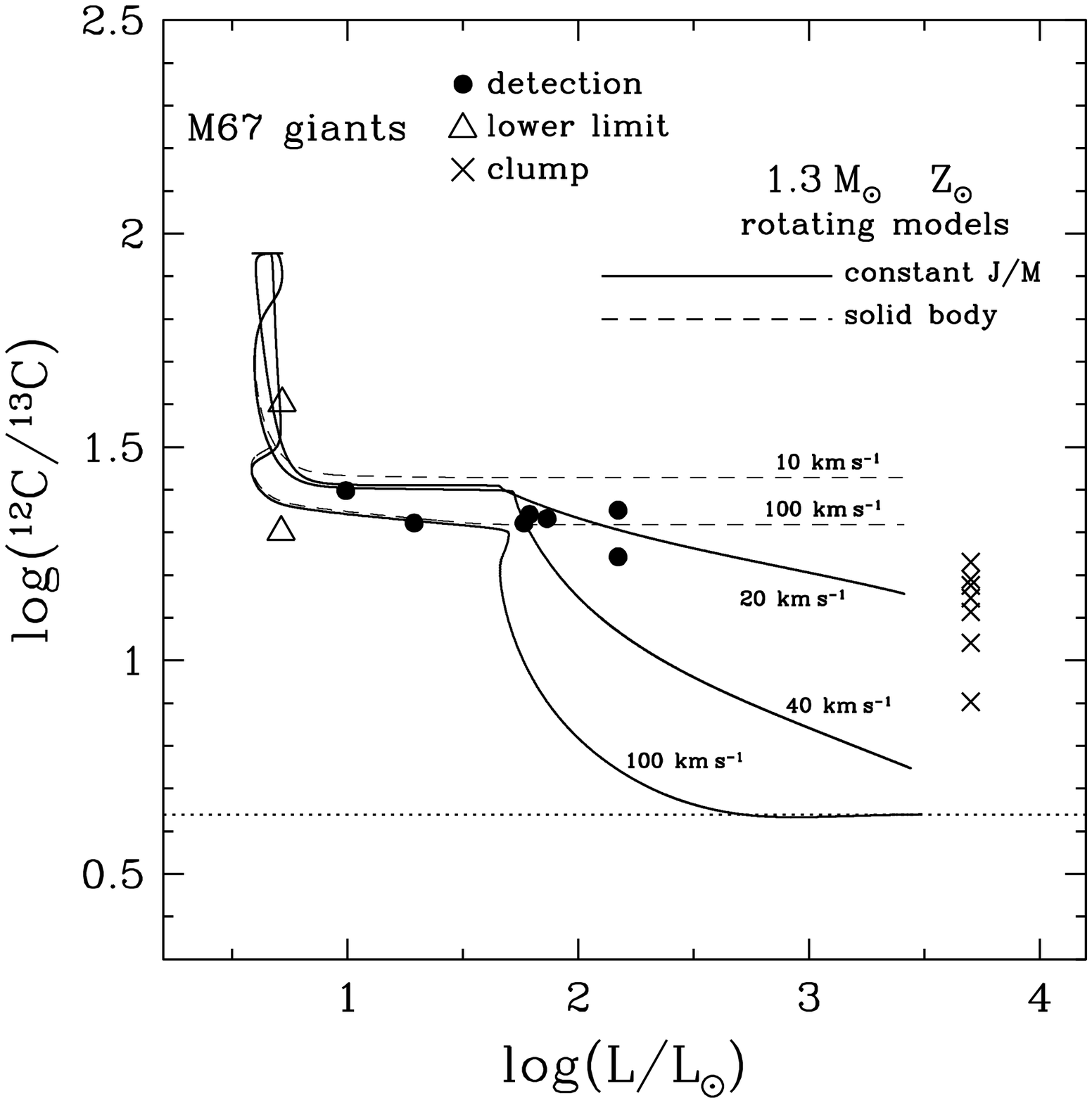}{1cm}{0}{500}{650}{0}{0}}
\vspace*{-2cm}
\caption{\label{fig:mix13msun} Post main sequence evolution of the
surface \isot ratio of a rotating 1.3\msun star of solar metallicity
as a function of the rotational velocity at turnoff, for the two
extreme cases of angular momentum distribution in the convective
envelope.  The dotted horizontal line indicates the nuclear
equilibrium value of \isot $\sim 4.5$.  The \isot data for M67 of
\citet{gil91} and \citet{tau00} have been averaged.  Large dots
represent giants with actual detections, and open triangles are giants
with only lower limits.  The clump stars are represented by crosses
and have been placed at $\log(L/L_{\odot})=3.7$ in order to avoid
crowding and also to indicate that, in an evolutionary sense, they
have already been through the first ascent RGB.  Their actual
luminosities are those shown in Figure \ref{fig:m67}.  }\end{figure}

\begin{figure}
\vspace*{-6cm}
{\hspace{-17cm}
\plotfiddle{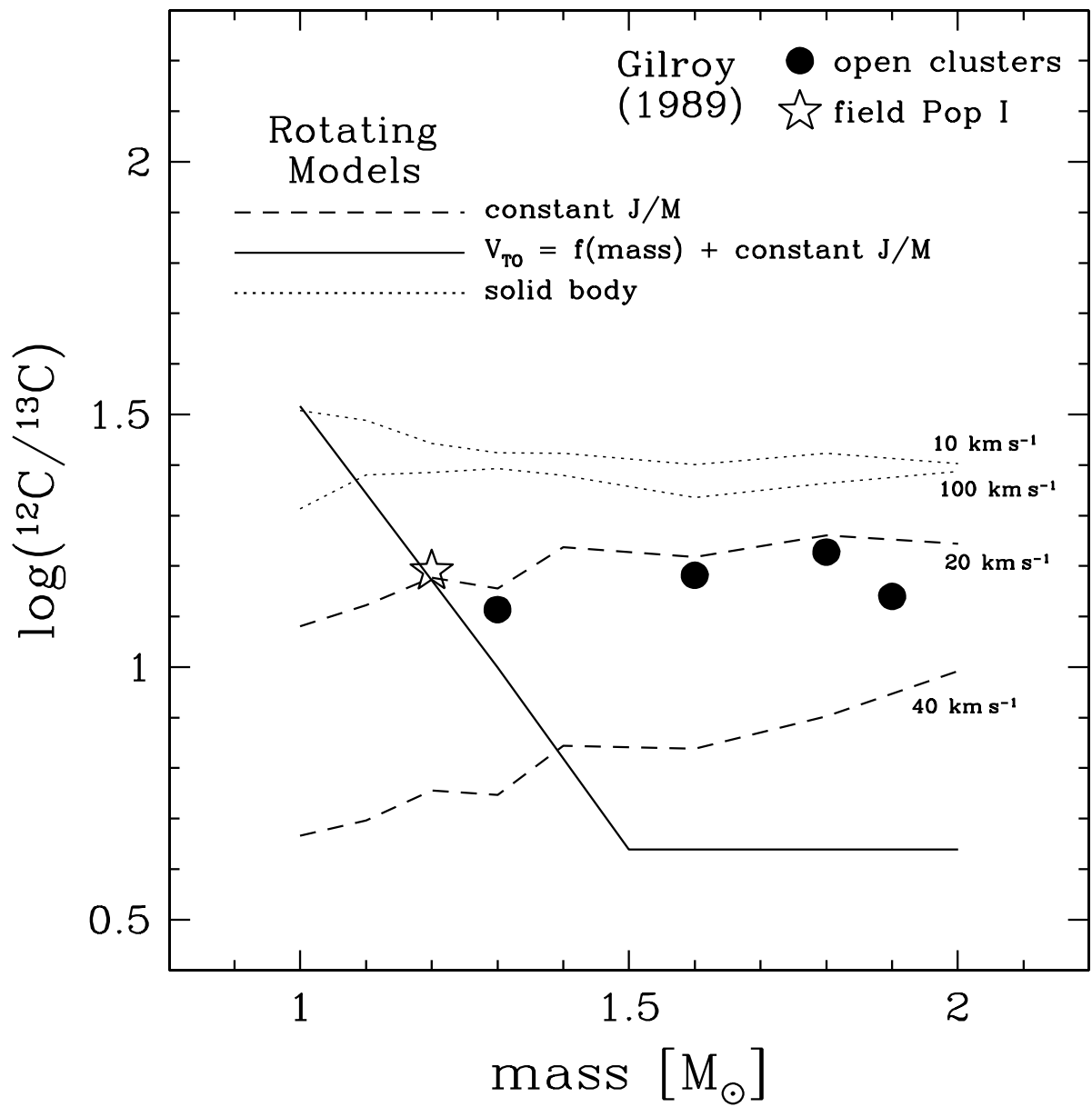}{1cm}{0}{500}{650}{0}{0}}
\vspace*{-2cm}
\caption{\label{fig:tipratios} Comparison between the \isot open
cluster data from \citet{gil89} and our rotating models.  The
theoretical lines represent the \isot ratio at the tip of the RGB,
just before the helium flash.  Dotted and dashed lines are,
respectively, models with solid body and differential rotation
enforced in the convective envelope.  For both types of angular
momentum law, the tip \isot ratios are plotted for two initial
rotation rates, as indicated.  The solid line represents models with
differential rotation in the envelope and initial rotation rates
selected according to the stellar mass as given by the \citet{kra67}
curve.}\end{figure}

Figure \ref{fig:m67} summarizes the \isot data in the old open cluster
M67 \citep{gil91,tau00} and illustrates the dependence of the first
dredge-up on the initial \isot ratio with M67-like standard models.
In Figure \ref{fig:mix13msun} we compare the luminosity evolution of
our rotating models to the M67 data, where the clump giants have been
placed at a luminosity higher than that of the RGB tip to indicate
that, in an evolutionary sense, they are posterior to the first ascent
giants.  Finally, the predicted RGB tip abundances as a function of
mass and rotation rate are compared with open cluster \isot ratio data
\citep{gil89} in Figure \ref{fig:tipratios}.

Three stages can be identified in the M67 data.  The subgiants (open
triangles) clearly have a higher \isot ratio than the more evolved
stars, and the lower RGB data (large dots) level off at \isot $\sim
22$.  These values are consistent with the theoretically expected
first dredge-up and lack of mixing on the lower RGB itself (i.e.,
prior to the RGB bump), regardless of rotation rate, angular momentum
distribution in the convective envelope, or any inhibiting effect of
$\mu$-barriers.  The intermediate luminosity giants
($\log(L/L_{\odot}) \sim 2.2$) have conflicting data; \citet{gil91}
reported lower ratios than for the faint giants, while \citet{tau00}
reported the same abundances for these stars as found for the fainter
giants.  The clump giants (crosses) show the lowest \isot ratios,
indicating the presence of a mixing episode either on the upper RGB
(past $\log(L/L_{\odot}) \sim 1.8$) or associated with the helium
flash.  The first explanation is more consistent with the trends seen
in more densely sampled Population II giant studies.  Models with
turnoff rotation rates similar to the predicted ones for progenitors
of M67 giants ($20$ to $40 \kms$) successfully reproduce the observed
trends, while more rapidly rotating models have too much mixing
(Figure \ref{fig:mix13msun}).

The comparison with \isot data as a function of mass in Figure
\ref{fig:tipratios} is at first glance similarly encouraging, with the
$20\kms$ turnoff rotation models reproducing the observed data.
However, the actual theoretical prediction is that shown by the solid
line since, as discussed in \S\,3.3, the mean rotation rates of
turnoff stars increase dramatically above $\sim 1.3$\msun, so the data
in this mass range are actually much less mixed than we would expect
based upon typical turnoff rotation rates of $150\kms$ above the break
in the \citet{kra67} mass-rotation relationship
(Fig. \ref{fig:hyades}).

The mean measured \isot ratio for the faint M67 giants (lower RGB) is
moderately below the theoretically predicted values, as can be seen in
Figure \ref{fig:m67}.  Some classes of models (fast rotation)
experience internal mixing on the subgiant branch, and it is tempting
to ascribe these lower \isot measurements with the existence of such
early mixing.  However, there is a significant uncertainty in the
initial conditions that is not usually included in theoretical models
of RGB mixing.  The predicted post dredge-up abundances are a
sensitive function of the assumed turnoff \isot ratio.  Although the
solar mix value is commonly assumed, there is no a priori reason why
all stars should share this local property.

Measurements of the \isot ratio toward molecular clouds in the solar
neighborhood show a wide range of values, ranging from as low as 40 to
as high as 150 \citep{goto03}, and there exists an unmistakable trend
of increasing \isot ratio from the Galactic center to the edge of the
Galaxy \citep{ism90}, clearly indicating different stages in the
evolution of carbon at different Galactic locations after a number of
generations of stars.  Until \isot ratios are obtained for dwarfs in
the distant stellar clusters under study (see \citealt{eugenio04}),
the only safe assumption regarding its initial or MS value is that
this should be systematically higher the older the population in
question is.

The dependence of the \isot ratio after first dredge-up is shown by
the four lines in Figure \ref{fig:m67}, which correspond to standard
models with varying initial ratios.  The level of the first dredge-up
among the M67 giants is best explained with initial \isot ratios lower
than the solar system value, but consistent with that inferred from
the large-scale Galactic trend at the position of this cluster.  This
low values, however, may have been higher 4 to 4.5 Gyr ago, when the
Sun and the M67 stars formed.

\subsection{Population II giants: the field}

While the open clusters (Figs. \ref{fig:mix13msun} and  
\ref{fig:tipratios}) represent the high-metallicity half of the
problem of abundance anomalies in red giants, the old field population
and the globular clusters comprise the much better explored, both
observationally and theoretically, metal-poor half.  In summary, the
field data show a non-negligible scatter in the initial abundances, no
changes in the abundance ratios during the lower RGB, and the presence
of non standard mixing operating on stars brighter than the lower RGB
and past the bump.

\begin{figure}
\vspace*{-1cm}
{\hspace{-17cm}
\plotfiddle{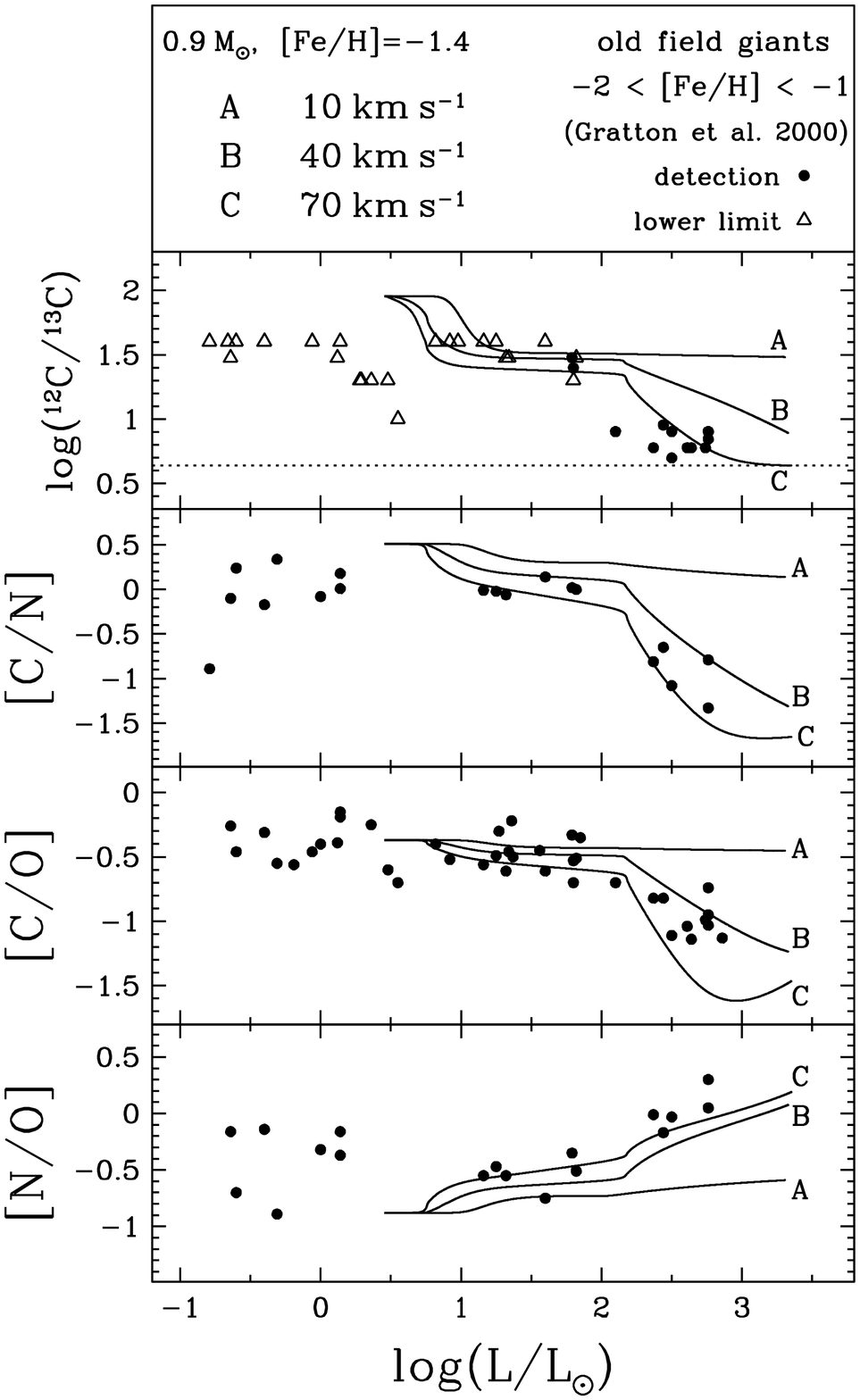}{1cm}{0}{450}{550}{0}{0}}
\vspace*{-0.5cm}
\caption{\label{fig:fielddata} Comparison between the \citet{gra00}
data for metal-poor field giants with well determined metallicities in
the range $-2 <\,\,\feh < -1$, and rotating models of a star of
0.9\msun and a metallicity of ${\rm [Fe/H]=-1.4}$.  Initial rotation
rates are 10, 40, and $70\kms$, corresponding to models with a turnoff
rotation rate of $4\kms$ and mixing velocities enhanced by a factor of
$(10/4)^{2}$, $(40/4)^{2}$, and $(70/4)^{2}$, respectively.  The
dotted horizontal line in the uppermost panel indicates the nuclear
equilibrium value of \isot $\sim 4.5$.  Compare with the \isot ratios
of globular cluster giants of Figure
\ref{fig:GCisotope}. }\end{figure}

One of the most comprehensive studies of old, metal-poor field giants
is that of \citet{gra00}, which selected well studied stars with
metallicities in the range $-2\, <\, \feh < -1$.  In Figure
\ref{fig:fielddata} we present a comparison between the \citet{gra00}
data and rotating models for a representative field-like star of
0.9\msun and $\feh = -1.4$.  In the panel with the \isot ratios, black
dots represent reliable measurements (i.e., both isotopes are
detected) and the open triangles are stars with only lower limits.
The horizontal branch stars of the sample are not included.  With
varying degrees, the data in all the four panels in Figure
\ref{fig:fielddata} show the signatures of some gradual CNO processing
as the luminosity increases: conversion of $^{12}$C into $^{13}$C,
overall carbon depletion, and nitrogen enhancement.

In the same spirit as with M67, we dissect the field data of Figure
\ref{fig:fielddata} into three different evolutionary stages.  The
first goes from the MS turnoff (or even before) to the completion of
the first dredge-up during the \sgb.  Since all the \isot data of this
phase are lower limits, the only thing that can be said about them is
that they are at least at the level or above the standard first
dredge-up expectation, except for the four MS stars with lower limits
below \isot $\sim 20$.  For the other three abundance ratios shown
([C/N], [C/O], and [N/O]) the large intrinsic scatter, present already
among the MS stars, obscures any possible underlying trend on this
early evolutionary stage ($\log(L/L_{\odot}) \gtrsim 1$).  This
indicates that a similarly large scatter in abundance ratios on more
evolved stages may be primordial, and need not require star-to-star
variations due to mixing.

The second evolutionary stage is better populated with data and
corresponds to the lower RGB, in the interval $1.2 \lesssim
\log(L/L_{\odot}) \lesssim 2$.  The abundance ratios do not change
along the lower RGB.  The third distinct group of stars are those on
the upper RGB, with luminosities brighter than $\log(L/L_{\odot}) \sim
2.1$, and which show in all panels systematically lower \isot and
[C/N] ratios than stars on the lower RGB.  These stars exhibit clear
evidence of in situ mixing on the upper RGB.

Finally, there is no evidence in the [O/Fe] measurements of
\citet{gra00} of any change in oxygen (as well as in the data of
Fulbright \& Johnson 2003), and thus the low levels of [C/O] and high
levels of [N/O] in the bottom two panels are the result of the
aforementioned carbon depletion and nitrogen enhancement,
respectively.  This implies that the extra mixing in these field
giants is not deep enough to reach regions of significant ON
processing, as is the case in some globular clusters (\S\,6.3).

The models in Figure \ref{fig:fielddata}, chosen to be representative
of the \citet{gra00} field data, started from a rigidly rotating MS
turnoff star with surface rotation rates of 10, 40, and $70\kms$, and,
given that we already know that models with rigidly rotating envelopes
do not mix, only models with constant specific angular momentum in the
convective envelope are considered.  Before going further, note that
the initial mixture used to generate these models is solar, which is
not the most optimal mixture to use for comparison to metal-poor field
stars.  Increasing evidence \citep{bes82,hen00,nor01,nor02,li03,isr04}
indicates that a significant fraction of these stars have overabundant
amounts of nitrogen (with respect to solar).  Thus, when inspecting
the panels of Figure \ref{fig:fielddata} involving nitrogen, one has
to bear in mind that the curves representing our theoretical results
could be shifted vertically by that initial overabundance.

The evolution of the theoretical surface ratios are very similar to
those of Figure \ref{fig:mix13msun}.  First, the post dredge-up level
of the models ($\log(L/L_{\odot}) \approx 1.3$) is dependent on the
initial rotation rate, which is the result of extra mixing in the deep
radiative interior prior to the occurrence of the maximum penetration
of the convective envelope (i.e., before the end of the first
dredge-up).  Furthermore, the post dredge-up ratios of the models are
consistent with the \isot ratios of field giants in this luminosity
range (upper panel), and also with the [C/N] and [N/O] data (recall
here that the likely higher initial nitrogen abundances would
vertically shift these lines), while the large scatter in the [C/O]
data prevents us to say anything about this ratio.

Second, along the lower RGB, $1.2 \lesssim \log(L/L_{\odot}) \lesssim
2.1$, all rotating models of field-like giants show little or no extra
mixing: the \isot ratio does not change at all, the total carbon and
nitrogen abundances do change although very slowly and are inversely
correlated, while oxygen remains unaltered.

Finally, the third distinct evolutionary stage corresponds to the
upper RGB, after the occurrence of the RGB bump, $\log(L/L_{\odot})
\approx 2.2$, and at which point the models with differential rotation
in the envelope experience strong extra mixing.  The amount of
non-standard mixing is a strong function of the initial rotation rate.
The case of $V_{\rm TO}=10\kms$ (which, we recall, actually
corresponds to an upper limit for the rotation rates of old field
stars) experiences very little extra mixing, and is not able to
reproduce the abundances of stars brighter than the RGB bump, which
all show \isot$\lesssim 10$.  For larger initial angular momentum
budgets, however, the extra mixing is vigorous, and the uppermost
panel shows that the low values of \isot are achieved by the model
with an initial rotation rate of $V_{\rm TO}=70\kms$ at approximately
the right luminosities.  The two middle panels of Figure
\ref{fig:fielddata} show that, for initial rotation rates between 40
and 70$\kms$, our models reproduce the right [C/N] and [C/O] ratios.
The bottom panel, showing the evolution of the [N/O] ratio, would seem
to require a higher degree of oxygen depletion, since most data points
fall above the model with $V_{\rm TO}=70\kms$.  However, if we had
started with a larger initial nitrogen abundance, as shown by a
fraction of metal-poor field stars, it is easy to verify that such a
discrepancy would disappear and the evolution of the [N/O] ratio of
the models would agree with the \citet{gra00} data.

Nevertheless, despite this apparent success of the models with larger
rotation rates at turnoff, we need not forget that metal-poor field
stars are observed to have rotation velocities of less than $10\kms$
(\S\,3.3).  Therefore, the maximal mixing scenario, with the currently
available estimates for the velocities of extra mixing, can not
reproduce the observed abundances of these old field stars with
rotation rates that are sensible for them.  

In conclusion, and considering that the amount of extra mixing scales
as the square of the rotation rate, we obtain that the diffusion
coefficients associated with the rotationally-induced hydrodynamic
instabilities as computed in our code (\S\,3.1 and Appendix) are too
small by a factor between FC $=(40/10)^{2} \sim 20$ and FC
$=(70/10)^{2} \sim 50$ to account for the surface abundance patterns
of metal-poor field giants.  Note also that this conclusion applies
for surface rotation rates of stars that are rotating rigidly at the
MS turnoff.  If, contrary to the solar case, we had started our models
from differentially rotating MS stars, then, for any fixed initial
surface rotation rate, the extra mixing would have been much more
vigorous than that of the models we show in Figure \ref{fig:fielddata}
simply because of the larger initial angular momentum budget of the
turnoff star (see \S\,3.2 and equation \ref{eqn:budget}).

\subsection{Population II giants: globular clusters}

\begin{figure}
\vspace*{-2cm}
{\hspace{-15cm}
\plotfiddle{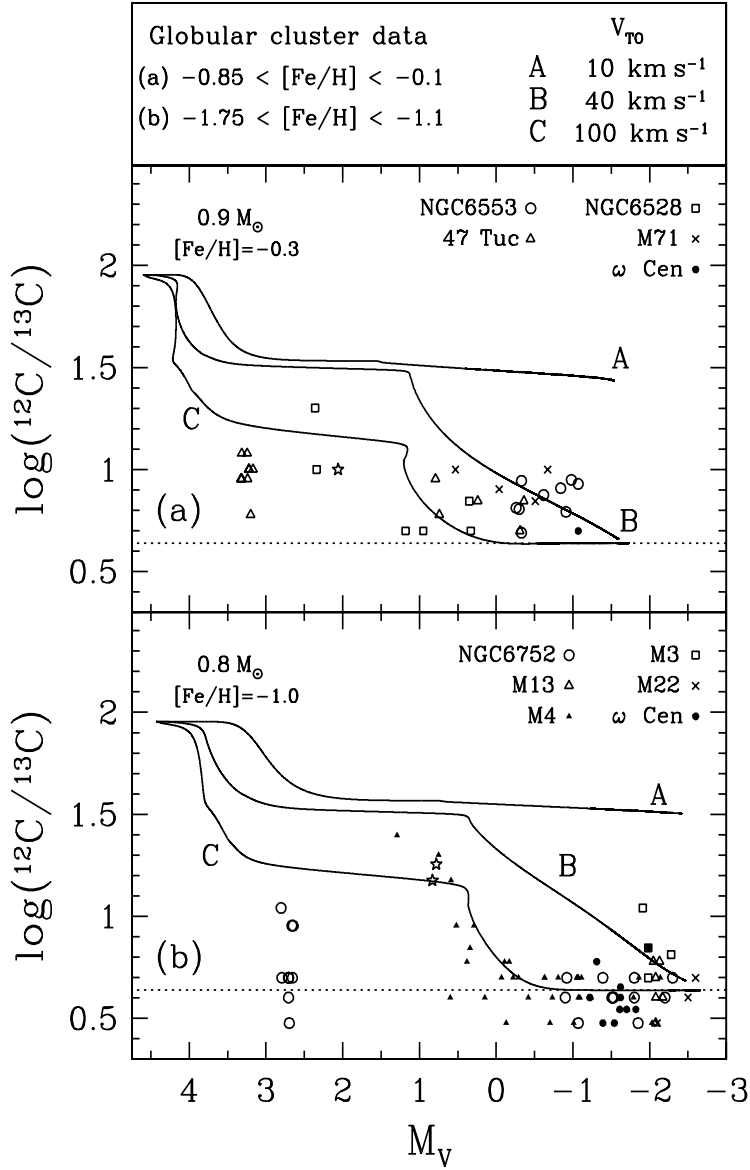}{1cm}{0}{350}{500}{0}{0}}
\vspace*{-0.8cm}
\caption{\label{fig:GCisotope} Comparison between surface \isot ratios
of globular clusters giants and our rotating models.  The data have
been separated in two ranges of metallicity, and plotted in comparison
to stellar models representative of the corresponding sample, for
initial rotation rates of 10, 40, and $100\kms$ at turnoff.  These
correspond to models with a turnoff rotation rate of $4\kms$ and
mixing velocities enhanced by a factor of $(10/4)^{2}$, $(40/4)^{2}$,
and $(100/4)^{2}$, respectively.  Data for NGC 6553, NGC 6528, and 47
Tuc are taken from \citet{she03}, except for the 47 Tuc giants
populating the lower RGB ($M_{\rm V}\sim 3.3$), which are from
\citet{eugenio04}; for M3 from \citet{pil03}; for $\omega$ Cen from
\citet{smith02}; for M71 from \citet{bri97}; for M13 from
\citet{she96b}; for M22 from \citet{sun91}; for NGC 6752, the
brightest giants are from \citet{sun91} and the lower RGB giants
($M_{\rm V} \sim 2.7$) from \citet{eugenio04}; and for M4 from both
\citet{she03} and \citet{sun91}.  The data point marked as a star in
panel (a) is a giant in NGC 6528 ($\feh=-0.1$) for which the
measurement is only a lower limit, as are the two stars in panel (b)
but representing red giants in M4 ($\feh=-1.1$).  }\end{figure}

\begin{figure}
\vspace*{-1cm}
{\hspace{-16cm}
\plotfiddle{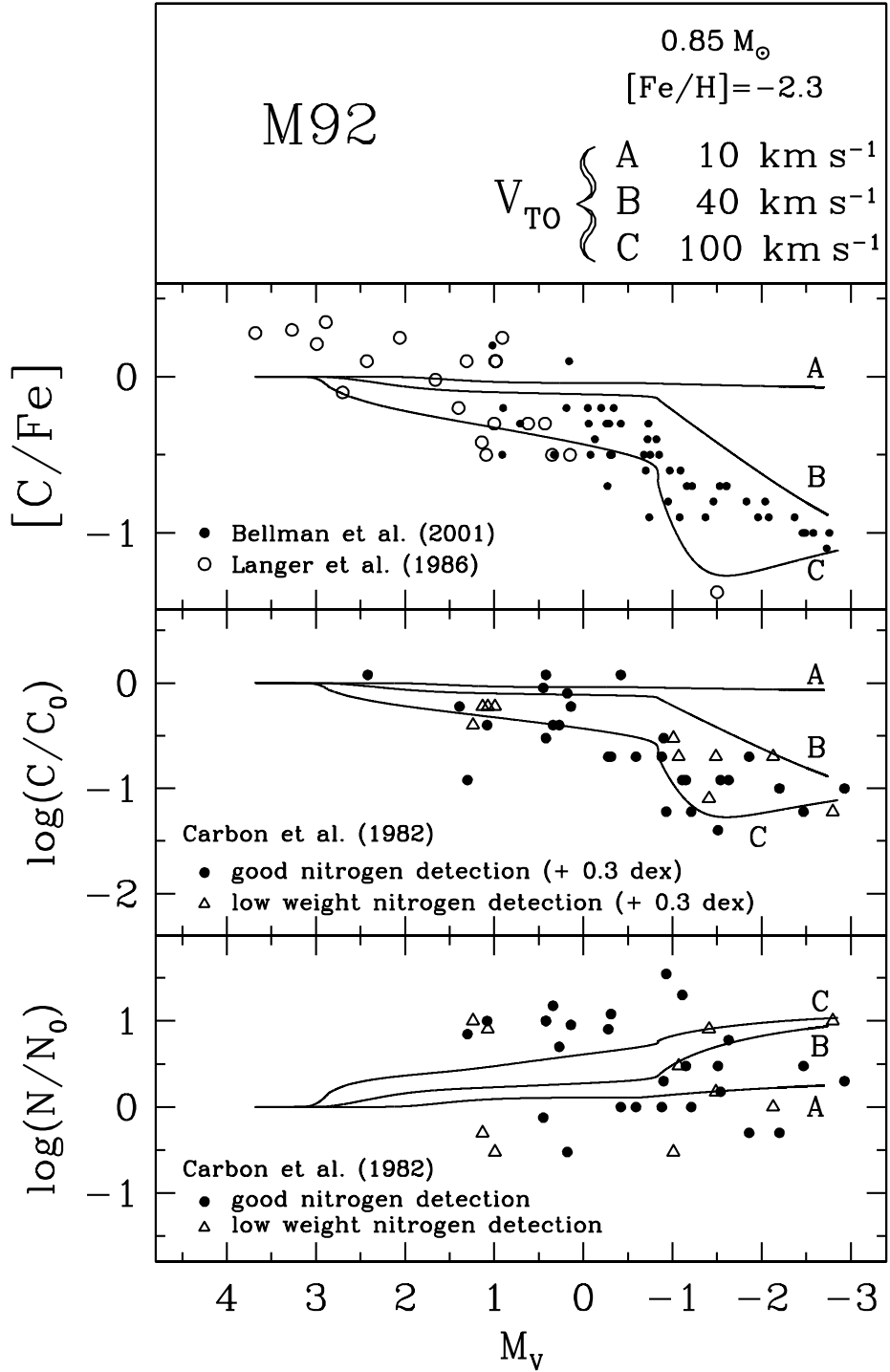}{1cm}{0}{450}{600}{0}{0}}
\vspace*{-1cm}
\caption{\label{fig:M92data} Carbon and nitrogen abundances of M92
giants as compared to the same rotating models of M92-like giants of
Figure \ref{fig:GCisotope}.  The models, which all started from a
solar mixture, have been arbitrarily set to zero in the vertical axes
of all panels.  Note, however, that initial mixtures more appropriate
to globular cluster stars would make the models shift up and
down. }\end{figure}

\begin{figure}
\vspace*{-3cm}
{\hspace{-16cm}
\plotfiddle{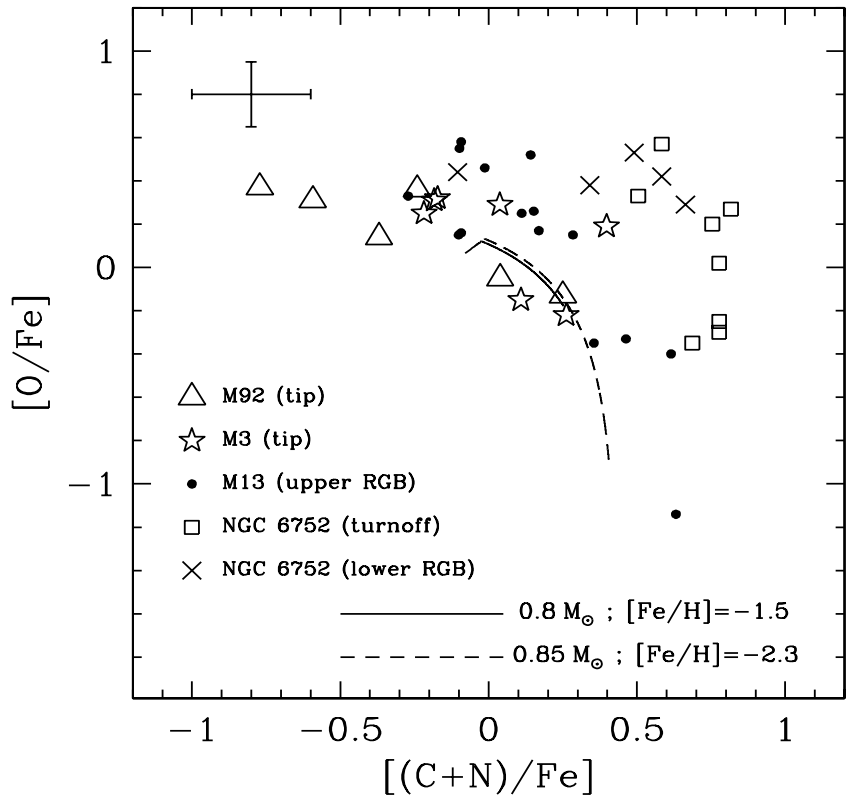}{1cm}{0}{400}{500}{0}{0}}
\vspace*{-1cm}
\caption{\label{fig:Odepletion1} Oxygen and (C+N) in globular cluster
giants.  The signature of ON-burning would be the depletion of oxygen
accompanied by a simultaneous enhancement of the sum (C+N).  Dashed
and solid lines indicate ON burning in two of our most metal-poor
rotating models.  }\end{figure}

\begin{figure}
\vspace*{-3cm}
{\hspace{-16cm}
\plotfiddle{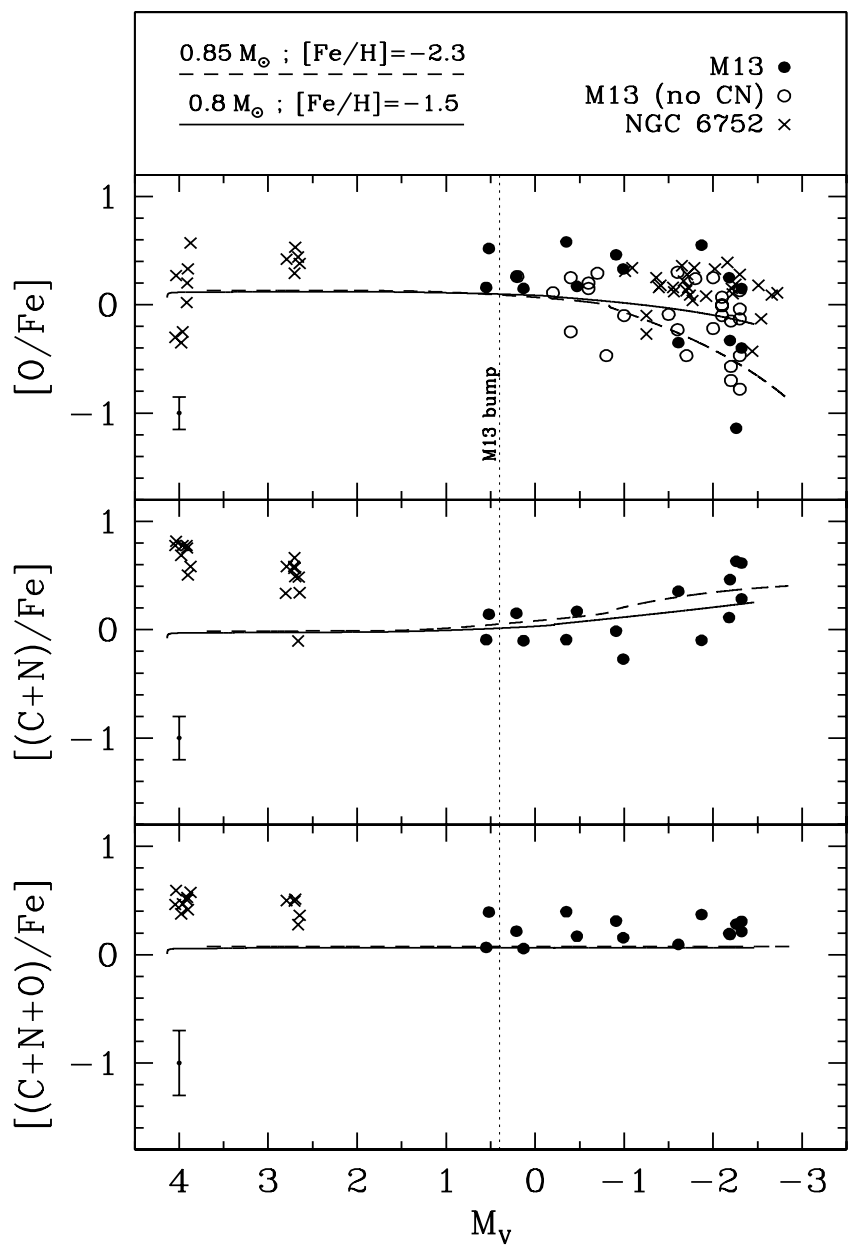}{1cm}{0}{400}{500}{0}{0}}
\vspace*{-1cm}
\caption{\label{fig:Odepletion2} Oxygen depletion among M13 and NGC
6752 giants, together with carbon and nitrogen data when available.
Also shown are the MS and lower RGB stars in NGC 6752 of
\citet{eugenio04}.  The same metal-poor rotating models of Figure
\ref{fig:Odepletion1} are shown.  The location of the RGB bump of M13
is indicated by the vertical dotted line at M$_{V} \sim 0.4$.
}\end{figure}

The case of globular clusters is explored in Figures
\ref{fig:GCisotope} to \ref{fig:Odepletion2}.  In the first two we
plot, respectively, data on the \isot ratio for several globular
clusters and carbon and nitrogen data for M92.  The issue of oxygen
depletion is explored in Figures \ref{fig:Odepletion1} and
\ref{fig:Odepletion2} using very recent data for M13 and NGC 6752, two
globular clusters with approximately the same metallicity.

The \isot data in Figure \ref{fig:GCisotope} have been separated in
two samples according to metallicity.  In the upper panel, (a), we
include clusters with $\feh > -0.7$, plus one $\omega$ Cen star of
$\feh=-0.85$, and the lower panel, (b), shows data from clusters with
$-1.6 <\,\,\feh < -1.1$ and $\omega$ Cen stars as metal-poor as
$\feh=-1.75$.  The data point marked as a star in panel (a) is a giant
in NGC 6528 ($\feh=-0.1$) for which the measurement is only a lower
limit, as are the two stars in panel (b) but representing red giants
in M4 ($\feh=-1.1$).  Error bars have been omitted to avoid confusion,
but the reader should bear in mind that all the faintest data points
in both panels have large uncertainties: in (a), the two NGC 6528
giants with $M_{\rm V}\sim 2.4$ have error bars that reach
$\log($\isot)$\,\sim 1.5$, and the same applies, in (b), for the three
M4 giants closest to the data points representing lower limits and
marked as stars.

Until very recently, almost all the \isot data for globular cluster
giants were restricted to luminous giants past the RGB bump and close
to the RGB tip.  Therefore, little could be said about the first two
stages in the post-MS mixing history amply discussed for M67 and the
field giants (the first dredge-up dilution and the lower-RGB, before
the RGB bump).  Recently, however, \citet{eugenio04} obtained the
first measurements of \isot in stars close to the base of the RGB (as
well as some MS turnoff stars) in 47 Tuc and NGC 6752.  These data,
although restricted to a very narrow range of luminosities, have
important implications for our following discussion.

Both groups of globular cluster giants in Figure \ref{fig:GCisotope}
show very low \isot ratios, but it is still possible to notice the
slight metallicity dependence of these data: the more metal-rich
sample has $\langle$\isot$\rangle\sim 6.3$, while the most metal-poor
sample has $\langle$\isot$\rangle\sim 4$.

Next, is there indication in the globular cluster data of Figure
\ref{fig:GCisotope} of any evolution of \isot with luminosity on the
RGB such as that observed among open cluster and metal-poor field
giants?  \citet{she03} found that to be the case among giants in the
globular cluster M4 (filled triangles in lower panel), which indeed
seem to show high values of \isot before and around the bump
luminosity ($M_{\rm V}\sim 0.5$), but which rapidly decrease toward
higher luminosities, with all the M4 giants past $M_{\rm V}\sim 0$
showing \isot ratios already close to nuclear equilibrium.  A similar
indication, although not as obvious as in M4, may be seen in the data
for NGC 6528 (squares in upper panel).  These two clusters, therefore,
indicate a continuous evolution of the surface \isot along the RGB.
Also, as can be seen in the models in this figure, such luminosities
correspond very well with the expected location of the RGB bump for
stars of approximately the right masses and metallicities.

At the same time, however, the recent data from \citet{eugenio04} for
47 Tuc and NGC 6752 tell a different story, by showing very low values
of \isot as well as a large scatter among giants at the base of the
RGB, that is, much earlier than the RGB bump.  This indicates that the
stars in these samples already had these low values of \isot before
becoming giants.  If that is the case, then these stars acquired their
current anomalous patterns of surface abundances via pollution from
more massive stars from the same or a previous generation, and the
problem then becomes to figure out at what point in their evolution
did the current giants incorporate this enriched material, either
while they were MS stars or, alternatively, during the star formation
process.

An alternative to this pollution scenario would be the action of very
strong extra mixing soon after the MS turnoff, but this possibility is
unlikely given the unrealistically large mixing velocities that would
be required.  Note, nevertheless, that three of the lower RGB stars in
NGC 6752 from \citet{eugenio04} show \isot $\sim 10$, while all the
stars from this same cluster but with higher luminosities studied by
\citet{sun91} display essentially nuclear equilibrium values.  This
may be indication of continuous extra mixing even among stars that
arrived to the RGB already with anomalous abundance ratios.  There is,
therefore, clear evidence in globular clusters for both evolutionary
and primordial variations in the abundances of carbon isotopes, with
neither of these scenarios necessarily excluding the other but,
rather, operating at the same time.

For comparison with the globular cluster \isot data in Figure
\ref{fig:GCisotope}, we generate rotating stellar models of 0.9\msun
with $\feh=-0.3$ and 0.8\msun with $\feh=-1.0$, for panels (a) and (b)
respectively.  The former corresponds to a metallicity intermediate
between that of M71 or 47 Tuc and that of NGC 6528, and the latter
corresponds closely to giants in M4, which has a turnoff mass $\sim
0.81$\msun and $\feh=-1.1$ \citep{richer97}.  Only models with
constant specific angular momentum in the convective envelopes are
shown.

Adequate turnoff rotation rates for the progenitors of globular
cluster giants are $V_{\rm TO} \sim 4\kms$.  However, with the
estimates of mixing velocities available today (implemented in YREC),
such a slow rotation does not produce almost any changes in the
surface abundances.  Therefore, we run models with turnoff rotation
rates of 10, 40, and $100\kms$, which correspond to models with
$V_{\rm TO} = 4\kms$ and diffusion velocities multiplied by a factor
of $(10/4)^{2}$, $(40/4)^{2}$, and $(100/4)^{2}$, respectively.
Similarly, models with fast initial rotation mimic models of slow
initial rotation but rapidly spinning cores at the turnoff (\S\,3.2).

As previously seen for models of M67 and field giants, different
initial rotation rates produce different amounts of extra mixing
immediately after the turnoff and during the SGB, so that the post
first dredge-up \isot ratio is a function of the initial angular
momentum budget.  Again, this is just a consequence of the not
inclusion in our models of any inhibiting effects of $\mu$-barriers.
Vigorous mixing during the SGB, as indicated by the model with $V_{\rm
TO}=100\kms$ (equally reproduced by a model of $V_{\rm TO}=4\kms$ plus
a factor of FC $=(100/4)^{2}= 625$ more efficient extra mixing, i.e.,
larger diffusion coefficients), is capable of producing a factor of 2
smaller \isot ratios after dredge-up.  However, if the starting \isot
ratio is indeed like that of the solar system, one would need even
stronger extra mixing than that experienced by the most mixed model in
Figure \ref{fig:GCisotope} during the \sgb in order to get close to
the \isot values shown by the lower RGB giants studied by
\citet{eugenio04}, which is very unlikely.

Instead, what these low luminosity giants might indicate is that the
starting mixture has a significantly lower \isot than that of the
solar system.  Indeed, rotating models with initial \isot as low as 30
(not shown in Figure \ref{fig:GCisotope} to avoid confusion), while
achieving lower levels of \isot during the first dredge-up, still do
not fall as low as the lower RGB stars in 47 Tuc and NGC 6752.  We
conclude, therefore, that the low \isot ratios of lower RGB giants in
47 Tuc and NGC 6752 are most likely primordial to the RGB, and their
origin poses an interesting problem yet to be solved.  This again
demonstrates that the adoption of a solar-system mixture of carbon
isotopes as a starting point for the models is a bad one.

Past the bump (at $M_{\rm V}\approx 1.2$ or $\log(L/L_{\odot}) \approx
1.45$ for the models in (a), and $M_{\rm V}\approx 0.4$ or
$\log(L/L_{\odot}) \approx 1.90$ for models in (b)), on the upper RGB,
stars are well mixed, based upon \isot as an indicator.

In Figure \ref{fig:M92data} we plot carbon and nitrogen abundances for
giants in M92, and compare them to our M92-like rotating models.  The
[C/Fe] data in the upper panel represented as black dots are from
\citet{bellman01}, who extended earlier work by \citet{carbon82} and
\citet{lan86} in order to determine the luminosity at which the carbon
depletion reported by these works sets in.  The open circles represent
stars from \citet{lan86} that were not observed by \citet{bellman01},
which mainly fill the faint end of the M92 RGB.  Since
\citet{bellman01} report only carbon abundances, we plot in the middle
and lower panels the carbon and nitrogen abundances of
\citet{carbon82}.  In these two panels the black dots represent all
those M92 stars for which the nitrogen abundance is reliable as
reported by \citet{carbon82}, and the open triangles represent those
with uncertain (low weight) nitrogen abundances.  The carbon
abundances from \citet{carbon82} have been shifted vertically by
$+0.3$ dex in order to bring the zero point of those measurements in
agreement with the (solar) initial mixture of our models, and we do
this because the goals of this investigation do not involve initial
abundances, but rather their subsequent evolution under the action of
extra mixing.

The carbon abundance data in the upper two panels of Figure
\ref{fig:M92data} unambiguously show a progressive depletion of carbon
with increasing luminosity along the RGB, although the large
dispersion at any given magnitude difficults a secure determination of
where this depletion starts.  Note that the data points with uncertain
nitrogen determinations do not alter the pattern of carbon depletion
in the \citet{carbon82} data.  Regarding nitrogen, the bottom panel
shows a dispersion larger than 1 dex at all luminosities, and thus one
can not distinguish if either the nitrogen abundances are not altered
along the RGB of M92 or, alternatively, if any pattern of nitrogen
enhancement is lost in a larger dispersion of initial abundances.

Once again, we conclude that in order to bring our maximal mixing
scenario in agreement with the data, either the diffusion velocities
associated with hydrodynamic instabilities would have to be
underestimated by about two orders of magnitude (FC $\sim 400$), or
turnoff stars in globular clusters need to have rapidly rotating
interiors, in contrast with the helioseismological data.  Finally,
note the good agreement between the luminosity of the RGB bump in our
models and that shown by the data, especially for the case of M4: the
model with $V_{\rm TO}=100\kms$ follows very closely the evolution of
the \isot ratio for the M4 giants from the lower RGB, across the onset
of extra mixing right after the RGB bump, and up to the RGB tip.

As final remarks on Figure \ref{fig:M92data}, let us address the
controversial issue of the location of the onset of carbon depletion
in M92.  In contrast with the \isot evolution of the field giants in
the \citet{gra00} sample (upper panel of Fig. \ref{fig:fielddata}) as
well as in M4 (panel (b) in Fig. \ref{fig:GCisotope}), the M92 giants
do not show a clear break point in luminosity before which stars have
high and approximately constant carbon abundances and after which they
start to decline.  Instead, what we see when going from brighter to
less luminous stars is that the average carbon abundance slowly but
progressively increases, without indication of leveling off at least
as far as the least luminous data points concern, which belong already
to M92's SGB.

\citet{bellman01} conclude that there is carbon depletion in M92 stars
at least from $M_{\rm V}=0.5-1.0$ and, as most of the literature on
the subject, regard this as a problem for the theory because such
luminosity is well below that of the RGB bump in M92, which has been
observationally determined to occur at $M_{\rm V}=-0.4$ \citep{fus90}
and is located at $M_{\rm V}\approx -0.8$ in the models of Figure
\ref{fig:M92data}.  

As stressed on various occasions in the present work, we think this
just reflects the theoretical prejudice that a $\mu$-barrier
completely inhibits even the slightest amount of extra mixing until it
gets erased or smoothed by the advancing HBS.  Such a chocking off of
the meridional currents responsible for the extra mixing might
certainly turn out to be the case, but we argue that, at the time of
writing this, there is no observational result that firmly establishes
that as a fact.  Thus, we let the extra mixing in our models to occur
whenever the conditions allow so, that is, whenever there is an energy
reservoir available to drive the mixing, as equation
(\ref{eqn:epsilon}) captures and is discussed elsewhere in this
section.  The end result is that, although in much smaller degree than
after the RGB bump, our models of M92-like giants do show extra mixing
as early as $M_{\rm V}=2.0$, which produces a much better agreement
with the observed pattern of carbon depletion than models that assume
complete inhibition of extra mixing before the RGB bump.

Furthermore, it is quite likely that the M92 giants we currently see
come from MS progenitors polluted by the ashes of previous
nucleosynthesis, and which therefore already had a large spread in
carbon abundances before becoming giants.  This is supported by the
recent finding of \citet{cohen05b} of a spread larger that 1 dex in
the carbon abundances among M15 subgiants.

Next we explore the question of oxygen depletion.  Metal-poor giants
in the field do not show any variation of the oxygen abundance with
position on the RGB, with $\langle$[O/Fe]$\rangle$$\,\approx 0.34$ and
a dispersion of only $\sigma\sim 0.15$ dex at all luminosities from
the MS turnoff to the tip of the RGB \citep{gra00}.  In globular
clusters, on the other hand, the issue is controversial.  This is not
surprising because the effect itself is expected to be small: since
oxygen burns at higher temperatures than carbon and nitrogen, the
appearance of ON-processed material at the surface requires even
deeper mixing than that needed to transport material enriched by
CN-burning.  In addition, mixing of oxygen-depleted material is
expected to be more evident with decreasing metallicity \citep{swe79}:
being hotter, regions of ON-burning are closer to the surface in
metal-poor stars than in more metal-rich ones.  Finally, primordial
variations in oxygen abundances, now known to exist, complicate the
issue even more.

An apparent anticorrelation between the nitrogen and oxygen abundances
in M92 giants, although restricted to a narrow range of luminosities
close to the RGB tip, has been interpreted as evidence of mixing of
ON-processed material \citep{pil88,sne91}.  Apart of being an
intrinsically small effect, the evidence for oxygen depletion becomes
controversial due to possibly large systematics between abundance
determinations from different investigators, which might arise from
the use of different spectral regions or differences in the spectral
analysis.  Thus, ideally, uncontroversial evidence for or against the
occurrence of oxygen depletion should come from measured abundances of
stars in the same cluster that, while sampling the extension of the
RGB as best as possible, are obtained through a homogeneous procedure.
However, such an example is still not available.

In order to investigate whether the extra mixing reaches regions of
significant ON-burning, we need first to isolate the effects of
CN-burning made evident by the pattern of carbon depletion in M92
(Fig. \ref{fig:M92data}).  This is achieved by comparing the oxygen
abundances with the sum of carbon and nitrogen, (C+N), which should
remain constant if only CN-burning is involved.  To this goal we use
recent data for the globular clusters M13 \citep{cohen05,bro91} and
NGC 6752 \citep{yon03,eugenio04}, as well as data for M92
\citep{sne91} and M3 \citep{smith96}.  M13 and NGC 6752 have
approximately the same metallicity (\feh$\sim -1.5$) and, more
importantly, they have giants with measured CNO abundances spanning a
relatively wide range of luminosities along their RGBs, allowing an
assessment of how these species evolve.

Figure \ref{fig:Odepletion1} shows a comparison between oxygen and
(C+N) abundances for all the stars with measured CNO abundances in the
clusters under consideration, and Figure \ref{fig:Odepletion2} shows
the run of oxygen and (C+N) as a function of position on the RGB.
Globally, the combined data in Figure \ref{fig:Odepletion1} indicate
an increasing spread in [O/Fe] when going toward higher (C+N), and
clusters like M92 and M13 (triangles and black dots, respectively) can
have their giants separated in two distinct groups, one with high
[O/Fe] and low (C+N), and another with low [O/Fe] and high (C+N).
Nevertheless, there is also a clear primordial spread of [O/Fe], as
evidenced by the MS turnoff stars of NGC 6752 (open squares).

In Figure \ref{fig:Odepletion2}, a fraction of the M13 giants close to
the tip of the RGB are oxygen-poor with respect to all the less
luminous giants in the same cluster, and it can be verified that this
oxygen depletion corresponds to a simultaneous enhancement of the sum
of (C+N) with increasing luminosity \citep{cohen05}.  Along with this
anticorrelated evolution of oxygen and (C+N) on the upper RGB of M13,
the total (C+N+O) remains constant, clearly indicating the appearance
of ON-processed material at the surface.  Although not shown in Figure
19 to avoid confusion and spanning a small range of luminosities, the
seven M3 bright giants of \citet{smith96} also provide an indication
of ON-processed material having been mixed to the surface, by
displaying (C+N+O) values clustered within a range of 0.3 dex, but, at
the same time, (C+N) values dispersed within 0.6 dex.

In the case of NGC 6752 (crosses), while the lower-RGB giants studied
by \citet{eugenio04} share, within the uncertainties, the same oxygen
abundance ($\langle$[O/Fe]$\rangle$\,$\sim 0.4$), the giants on the
upper RGB \citep{min96} clearly show a lower average [O/Fe] and a
substantial scatter, suggesting some depletion of oxygen along its
RGB.  Nevertheless, NGC 6752 already displays an anticorrelation of
oxygen and CN among turnoff and lower RGB stars, and therefore any
further oxygen burning on the upper RGB of this cluster occurs on top
of these earlier patterns.

The models in Figures \ref{fig:Odepletion1} and \ref{fig:Odepletion2},
which use mixing coefficients enhanced by a factor needed to reproduce
the nuclear equilibrium values of \isot on the upper RGB (FC $=400$),
are both able to deplete oxygen and produce an anticorrelation with
(C+N) similar to that suggested by the data.  They also show a strong
metallicity dependence, with the M92-like model (0.85\msun,
\feh$=-2.3$) depleting about 5 times more oxygen than the model for
giants in M13 and NGC 6752 (0.80\msun, \feh$=-1.6$).  However, the
models, in general, do not span the entire ranges of [O/Fe] and (C+N)
that the data occupy.  This suggests that a significant part of the
spread in (C+N) in M92 is of primordial origin, a conclusion also
supported by the pattern of carbon depletion in Figure
\ref{fig:M92data}.  The model with \feh$=-1.5$ does not deplete as
much oxygen as is suggested by the giants of both M3 and M13.  We
note, nevertheless, that M13 is the globular cluster whose giants show
the largest observed spread of oxygen abundances (see Fig. 12 of
\citealt{ram02}).

\subsection{The interaction between rotation and convection}

The most straightforward conclusion from our Population II results is
that the efficiency of mixing in RGB stars is underestimated in our
models.  However, a simple rescaling of the diffusion coefficients to
match the Population II abundances would certainly overmix the
Population I models even at low mass, and the higher mass Population I
giants are already overmixed relative to the data. The root cause is
simple: in our models, the major factor governing the degree of
rotational mixing in the radiative interior of red giants is the
degree of rotation at the base of the surface convection zone.
Empirically, this implies that the degree of rotational mixing does
not increase with increased absolute angular momentum content to the
degree implied in the models.  We begin by justifying the physical
basis for an interaction between convection and rotation, in the sense
that the degree of differential rotation in convective regions
decreases as the rotation rate increases.  We then present some
preliminary calculations on the magnitude of the change needed to
explain the data.

There is strong evidence that the angular momentum distribution in
convective regions depends on rotation and evolutionary state.  In the
solar convection zone the rotation velocity is comparable to, or
larger than, typical convective velocities inferred from mixing length
theory; the latitude-averaged rotation period is independent of depth,
consistent with a strong interaction between convection and rotation.
By contrast, the survival of rapid rotation in metal-poor HB stars
requires strong differential rotation with depth in the convective
envelopes of their RGB precursors.  This is not surprising; the
convective envelopes of halo giants have a large moment of inertia and
small starting angular momentum content, so the rotation velocities
will be much less than the typical convective velocities (and there
are many more pressure scale heights in the convective envelope of a
giant than there are in the solar case).  We therefore can infer that
the angular momentum distribution in convective regions can vary
between solid body and constant $J/M$ as the number of pressure scale
heights in the convection zone and the relative rotational and
convective velocities differ.  The higher mass Population I giants
could therefore represent an intermediate case, with a shallower
rotation velocity profile as a function of depth than the one expected
from a simple extrapolation of the behavior of the very slowly
rotating low mass giants.

We estimate the magnitude of the change needed as follows.  We make
use of the M67 data discussed in \S\,6.1 to determine how steep the
angular momentum gradient in the convective envelope of a M67-like
giant must be in order to reproduce the difference between the \isot
levels of the lower RGB giants and the clump giants in M67 while, at
the same time, using diffusion coefficients (i.e., strength of extra
mixing) scaled by the factor of $\sim 400$ needed to account for the
field and globular cluster \isot ratios (\S\,6.2 and \S\,6.3).

\begin{figure}
\vspace*{-3cm}
{\hspace{-16cm}
\plotfiddle{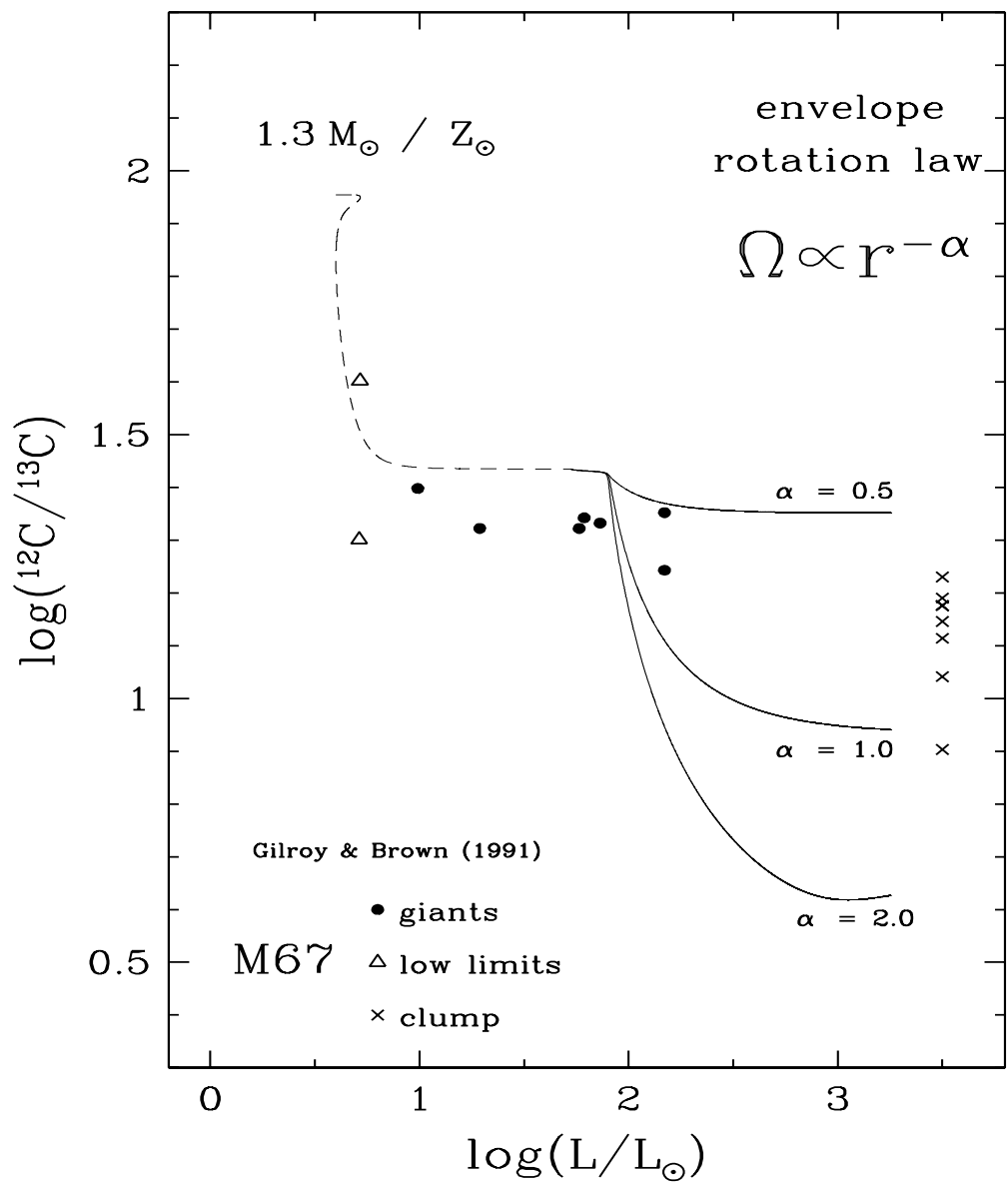}{1cm}{0}{450}{500}{0}{0}}
\vspace*{-1cm}
\caption{\label{fig:CZprofile} Dependence of canonical extra mixing
(solid lines) on the angular momentum distribution of the convective
envelope, as parametrized by $\alpha$.  The data shown are the same as
in Figure \ref{fig:mix13msun}.  The dashed line is a typical rotating
model with very mild mixing and an initial rotation rate of $30\kms$
at turnoff, appropriate for progenitors of current M67 giants.  The
solid lines are models with varying degrees of differential rotation
in their convective envelopes.  A much shallower profile than that
required for globular cluster giants is needed to account for the
\isot levels of clump stars in M67.  See \S\,6.4 for a detailed
discussion.  }\end{figure}

Our results are shown in Figure \ref{fig:CZprofile}, where we plot a
series of rotating models for a M67-like giant with canonical mixing
(that is, only from the RGB bump\footnote{The evolution through the
\sgb and lower RGB until the location of the RGB bump (dashed line) is
done under the same conditions as those of the models in Figure
\ref{fig:mix13msun}, that is, with the current estimations for the
diffusion coefficients (${\rm FC = 1}$).  Note that our using of
different values of FC before and after the RGB bump is the sensible
thing to do, because of the different origin of rotational mixing in
these two regimes: while mixing before the bump is a consequence of
meridional circulation and depends on the degree of departure from
spherical symmetry (absolute rotation rate), upper-RGB canonical extra
mixing operates only in the presence of large angular velocity
gradients.  Therefore, it is the latter mode of extra mixing the one
that is affected by the details of the angular momentum distribution
in the envelope, and around which there exists considerable
uncertainty regarding the actual size of the diffusion velocities
associated to it.  It is this type of extra mixing that we are
calibrating using the factor FC in order to produce the mixing levels
of bright globular cluster giants.}) enhanced by FC $= 400$ and
differentially rotating convection zones with varying angular velocity
power laws.  The initial surface rotation rate, at turnoff, is
$30\kms$.

What concerns us here is the behavior past the RGB bump, when
canonical extra mixing is strong.  Starting from the bump model left
by the run described above (end of dashed line in
Fig. \ref{fig:CZprofile}), we run models with the new diffusion
coefficients (FC $= 400$) and with varying angular velocity profiles
in the convective envelope, parametrized by $\Omega_{\rm CZ} \propto
r^{-\alpha}$, and shown as the solid lines in Figure
\ref{fig:CZprofile}.

The model with constant specific angular momentum ($\alpha = -2$), as
expected from our results of \S\,6.1, mixes too much in comparison to
what is needed to account for the \isot levels of the clump stars in
M67 (which, we recall, are supposed to represent the abundance ratios
of tip giants in this cluster).  The models with shallower rotation
profiles in the envelope, however, mix less efficiently, as is
expected from the discussion in \S\,5.1, and it can be verified that
values of $\alpha$ between 0.5 and 1.0 can reproduce the \isot ratios
of clump stars.

In conclusion, we have shown that, assuming that the diffusion
coefficients associated to extra mixing have been underestimated by
about two orders of magnitude (which can be explained by more
instabilities, higher order terms for angular momentum and composition
transport, or additional physical mechanisms needed to be taken into
account), one can reproduce the different mixing levels seen in Pop I
and Pop II giants with the same models, underlying physics, and with
their corresponding initial conditions (surface rotation rates and
internal angular momentum distributions), provided the convective
envelopes of these two different populations respond differently to
the degree of absolute rotation.

\section{Summary and conclusions}

The problem addressed in the present work is an old one and
consequently has been the subject of numerous investigations: to what
extent can the CNO abundance anomalies seen in low-mass red giants be
explained by extra mixing driven by rotationally induced hydrodynamic
instabilities?  This old question, however, has always been explored
with at least one of two extremely important requirements either
ignored or unsatisfactorily addressed.  The first one is that whatever
solution is put forward, it must be able to explain, using the same
machinery and set of assumptions, the data on abundance patterns seen
in Population II giants as well as those of Population I.  That is,
the problem of abundance anomalies is not restricted to metal-poor
stars in the field and globular clusters, but also extends to the more
metal-rich red giants in open clusters, which must also be addressed.
Second, any successful model must make use of initial conditions and
assumptions in compliance with existent observations.  Particularly,
initial rotation rates must be adopted following data on rotation
rates seen among stars of the type under study, and any assumptions on
the distribution of angular momentum interior to the stars must
similarly be based on some sort of empirical evidence.  These two
broad requirements are essential for a satisfactory solution to the
problem of abundance anomalies in red giants, and constitute the most
important long term goals of our efforts in the present work.

\subsection{Overall properties of RGB models with rotational mixing}

In agreement with previous work, models with rigidly rotating
convective envelopes do not experience any extra mixing, while models
with differentially rotating envelopes do (\S\,5.1).  Observations of
fast rotation of HB stars in globular clusters strongly argue for a
large reservoir of angular momentum in the interior of their
progenitor red giants, which is only possible if these slowly rotating
stars have their convective envelopes rotating differentially with
depth \citep{sip00}.

For giants with differentially rotating envelopes, the amount of extra
mixing experienced on the RGB is highly dependent on the choice of the
angular momentum profile of the progenitor MS turnoff star.  The
reason is that the quantity that ultimately determines the amount of
rotational energy available to drive extra mixing is the angular
momentum content of the material incorporated into the convective
envelope during the standard dredge-up and that will later fall from
the envelope into the radiative region above the HBS where extra
mixing takes place.  At a fixed surface rotation rate, this quantity
is larger for a star that reaches the MS turnoff with a differentially
rotating interior (i.e., its radiative core more rapidly spinning than
its envelope) than for a turnoff star with a solid body rotational
profile such as that of the Sun (eq. \ref{eqn:budget}).  As a
consequence, the material that during the RGB ascent (past the first
dredge-up) falls from the convective envelope into the radiative
region where extra mixing takes place has a much higher angular
velocity when the progenitor MS star had a rapidly spinning core than
when its interior rotated at the same rate as its surface, and thus
the former case will experience diffusion velocities much larger than
the latter.  However, there is no reason to expect that old MS turnoff
stars would have rapidly spinning cores, while, actually, there are
reasons to expect the opposite, with the flat rotational profile of
the Sun among the strongest.  Thus, we advocate that rotating RGB
models should start from rigidly rotating turnoff stars, at least
until either direct observations or indirect empirical evidence
suggest otherwise.

We find that rotational mixing, albeit existent, is inefficient on the
SGB and lower RGB, independently of any inhibiting effects of
$\mu$-gradients.  Therefore, we argue against the widespread notion
that the weakness (or absence) of well mixed stars on the lower RGB of
globular clusters (i.e., before the bump) is evidence for the
inhibiting effects of the $\mu$-barriers.  In our view, the ability of
the star to experience extra mixing is directly proportional to its
luminosity, so that when it is on the upper RGB it experiences much
more mixing than during the lower RGB, when the available energy
(luminosity) is much smaller.  Whether $\mu$-barriers do or do not
inhibit mixing is an effect that goes on top of this more fundamental
result.

As a consequence, the lack of CNO anomalies in giants below the RGB
bump luminosity in globular clusters should not be taken as evidence
for the inhibition of mixing by $\mu$-barriers, because mixing is
inefficient during those stages anyway.  Moreover, and for this very
same reason, even if $\mu$-barriers do indeed inhibit mixing, the CNO
abundances should not be used to calibrate the effects of
$\mu$-barriers on mixing, as has been done in the past.

Nevertheless, our non-inclusion of inhibiting effects by
$\mu$-barriers has the effect of rapidly smoothing, and eventually
erasing, the RGB bump, in contradiction with the observational
situation.  We therefore conclude that the very existence of the RGB
bump provides the first sound evidence for an inhibiting effect of
mixing by $\mu$-barriers, rather than the apparent lack of CNO mixing
before the RGB bump.  However, the peak in the LF that signals the
bump is not a good diagnostic of the strength of this inhibition, and
simply indicates the previous existence of an effective wall for extra
mixing.  Better tests will be needed to attempt an empirical
calibration of the effect of $\mu$-barriers, one of the most promising
being provided by HB stars \citep{sip00,behr03}.

For giants with differentially rotating envelopes, the degree of extra
mixing is primarily determined by the initial rotation rate.  Mass and
metallicity effects are of second order.  For fixed stellar mass and
initial rotation rate, mixing is more efficient with decreasing
metallicity.  For fixed metallicity and rotation rate, lower mass
stars experience more efficient mixing, but the effect is small.

Comparison between the CNO data for M13 and NGC 6752 (Figs.
\ref{fig:Odepletion1} and \ref{fig:Odepletion2}) argues for different
initial mixtures of these elements from cluster to cluster.  This,
added to the different values of \isot measured at different places in
the Galaxy (\S\,6.1), indicates that the usually adopted assumption of
an initial \isot ratio close to the solar value might be far from
adequate.  This is important in the context of models of low mass
giants because the exact post first dredge-up value of \isot depends
on the starting adopted ratio.

\subsection{A semi-empirical picture of rotational mixing}

Maximal mixing models of giants with differentially rotating
convective envelopes which started from a rigidly rotating turnoff
star are able to reproduce the surface \isot data of giants in open
clusters with initial rotation rates adequate to these stars, but fail
to do so in the case of metal-poor giants both in the field and
globular clusters.  Successful reproduction of the latter by these
same models would require either prohibitively large initial rotation
rates, or else progenitor MS stars with rapidly spinning cores,
requirements which are either not supported by current observational
evidence or not yet probed with any current technique.

To reproduce the data, our models of field and globular cluster giants
would require about 7 and 20 times faster initial rotation rates,
respectively, than has been determined by observations.  This means
that, in order for rotationally induced instabilities to account for
these data with reasonable initial conditions, one of the following
two alternatives must be true: (1) the strength of mixing (i.e., the
diffusion coefficients) associated to angular velocity gradients has
been underestimated by factors of at least $7^2\sim 50$ and $20^2\sim
400$, for field and globular cluster giants respectively, and so a
serious revision of these theoretical computations is needed, or (2)
higher-order terms in the equations governing the evolution of the
angular momentum and mixing inside the star, which have been neglected
in our models, are actually important and need to be implemented.
Among these we can mention the effects of gravity waves and magnetic
fields, as well as the additional terms contemplated in the
advection-diffusion scheme of \citet{chabzahn92} and \citet{zahn92}.
These mechanisms are already beyond the scope of the present work, and
thus we elaborated only on the first alternative.

The assumption that the mixing velocities have indeed been
underestimated by a large factor, while solving the problem of field
and globular cluster giants, would constitute, however, a new problem
for the more metal-rich giants, since larger diffusion coefficients
would now produce much more mixing than needed to account for the open
cluster data.  This is an inescapable result if one still requires, as
we advocate in this work, that the same models are to be applied to
both Population I and Population II giants, and, furthermore, starting
with sensible surface rotation rates for each stellar type.  

There is, nevertheless, a possible solution to this dilemma, which,
however, involves the interaction between convection and rotation.
While convincing evidence for differential rotation in the convective
envelopes of RGB stars exists, both from the observed levels of
surface rotation of HB stars \citep{sip00} as well as from the fact
that giants with rigidly rotating envelopes are not capable of any
mixing, what remains as an open question is the degree of such
differential rotation, that is, how steep is the angular velocity
gradient between the stellar surface and the base of the convective
envelope.  We found in \S\,5.1, by exploring the two most extreme
cases, that the shallower this gradient is, the less extra mixing is
experienced by the star.  Then, with the diffusion coefficients
calibrated to match the globular cluster data with slow initial
surface rotation, we showed that the only way to also account for the
mixing levels seen in open cluster giants with their appropriate
rotation rates is if they have convective envelopes with a
significantly shallower angular velocity gradient than their
metal-poor counterparts.  If that were indeed the case, then our
models of rotating giants with maximal mixing would be able to
reproduce the patterns of CNO abundances shown both by Population I
and Population II giants, with the same physics and using initial
conditions in complete agreement with the known properties of these
stars.

This conclusion, therefore, leads us to speculate if indeed we have
indirectly come across with important clues on the not well understood
interaction between convection and rotation, that is, what the
different levels of mixing in Pop II and Pop I giants might be telling
us is simply that small levels of rotation (field and globular cluster
stars) would allow the development of strong differential rotation in
convective regions, while fast absolute rotation (the earlier-type
stars) would lead to convective regions with flatter rotation curves.
In this sense, the different levels of mixing seen in giants of
different populations would be just the result of the Kraft curve
combined with the interaction between rotation and convection.  Our
results thus highlight the foremost importance of the study of the
interaction between convection and rotation in stars.

Finally, a note on the kind of data that are needed to better
constrain the theory and improve our understanding of mixing and the
chemical evolution of stellar clusters.  First, we need more data on
red giants in open clusters with turnoff masses between 1 and 2\msun.
Open clusters of different ages have the crucial advantage of offering
information on the mixing properties of stars of different masses, but
our only source of information of this kind remains the work of
\citet{gil89} and \citet{gil91}, which contain only 4 clusters in this
range of turnoff mass (Fig. \ref{fig:tipratios}) and, furthermore,
with few giants on each case.  Similarly, field metal-rich (Pop I)
giants with available parallaxes (so that their evolutionary state can
be determined) are certainly available but also unexplored, and they
should provide another window to understand the mixing properties of
stars and for refinement of the models.  Second, applicable to all
environments, the observing strategies should be ideally designed such
that abundances are obtained as homogeneously as possible, so that
possible systematics between samples from different clusters or for
different evolutionary states are minimized.

%

\begin{acknowledgements}

We are grateful to Ana Palacios and Corinne Charbonnel for valuable
discussions during scientific meetings at the shores of Canc\' un and
Tuscany, which led to significant improvements on this paper.  We also
thank Eugenio Carretta for providing us with data on the faint giants
in 47 Tuc and NGC 6752 prior to publication.  We are also grateful to
the anonymous referee, whose questions and suggestions led to
improvements to the manuscript.  This work was supported in part by
grant AST 02-06008 from the National Science Foundation to the Ohio
State University Research Foundation.

\end{acknowledgements}

\begin{appendix}

\section{Diffusion coefficients for secular instabilities}

For the secular shear we use the expression of \citet{zahn92},

\begin{equation}
D_{\rm SS} = \frac{8}{45} Ri_{\rm c}\,K \biggl(\frac{r}{N}\frac{d\Omega}{dr}\biggr)^2, 
\end{equation}

\noindent where $Ri_{\rm c}\lesssim 1/4$ is the critical Richardson
number and $K$ is the thermal diffusivity.  The buoyancy frequency $N$
is given by $N^2 = g(\nabla_{\rm ad}-\nabla)/H_p$, where $\nabla_{\rm
ad}$ and $\nabla$ are, respectively, the adiabatic and actual
temperature gradients, $g$ is the gravity, and $H_p$ is the pressure
scale height.  In the classical case the critical Richardson number is
1/4, which we adopt for the current work.  The instability is
triggered if $D_{\rm SS} > \nu Re_{\rm c}/3$, where $\nu$ is the
viscosity and $Re_{\rm c}$ is the critical Reynolds number; we take a
value of 1000 from laboratory experiments on Couette flow.

For the GSF instability we use the expression of \citet{kip80}

\begin{equation}
D_{\rm GSF} = \frac{K_{0}}{2c_{p}\rho(\nabla_{\rm ad} - \nabla)}\frac{\tilde{g}}{g}\frac{H_{p}}{H_{\Omega}},
\end{equation}

\noindent where $K_{0} = 4acT^3/3\kappa\rho$, $c_{p}$ is the heat
capacity at constant pressure, $\tilde{g}/g$ is the departure from
spherical symmetry, and $H_{\Omega}$ is the angular velocity scale
height.  To determine $\tilde{g}/g$ both here and in the expression
for meridional circulation below, we use equation (3.23) from
\citet{zahn92} (an expression that includes the quadrupole term in the
potential).

Our diffusion coefficient for meridional circulation has the form
$D_{\rm eff} = f_{\rm c}rU(r)$, where $f_{\rm c}$ is an efficiency
factor for chemical mixing relative to angular momentum transport and
the velocity $U(r)$ is taken, with some additional approximations,
from equation (4.38) of \citet{mae98} (which updates equation 3.39 of
\citealt{zahn92}).

The \citet{mae98} treatment for meridional circulation derives driving
terms from variations in gravity, density, temperature, and mean
molecular weight on isobars.  Strong horizontal turbulence is invoked
to turn a two dimensional velocity field into a one dimensional
diffusion equation for chemical mixing and a one dimensional diffusion
$+$ advection equation for angular momentum transport.  We make three
simplifying assumptions relative to the extremely complex prescription
therein.  First, we retain only terms of order $\Omega$ and
$d\Omega/dr$; second, we neglect the inhibiting or enhancing effect of
$\mu$ gradients; and third, we assume that the efficiency factor
$f_{\rm c}$ for chemical mixing relative to angular momentum transport
is unity.  The latter two assumptions are consistent with our maximal
mixing approach, namely considering the maximum degree of mixing
consistent with a given angular velocity distribution.  The higher
order terms in the \citet{mae98} treatment, in addition to being
extremely challenging numerically, duplicate the physical driving
mechanisms of the GSF instability that are already included in our
treatment and are thus neglected for the current purpose along with
the time-dependent driving terms from entropy variations from
\citet{mae98}.  Numerical experiments indicate that inclusion of these
higher order terms largely leads to cancellation effects, with a final
angular velocity comparable to that obtained with a simpler first
order treatment.

The expression we use is

\begin{equation}
U(r) = \frac{P}{\rho\,c_{p}T}\biggl(\frac{1}{\nabla_{\rm ad}-\nabla-\nabla_{\mu}}\biggr)\frac{\epsilon_m(r)}{g}\,E_{\Omega},
\end{equation}

\noindent where $\nabla_{\mu} = d\ln\mu/dP$, $\epsilon_m(r) =
L(r)/M(r)$, and $E_{\Omega}$ is given by a simplified version of
equation (4.42) of \citet{mae98},

\begin{equation}
E_{\Omega} = 2\biggl(1-\frac{\Omega^2}{2\pi\,G\rho}-\frac{\epsilon}{\epsilon_m}\biggr)\frac{\tilde{g}}{g} - \frac{3M}{4\pi\rho r^3}\biggl(\frac{2}{3}-\frac{2H_{\rm T}}{r}\biggr)\Theta - \frac{\epsilon}{\epsilon_m}\biggl(\frac{\epsilon_{\ast}}{\epsilon}\frac{d\epsilon_{\ast}}{dT} - \frac{dK_{0}}{dT} + 1 - \frac{\epsilon_{\ast}}{\epsilon}\biggr)\Theta,
\end{equation}

\noindent In this last expression, the specific rate $\epsilon$ refers
to the sum of nuclear and gravitational energies, while
$\epsilon_{\ast}$ includes the nuclear energy only.  Finally, $H_{\rm
T}$ is the temperature scale height, and the factor $\Theta$
represents density fluctuations and is given by,

\begin{equation}
\Theta = \frac{\tilde{\rho}}{\rho} = \frac{1}{3}\frac{r^2}{g}\frac{d(\Omega^2)}{dr}.
\end{equation}

\end{appendix}

\begin{deluxetable}{ccccccrrrrrrrrrr}
\footnotesize \tablecaption{Properties of the starting models (MS
turnoff).  \label{tbl-1}} \tablewidth{0pt} \tablehead{
\colhead{M/M$_{\odot}$} & \colhead{[Fe/H]} & \colhead{ age} &
\colhead{log(L/L$_{\odot}$)} & \colhead{$M_{\rm V}$} & \colhead{ R/R$
_{\odot}$} \\ & \colhead{} & \colhead{(Gyr)} } 
\startdata 
0.8 &-1.0 &11.2 &0.184  &4.41 &1.088 \nl 
0.8 &-2.3 &10.5 &0.484  &3.88 &1.135 \nl
0.85 &-1.0 &9.00 &0.275 &4.18 &1.154 \nl 
0.85 &-2.3 &8.73 &0.483 &3.68 &1.147 \nl 
0.9 &-0.3 &10.3 &0.108  &4.58 &1.103 \nl 
0.9 &-0.7 &8.07 &0.269  &4.27 &1.171 \nl 
0.9 &-1.4 &7.04 &0.456  &3.76 &1.184 \nl 
1.0 & 0.0 &7.96 &0.130  &4.58 &1.137 \nl 
1.1 & 0.0 &5.95 &0.304  &4.13 &1.322 \nl 
1.2 & 0.0 &4.58 &0.517  &3.58 &1.614 \nl 
1.3 & 0.0 &3.46 &0.595  &3.38 &1.690 \nl 
1.4 & 0.0 &2.75 &0.730  &2.90 &1.937 \nl 
1.5 & 0.0 &2.22 &0.841  &2.82 &2.134 \nl 
1.6 & 0.0 &1.79 &0.951  &2.46 &2.277 \nl
1.8 & 0.0 &1.28 &1.164  &1.91 &2.691 \nl 
2.0 & 0.0 &0.95 &1.394  &1.30 &2.949 \nl


\enddata


\tablenotetext{a}{Numbers at X$_{\rm core} \approx$ 0.01 (``turnoff'').}
\tablenotetext{b}{Starting models are standard, i.e., do not include rotation.}

\end{deluxetable}


\clearpage

\clearpage

\clearpage

\clearpage

\clearpage

\clearpage

\clearpage

\clearpage

\clearpage

\clearpage

\clearpage

\clearpage

\clearpage

\clearpage

\clearpage

\clearpage

\clearpage

\clearpage

\clearpage

\clearpage

\clearpage

\clearpage

\end{document}